\documentclass[reprint,superscriptaddress,nofootinbib,amsmath,amssymb,aps,pra,longbibliography,floatfix]{revtex4-2}
\pdfoutput=1
\usepackage[utf8]{inputenc}

\usepackage[scaled=0.95]{helvet}
\usepackage[unicode=true,bookmarks=true,bookmarksnumbered=false,bookmarksopen=false,breaklinks=false,backref=false,hidelinks]{hyperref}

\usepackage{xcolor}
\definecolor{myred}{rgb}{0.8,0.3,0.2}
\definecolor{myredfill}{rgb}{1,0.93,0.93}

\definecolor{myred2}{rgb}{0.6,0.5,0.3}
\definecolor{myredfill2}{rgb}{0.93,0.83,0.83}

\definecolor{myblue}{rgb}{0.05,0.25,0.7}
\definecolor{mybluefill}{rgb}{0.92,0.98,1}
\definecolor{mygreen}{rgb}{0.3,0.6,0.4}
\definecolor{mygreenfill}{rgb}{0.93,1,0.97}
\definecolor{mygrey}{rgb}{0.4,0.4,0.4}
\definecolor{mygreyfill}{rgb}{0.95,0.95,0.95}
\definecolor{mypurple}{rgb}{0.6,0.2,0.6}
\definecolor{mypurplefill}{rgb}{0.99,0.94,0.99}
\definecolor{myyellow}{rgb}{0.9,0.9,0.3}
\definecolor{myyellowfill}{rgb}{1,1,0.92}
\definecolor{myorange}{rgb}{0.9,0.7,0.3}
\definecolor{myorangefill}{rgb}{1,0.95,0.92}
\definecolor{mybrown}{rgb}{0.4,0.25,0.1}
\definecolor{mybrownfill}{rgb}{1,0.9,0.8}

\definecolor{mediumgrey}{rgb}{0.7,0.7,0.7}

\definecolor{linkred}{rgb}{0.6,0.1,0.1}
\definecolor{citeblue}{rgb}{0.2,0.35,0.75}
\definecolor{urlblue}{rgb}{0.2,0.25,0.45}

\hypersetup{hidelinks}

\usepackage{bbm}
\usepackage{stmaryrd}
\usepackage{tikz-cd}
\usetikzlibrary{decorations.pathreplacing,calligraphy}
\usepackage{amsthm}
\usepackage{mathrsfs}
\usepackage{hhline}
\usetikzlibrary{backgrounds}
\usepackage{graphicx}
\usepackage{braket}
\usepackage[linesnumbered,ruled,vlined]{algorithm2e}
\usepackage{mathtools}
\usetikzlibrary{decorations.pathreplacing,angles,quotes}

\newtheorem{theorem}{Theorem}
\newtheorem{definition}{Definition}
\newtheorem{lemma}{Lemma}
\newtheorem*{lemma*}{Lemma}
\newtheorem{corollary}{Corollary}

\newtheorem{proposition}{Proposition}

\begin{document}

\title{Isolating Cause in a Quantum World}

\author{Graeme D. Berk}
\email{graemedean.berk@ntu.edu.sg}
\affiliation{\mbox{Nanyang Quantum Hub, School of Physical and Mathematical Sciences, Nanyang Technological University.}}
\affiliation{\mbox{Centre for Quantum Technologies, Nanyang Technological University.}}
\author{Jayne Thompson}
\email{thompson.jayne2@gmail.com}
\affiliation{\mbox{School of Computing and Data Sciences, Nanyang Technological University, Singapore 639673.}}
\affiliation{\mbox{Centre for Quantum Technologies, Nanyang Technological University.}}
\author{Mile Gu}
\email{mgu@quantumcomplexity.org}
\affiliation{\mbox{Nanyang Quantum Hub, School of Physical and Mathematical Sciences, Nanyang Technological University.}}
\affiliation{\mbox{Centre for Quantum Technologies, Nanyang Technological University.}}
\affiliation{\mbox{CNRS-UNS-NUS-NTU International Joint Research Unit, UMI 3654, Singapore 117543.}}

\date{\today}

\begin{abstract} 
The task of inferring causes from a sequence of observations has a long history in quantitative science. By isolating which information from the past is required to replicate the future, we gain insight into potential underlying causal structure. In complexity science, this has motivated the identification of oracular information---information stored in a model's memory that is relevant to its output, yet not contained in the past---and causal models, which contain no oracular information and precisely capture the potential causes of future events. The world, however, is fundamentally quantum. How do these concepts generalise to quantum processes, where observation is invasive and can take many forms? Here, we provide a means to identify oracular information in the quantum regime by introducing \emph{per-realisation faithfulness}, a steering-like condition in which no party with access to a purification of the model's memory can perform recorded measurements that modify the forecasts of the model. Next, we draw on ideas from $\epsilon$-machines, the provably memory-minimal causal models of classical processes, to develop a lower bound on the memory dimension of statistically faithful models of a quantum process. We provide conditions that guarantee saturation of this bound, and hence minimality.
\end{abstract}

\maketitle

\section{Introduction} 
Motivated by both resource constraints~\cite{fiderer2026informationthermodynamics} and Occam's razor, we ask: how much information from the past do we need to explain the future? In complexity science, the field of computational mechanics studies this question in the language of classical stochastic processes~\cite{inferringstatisticalcomplexity,computationalmechanics,crutchfield2001regularitiesunseenrandomnessobserved,analgorithmforpatterndiscovery,synchronizationandcontrolinintrinsic,predictionandgeneration}. There, a model maps information from the past into a memory system in the present,
such that subsequent operations on that memory replicate the conditional future behaviour with perfect statistical fidelity. The idea of minimal cause can then be formalised by introducing oracular information~\cite{synchronizationandcontrolinintrinsic,PhysRevX.8.031013}, information the memory system contains about the future beyond what past observations provide. Occam's preference for minimal cause manifests as a search for the memory-minimal model among causal models---models that contain no oracular information. 

The resulting model---known as the $\epsilon$-machine---is the provably minimal classical causal model of classical processes~\cite{computationalmechanics}. These $\epsilon$-machines are built on the principle of \emph{causal equivalence}: if two observed histories yield the same conditional future statistics, the model assigns them to the same memory state. Such models were also shown to asymptotically synchronise, whereby an agent's uncertainty about the model's memory approaches $0$ as they learn progressively more about the past~\cite{asymptoticsynchronization}. This provides a connection between the state of an $\epsilon$-machine and the belief states of an agent who attempts to infer future behaviour from the past. These developments have provided insights into dynamical systems, from discerning between anaesthetised and wakeful states in neurological signals~\cite{generalanesthesia} to interpreting crystallographic and geomagnetic data~\cite{VARN201547,PhysRevE.67.016203}, to characterising chaos~\cite{inferringstatisticalcomplexity} and asymmetries between cause and effect~\cite{crutchfield2009time,PhysRevX.8.031013}.

But what about the fundamental limits on modelling \emph{quantum} processes on quantum computers? Modelling quantum systems is both technologically important~\cite{Reiher_2017,Bauer_2020} and among the most promising areas for achieving large quantum advantages~\cite{Daley2022,PRXQuantum.2.010103}. Yet translating these ideas to the quantum domain requires significant conceptual advances. Unlike a classical process in which there is only one way to observe output random variables at each time-step, a quantum process can be observed with respect to a continuum of observables. Moreover, the choice of what to observe, or indeed, whether an observation takes place at all, fundamentally changes future behaviour. Classical causal models of such processes thus struggle to capture the full picture, limiting themselves mostly to predicting statistics for a fixed choice of observations~\cite{venegas2020measurement,suen2022surveying}. Going beyond this would certainly require quantum analogues of such causal models. Yet current work is limited primarily to quantum models of classical processes~\cite{quantummechanicscanreducethecomplexity,dimensionalquantummemoryadvantage,provablyunboundedmemoryadvantage,extremequantumadvantage,extremedimensionality}; a recent exception studies the special case of input-free sequences of correlated quantum states~\cite{elliott2026causalmemorystructuralcomplexity}.

Here, we use the process tensor formalism~\cite{nonmarkovianquantumprocesses,quantumstochasticprocesses,anintroductiontooperationalquantumdynamics}, which captures the operational behaviour of a non-Markovian quantum process under any possible choice of experimental interventions. These objects, also known as quantum combs~\cite{theoreticalframeworkforquantumnetworks} or quantum strategies~\cite{Gutoski_2007}, describe quantum circuit fragments~\cite{xiao2023uncertainty,ried2015causal} and quantum agents~\cite{dunjko2016quantumenhanced}, and are seeing growing use in diverse settings where non-Markovianity is relevant, from memory effects in error correction and adaptive sensing to diagnosis of quantum chaos~\cite{tanggara2024strategiccodeunifiedspatiotemporal,PhysRevLett.123.110501,fgc4-hgk1}. Our goal is to develop a general mathematical framework for causal modelling of such processes: to describe quantum models that, when correctly initialised, can output a conditional quantum state that captures the conditional future statistics of the target quantum process irrespective of how it is to be measured in the future. 

\begin{figure}[t]

\centering
\begin{tikzpicture}[scale=0.65,
    modelheader/.style={anchor=west,inner sep=0pt,
        font=\fontfamily{phv}\bfseries\fontsize{10.5}{12}\selectfont}]

\draw[myred,fill=myredfill,ultra thick,solid,rounded corners=5] (-1,-8) rectangle (12,6);

\draw[myblue,fill=mybluefill,ultra thick,solid,rounded corners=5] (-0.75,-7.75) rectangle (11.75,0.4);

\draw[mygreen,fill=mygreenfill,ultra thick,solid,rounded corners=5] (-0.5,-7.5) rectangle (11.5,-4.7);

\draw[mypurple,fill=myredfill, thick,dashed,rounded corners=4] (9,1) -- (11.1,1) -- (11.1,5.5) -- (6.9,5.5) -- (6.9,3.5) -- (8.6,3.5) --(8.6,1) -- (10,1);

\node[modelheader] at (-0.15,5.1) {Faithful (F)};

\draw[] (0,3) node {$\overleftarrow{x}$};
\draw[->, black, very thick,solid] (1,3) -- (4,3);
\draw[->, black, very thick,solid] (1,3) -- (4,1.5);
\draw[] (5-2.5,3+0.5) node {\footnotesize $p(\mu_1|\overleftarrow{x})$};
\draw[] (5,3) node {$\mu_1$};
\draw[] (5-2.8,3+0.5-2+0.4) node [rotate=-30] {\footnotesize $p(\mu_2|\overleftarrow{x})$};
\draw[] (5,1.5) node {$\mu_2$};
\draw[->, black, very thick,solid] (6,3) -- (8.5,3);
\draw[->, black, very thick,solid] (6,1.5) -- (8.5,1.5);
\draw[] (10,3) node {$p(\overrightarrow{X}|\mu_1)$};
\draw[] (10,1.5) node {$p(\overrightarrow{X}|\mu_2)$};

\draw[] (9,5) node {\footnotesize $\sum_i p(\overrightarrow{X}|\mu_i)p(\mu_i|\overleftarrow{x})$};
\draw[] (9,4) node {\footnotesize $=p(\overrightarrow{X}|\overleftarrow{x})$};


\node[modelheader] at (-0.15,-0.5) {Non-Oracular Faithful (nOF)};
\draw[] (0,3-5) node {$\overleftarrow{x}$};
\draw[->, black, very thick,solid] (1,3-5) -- (4,3-5);
\draw[->, black, very thick,solid] (1,3-5) -- (4,1.5-5);
\draw[] (5-2.5,3+0.5-5) node {\footnotesize $p(\mu_1|\overleftarrow{x})$};
\draw[] (5,3-5) node {$\mu_1$};
\draw[] (5-2.8,3+0.5-2+0.4-5) node [rotate=-30] {\footnotesize $p(\mu_2|\overleftarrow{x})$};
\draw[] (5,1.5-5) node {$\mu_2$};
\draw[->, black, very thick,solid] (6,3-5) -- (8.5,3-5);
\draw[->, black, very thick,solid] (6,1.5-5) -- (8.5,2.5-5);
\draw[] (10,3-5) node {$p(\overrightarrow{X}|\overleftarrow{x})$};


\node[modelheader] at (-0.15,-5.5) {Synchronising Faithful (SF)};
\draw[] (0,3-5-5) node {$\overleftarrow{x}$};
\draw[->, black, very thick,solid] (1,3-5-5) -- (4,3-5-5);
\draw[] (5,3-5-5) node {$\mu_1$};
\draw[->, black, very thick,solid] (6,3-5-5) -- (8.5,3-5-5);
\draw[] (10,3-5-5) node {$p(\overrightarrow{X}|\overleftarrow{x})$};


\draw[] (0,-10+1.5) node {Past};
\draw[] (5,-10+1.5) node {Memory};
\draw[] (10,-10+1.5) node {Future};

\draw[] (2.5,-10+1.5) node {$\tilde{e}_{\text{cl}}$};
\draw[] (7.5,-10+1.5) node {$\Phi_{\text{cl}}$};

\end{tikzpicture}

    \caption{{\bf Model classes.} A sequence model is generally a representation of a process that stores information about the past in a memory system, such that subsequent operations on the memory generate future outputs. A model is \emph{faithful} ($\texttt{F}$, red) if it reproduces the process's conditional future statistics after averaging over its possible memory states. In the classical setting illustrated here, the encoding map $\tilde{e}_{\text{cl}}$ assigns a past sequence $\overleftarrow{x}$ to a memory random variable $M$, taking values $\mu_i$ with probabilities $p(\mu_i|\overleftarrow{x})$. The prediction function $\Phi_{\text{cl}}$ maps each memory state $\mu_i$ to a probability distribution $p(\overrightarrow{X}|\mu_i)$ for the future sequence $\overrightarrow{X}$. Faithfulness requires $\sum_i p(\overrightarrow{X}|\mu_i)p(\mu_i|\overleftarrow{x})=p(\overrightarrow{X}|\overleftarrow{x})$. A model is \emph{non-oracular faithful} ($\texttt{nOF}$, blue) if it is faithful and its memory contains no information about its future output beyond that supplied by the past: every memory state reachable from the same past yields the same correct future distribution. A model is \emph{synchronising faithful} ($\texttt{SF}$, green) if it is faithful and the past determines a unique memory state, and hence a unique future distribution. The nested regions show $\texttt{SF}\subset\texttt{nOF}\subset\texttt{F}$; a unique future distribution need not imply a deterministic future outcome. These classical diagrams motivate the quantum model classes studied here. In the quantum setting, mixed memory states admit non-unique ensemble decompositions, and non-oracularity is characterised operationally by the inability of measurements on a purification of the memory to steer the model's conditional future forecasts.}
    \label{fig:three_model_classes}
\end{figure}
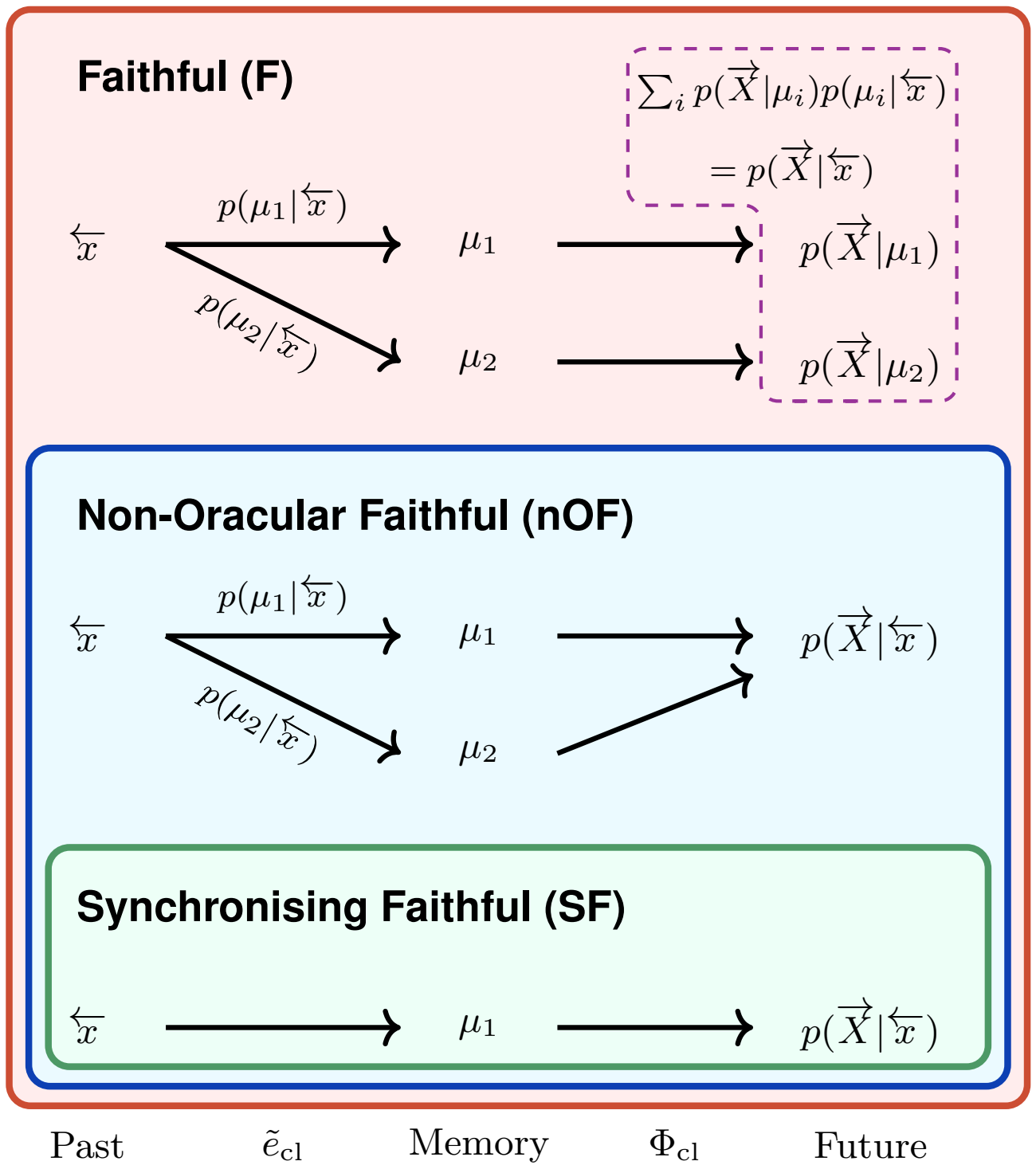

Our contributions include identifying how concepts from computational mechanics---oracular information, synchronisation, and causal equivalence---translate or change in the quantum regime. This allows us to define a hierarchy of quantum model classes beyond statistical faithfulness (see Fig.~\ref{fig:three_model_classes}), providing conditions for quantum analogues of causal models that contain no oracular information and synchronising models whose conditional memory states asymptotically become pure under a specified interaction policy. We establish the equivalence between non-oracularity and per-realisation faithfulness---the latter of which we characterise as the inability of recorded measurements on a purifying subsystem of the memory to steer the model's conditional future predictions. Along the way, we examine how quantum superpositions affect causal equivalence. We then derive the lower bound $\sqrt{D} \leq d$ on the minimal memory Hilbert space dimension $d$ of any faithful quantum model of a stationary process, where $D$ is the dimension of the real linear span of its conditional future process tensors under all possible histories. Finally, we identify conditions under which this bound can be saturated. 

The manuscript is organised as follows: Sec.~\ref{sec:quantumcomputationalmechanics} introduces our framework for quantum processes, and Sec.~\ref{sec:quantummodels} describes models of those processes. We introduce our model class hierarchy in Sec.~\ref{sec:quantumprediction} and establish the correspondence of Thm.~\ref{thm:predictioncorrespondence}. Sec.~\ref{sec:fundamentallimitations} provides our quantum generalisation of causal equivalence, our memory bound of Thm.~\ref{thm:lbfm}, the minimality theorem of Thm.~\ref{thm:probdistinct}, and a simple numerical experiment.

\section{Framework} \label{sec:quantumcomputationalmechanics}
We begin with a review of quantum processes and some of their associated concepts useful for generalising computational mechanics to the quantum domain. We recommend that time-relaxed readers begin by reviewing our notes on classical computational mechanics in App.~\ref{app:classicalcomputationalmechanics}, although this is not required to understand the results of this work. Doing so will provide a gentler introduction, and a reference point for our quantum generalisation.

\subsection{Quantum Processes} \label{sec:physicalscenario} 
Consider the scenario where an experimenter acts on the accessible part of an open system $s$ with a Hilbert space $\mathcal{H}^{s}$ of dimension $d_s$, at a set of \emph{episodes} $t-n' ,\dots,t,\dots,t+n-1$; there are $n'$ episodes of the past ($<t$) and $n$ episodes of the future ($\geq t$) about some arbitrary cut $t$. We work in the regime of arbitrarily long but finite time windows unless taking an explicit limit. We assume the existence of an underlying quantum process $\mathbf{T}$ consistently describing the observable behaviour for all such time windows.

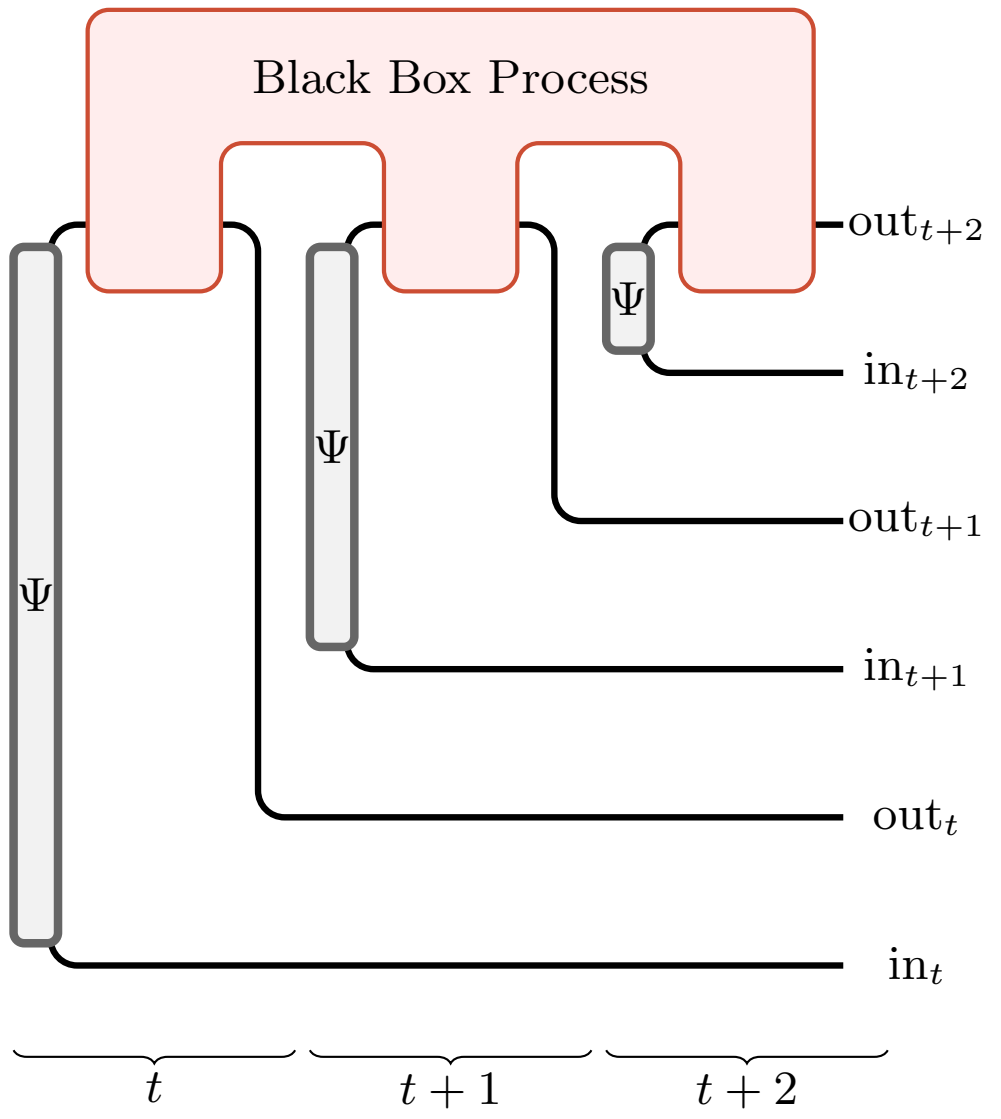
\begin{figure}[tbp]

\centering
\begin{tikzpicture}[scale=0.49]

\draw[black, very thick,solid] (9.7,1) -- (10.3,1);


\draw[black, very thick,solid,rounded corners=5] (4.3-4,1) -- (4.1-0.5-4,1) -- (4.1-0.5-4,1-6-4) -- (10.3,1-6-4);
\draw[black, very thick,solid,rounded corners=5] (5.7-4,1) -- (5.9+0.5-4,1) -- (5.9+0.5-4,1-4-4) -- (10.3,1-4-4);

\draw[black, very thick,solid,rounded corners=5] (4.3,1) -- (4.1-0.5,1) -- (4.1-0.5,1-6) -- (10.3,1-6);
\draw[black, very thick,solid,rounded corners=5] (5.7,1) -- (5.9+0.5,1) -- (5.9+0.5,1-4) -- (10.3,1-4);

\draw[black, very thick,solid,rounded corners=5] (4.3+4,1) -- (4.1-0.5+4,1) -- (4.1-0.5+4,1-2) -- (10.3,1-2);



\draw[mygrey,fill=mygreyfill,ultra thick,solid,rounded corners=2] (-1.7+0.8,3.7-3) rectangle (-0.3,2.3-11);
\draw[mygrey,fill=mygreyfill,ultra thick,solid,rounded corners=2] (-1.7+4+0.8,3.7-3) rectangle (-0.3+4,2.3-7);
\draw[mygrey,fill=mygreyfill,ultra thick,solid,rounded corners=2] (-1.7+0.8+8,3.7-3) rectangle (-0.3+8,2.3-3);





\draw[decoration={brace,mirror,raise=2pt},decorate]
  (-0.9+4-4,-6-4) -- node[below=3pt] {$t$} (2.9+4-4,-6-4);
\draw[decoration={brace,mirror,raise=2pt},decorate]
  (-0.9+8-4,-6-4) -- node[below=3pt] {$t+1$} (2.9+8-4,-6-4);
  \draw[decoration={brace,mirror,raise=2pt},decorate]
  (-0.9+8,-6-4) -- node[below=3pt] {$t+2$} (2.9+8,-6-4);








\draw[] (-0.6,-4) node {\small $\Psi$};
\draw[] (3.4,-4+2) node {\small $\Psi$};
\draw[] (7.4,-4+2+2) node {\small $\Psi$};


\draw[] (11.3,1) node {\small $\text{out}_{t+2}$};
\draw[] (11.3,1-2) node {\small $\text{in}_{t+2}$};

\draw[] (11.3,1-4) node {\small $\text{out}_{t+1}$};
\draw[] (11.3,1-2-4) node {\small $\text{in}_{t+1}$};

\draw[] (11.3,1-4-4) node {\small $\text{out}_{t}$};
\draw[] (11.3,1-2-4-4) node {\small $\text{in}_{t}$};


\draw[myred,fill=myredfill, thick,solid,rounded corners=4] (6.5,2.1) -- (5.9,2.1) -- (5.9,0.1)  -- (4.1,0.1) -- (4.1,2.1) -- (1.9,2.1) -- (1.9,0.1) -- (0.1,0.1) -- (0.1,3.9)  -- (9.9,3.9) -- (9.9,0.1) -- (8.1,0.1) -- (8.1,2.1) -- (6.5,2.1) ;

\draw[] (5,3) node [rotate=0] {Black Box Process};

\end{tikzpicture}

    \caption{{\bf The Choi state of a three-step process tensor.} For each episode of evolution $t,t+1,t+2$, one half of a maximally entangled pair $\Psi$ on two copies of $\mathcal{H}^s$ is sent into a process $\mathbf{T}$. The resultant multipartite state is the Choi state of $\mathbf{T}$. This Choi state can be characterised by multitime process tomography, independently of any model, as in Eq.~\eqref{eq:infinitetomographic}. However, once a model is specified, the Choi state of a process can be written as Eq.~\eqref{eq:quantumprocessdef}.}
    \label{fig:process_choi_state}
\end{figure}

Within a given window, the observable behaviour can be fully characterised, via the Choi isomorphism~\cite{anintroductiontooperationalquantumdynamics}, by a density matrix acting on $2n+2n'$ copies of $\mathcal{H}^{s}$---one for each input and one for each output at every episode. This state is obtained by taking a maximally entangled state $\Psi$ on two copies of $s$, and sending one half into the process $\mathbf{T}$ at each episode, while leaving the other half untouched (see Fig.~\ref{fig:process_choi_state}). The resultant state is the $n+n'$ time Choi state $\mathbf{T}_{t-n':t+n-1} \in \mathcal{D}\big( \mathcal{H}^{\text{in}_{t-n'}} \otimes \mathcal{H}^{\text{out}_{t-n'}} \otimes \dots \otimes \mathcal{H}^{\text{in}_{t+n-1}} \otimes \mathcal{H}^{\text{out}_{t+n-1}}\big)$ of $\mathbf{T}$ in this finite time window. Note that we use the convention of unit-normalised Choi states throughout this paper for both processes and policies.

\subsection{Policies}
Quantum observation is fundamentally non-passive. No probability distribution over outcomes exists until one specifies how the process will be probed. If one treats the two as a single joint object, the result is quantum models of a classical process, which is a separate and far more mature area~\cite{quantummechanicscanreducethecomplexity,srinivasan2017learninghiddenquantummarkov,quantumpartiallyobservable,expressivenessandlearning,provablyunboundedmemoryadvantage,Thompson2017,experimentallymodelling,extremequantumadvantage,extremedimensionality,dimensionalquantummemoryadvantage} compared to what we seek to study in this work---quantum models of quantum processes. 

The experimenter's side of the interaction is their policy $\mathbf{A}=\{ \mathbf{A}_{\overleftrightarrow{z}} \}_{\overleftrightarrow{z} \in \text{W}(\mathsf{Z}) }$, which is a set of instructions for actions to be taken---potentially including state preparations, quantum channels, and measurements. Here, $\text{W}(\mathsf{Z})$ denotes all length $n+n'$ words from the outcome alphabet $\mathsf{Z}$, and a specific action-outcome sequence $\mathbf{A}_{\overleftrightarrow{z}}$ is indexed by the trajectory of outcomes (word) obtained $\overleftrightarrow{z}=z_{t-n'},\dots ,z_{t},\dots,z_{t+n-1}$. In full quantum generality, the label $\overleftrightarrow{z}$ denoting experimental outcomes may not be well-defined as input-output pairs ($x_{\tau},y_{\tau}$), e.g. the case of Bell basis measurements on episodes of the Choi state. For this reason, we use a single joint label $z_{\tau}$ to denote both jointly.

\begin{figure}[tbp]

\centering
\begin{minipage}{\linewidth}

\resizebox{\linewidth}{!}{%
\begin{tikzpicture}[scale=0.5]

\draw[myblue,fill=mybluefill,thick,solid,rounded corners=4] (6.5,1.9) --  (6.1,1.9) -- (6.1,-0.1)  -- (3.9,-0.1) -- (3.9,1.9) -- (2.1,1.9) --  (2.1,-0.1) --  (-0.1,-0.1) --  (-0.1,1.9) --  (-1.9,1.9) -- (-1.9,-1.9) --  (0.1,-1.9) -- (15.5,-1.9) -- (15.5,0.1) -- (15.9,0.1) -- (15.9,1.9) -- (14.1,1.9) -- (14.1,-0.1) -- (11.9,-0.1) -- (11.9,1.9) -- (10.1,1.9) -- (10.1,-0.1) -- (7.9,-0.1) -- (7.9,1.9) -- (6.5,1.9);

\draw[black, very thick,solid] (-0.5,1) -- (0.3,1);
\draw[black, very thick,solid] (13.7,3) -- (14.5,3);
\draw[black, very thick,solid] (13.7,1) -- (14.5,1);
\draw[black, very thick,solid] (-2,3) -- (4.3,3);
\draw[black, very thick,solid] (9.7,3) -- (12.3,3);
\draw[black, very thick,solid] (5.7,3) -- (6.3,3);
\draw[black, very thick,solid] (1.7,1) -- (2.3,1);
\draw[black, very thick,solid] (3.7,1) -- (4.3,1);
\draw[black, very thick,solid] (9.7,1) -- (10.3,1);
\draw[black, very thick,solid] (11.7,1) -- (14.5,1);
\draw[black, very thick,solid] (6.3,3) -- (7.7,3);
\draw[black, very thick,solid] (7.7,3) -- (8.3,3);
\draw[black, very thick,solid] (5.7,1) -- (6.3,1);
\draw[black, very thick,solid] (7.7,1) -- (8.3,1);
\draw[black, very thick,solid] (-0.3,-1) -- (2.3,-1);
\draw[black, very thick,solid] (3.7,-1) -- (6.3,-1);
\draw[black, very thick,solid] (7.7,-1) -- (10.3,-1);
\draw[black, very thick,solid] (11.7,-1) -- (14.5,-1);



\draw[myred,fill=myredfill, thick,solid,rounded corners=4] (6.5,2.1) -- (5.9,2.1) -- (5.9,0.1)  -- (4.1,0.1) -- (4.1,2.1) -- (1.9,2.1) -- (1.9,0.1) --  (0.1,0.1) -- (0.1,2.1) --    (-1.5,4.1-4+2) --  (-1.5,3.9) -- (15.5,3.9) -- (15.5,2.1) -- (13.9,2.1) -- (13.9,0.1) -- (12.1,0.1) -- (12.1,2.1) -- (9.9,2.1) -- (9.9,0.1) -- (8.1,0.1) -- (8.1,2.1) -- (6.5,2.1) ;


\draw[black,thick, densely dotted] (6.8,0) -- (7.2,0);
\draw[myblue,fill=mybluefill,very thick,solid,rounded corners=2] (2.3+4,1.7) rectangle (3.7+4-0.9,-1.7);
\draw[myblue,fill=mybluefill,very thick,solid,rounded corners=2] (2.3+4+0.9,1.7) rectangle (3.7+4,-1.7);

\draw[black,thick, densely dotted] (6.8-4,0) -- (7.2-4,0);
\draw[myblue,fill=mybluefill,very thick,solid,rounded corners=2] (2.3+4-4,1.7) rectangle (3.7+4-0.9-4,-1.7);
\draw[myblue,fill=mybluefill,very thick,solid,rounded corners=2] (2.3+4+0.9-4,1.7) rectangle (3.7+4-4,-1.7);

\draw[black,thick, densely dotted] (6.8-8,0) -- (7.2-8,0);
\draw[myblue,fill=mybluefill,very thick,solid,rounded corners=2] (2.3+4-8,1.7) rectangle (3.7+4-0.9-8,-1.7);
\draw[myblue,fill=mybluefill,very thick,solid,rounded corners=2] (2.3+4+0.9-8,1.7) rectangle (3.7+4-8,-1.7);

\draw[black,thick, densely dotted] (6.8+4,0) -- (7.2+4,0);
\draw[myblue,fill=mybluefill,very thick,solid,rounded corners=2] (2.3+4+4,1.7) rectangle (3.7+4-0.9+4,-1.7);
\draw[myblue,fill=mybluefill,very thick,solid,rounded corners=2] (2.3+4+0.9+4,1.7) rectangle (3.7+4+4,-1.7);

\draw[black,thick, densely dotted] (6.8+8,0) -- (7.2+8,0);

\draw[myblue,fill=mybluefill,very thick,solid,rounded corners=2] (2.3+4+8,1.7) rectangle (3.7+4-0.9+8,-1.7);
\draw[myblue,fill=mybluefill,very thick,solid,rounded corners=2] (2.3+4+0.9+8,1.7) rectangle (3.7+4+8,-1.7);

\draw[] (7,3) node {\large Black Box Process $\mathbf{T}$};
\draw[] (7,5) node {\bf \large Observed Statistics};

\draw[decoration={brace,mirror,raise=2pt},decorate]
  (-0.9,-2) -- node[below=3pt] {$\mathcal{A}^{sa}_{z_{t-1}}$} (2.9,-2);
\draw[decoration={brace,mirror,raise=2pt},decorate]
  (-0.9+4,-2) -- node[below=3pt] {$\mathcal{A}^{sa}_{z_{t}}$} (2.9+4,-2);
  \draw[decoration={brace,mirror,raise=2pt},decorate]
  (-0.9+8,-2) -- node[below=3pt] {$\mathcal{A}^{sa}_{z_{t+1}}$} (2.9+8,-2);
  \draw[decoration={brace,mirror,raise=2pt},decorate]
  (-0.9+12,-2) -- node[below=3pt] {$\mathcal{A}^{sa}_{z_{t+2}}$} (2.9+12,-2);

\draw[white,fill=white] (-2,-2) rectangle (-1,4.2);

\draw[white,fill=white] (15,-2) rectangle (16,4.2);

\draw[] (-2,1) node {$s$};
\draw[] (-2,-1) node {$a$};

\end{tikzpicture}
}
\end{minipage}
\par\medskip
\begin{minipage}{\linewidth}
\resizebox{\linewidth}{!}{%
\begin{tikzpicture}[scale=0.5]

\draw[myblue,fill=mybluefill,thick,solid,rounded corners=4] (6.5,1.9) --  (6.1,1.9) -- (6.1,-0.1)  -- (3.9,-0.1) -- (3.9,1.9) -- (2.1,1.9) --  (2.1,-0.1) --  (-0.1,-0.1) --  (-0.1,1.9) --  (-1.9,1.9) -- (-1.9,-1.9) --  (0.1,-1.9) -- (15.5,-1.9) -- (15.5,0.1) -- (15.9,0.1) -- (15.9,1.9) -- (14.1,1.9) -- (14.1,-0.1) -- (11.9,-0.1) -- (11.9,1.9) -- (10.1,1.9) -- (10.1,-0.1) -- (7.9,-0.1) -- (7.9,1.9) -- (6.5,1.9);

\draw[black, very thick,solid] (-0.5,1) -- (0.3,1);
\draw[black, very thick,solid] (13.7,3) -- (14.5,3);
\draw[black, very thick,solid] (13.7,1) -- (14.5,1);
\draw[black, very thick,solid] (-2,3) -- (4.3,3);
\draw[black, very thick,solid] (9.7,3) -- (12.3,3);
\draw[black, very thick,solid] (5.7,3) -- (6.3,3);
\draw[black, very thick,solid] (1.7,1) -- (2.3,1);
\draw[black, very thick,solid] (3.7,1) -- (4.3,1);
\draw[black, very thick,solid] (9.7,1) -- (10.3,1);
\draw[black, very thick,solid] (11.7,1) -- (14.5,1);
\draw[black, very thick,solid] (6.3,3) -- (7.7,3);
\draw[black, very thick,solid] (7.7,3) -- (8.3,3);
\draw[black, very thick,solid] (5.7,1) -- (6.3,1);
\draw[black, very thick,solid] (7.7,1) -- (8.3,1);
\draw[black, very thick,solid] (-0.3,-1) -- (2.3,-1);
\draw[black, very thick,solid] (3.7,-1) -- (6.3,-1);
\draw[black, very thick,solid] (7.7,-1) -- (10.3,-1);
\draw[black, very thick,solid] (11.7,-1) -- (14.5,-1);

\draw[myred,fill=myredfill, thick,solid,rounded corners=4] (6.5,2.1) -- (5.9,2.1) -- (5.9,0.1)  -- (4.1,0.1) -- (4.1,2.1) -- (1.9,2.1) -- (1.9,0.1) --  (0.1,0.1) -- (0.1,2.1) --    (-1.5,4.1-4+2) --  (-1.5,3.9) -- (15.5,3.9) -- (15.5,2.1) -- (13.9,2.1) -- (13.9,0.1) -- (12.1,0.1) -- (12.1,2.1) -- (9.9,2.1) -- (9.9,0.1) -- (8.1,0.1) -- (8.1,2.1) -- (6.5,2.1) ;

\draw[black, very thick,solid] (-2,3) -- (16,3);

\draw[myred,fill=myredfill,ultra thick,solid,rounded corners=2] (0.3,3.7) rectangle (1.7,0.3);
\draw[myred,fill=myredfill,ultra thick,solid,rounded corners=2] (4.3,3.7) rectangle (5.7,0.3);
\draw[myred,fill=myredfill,ultra thick,solid,rounded corners=2] (8.3,3.7) rectangle (9.7,0.3);
\draw[myred,fill=myredfill,ultra thick,solid,rounded corners=2] (12.3,3.7) rectangle (13.7,0.3);



\draw[black,thick, densely dotted] (6.8,0) -- (7.2,0);
\draw[myblue,fill=mybluefill,very thick,solid,rounded corners=2] (2.3+4,1.7) rectangle (3.7+4-0.9,-1.7);
\draw[myblue,fill=mybluefill,very thick,solid,rounded corners=2] (2.3+4+0.9,1.7) rectangle (3.7+4,-1.7);

\draw[black,thick, densely dotted] (6.8-4,0) -- (7.2-4,0);
\draw[myblue,fill=mybluefill,very thick,solid,rounded corners=2] (2.3+4-4,1.7) rectangle (3.7+4-0.9-4,-1.7);
\draw[myblue,fill=mybluefill,very thick,solid,rounded corners=2] (2.3+4+0.9-4,1.7) rectangle (3.7+4-4,-1.7);

\draw[black,thick, densely dotted] (6.8-8,0) -- (7.2-8,0);
\draw[myblue,fill=mybluefill,very thick,solid,rounded corners=2] (2.3+4-8,1.7) rectangle (3.7+4-0.9-8,-1.7);
\draw[myblue,fill=mybluefill,very thick,solid,rounded corners=2] (2.3+4+0.9-8,1.7) rectangle (3.7+4-8,-1.7);

\draw[black,thick, densely dotted] (6.8+4,0) -- (7.2+4,0);
\draw[myblue,fill=mybluefill,very thick,solid,rounded corners=2] (2.3+4+4,1.7) rectangle (3.7+4-0.9+4,-1.7);
\draw[myblue,fill=mybluefill,very thick,solid,rounded corners=2] (2.3+4+0.9+4,1.7) rectangle (3.7+4+4,-1.7);

\draw[black,thick, densely dotted] (6.8+8,0) -- (7.2+8,0);

\draw[myblue,fill=mybluefill,very thick,solid,rounded corners=2] (2.3+4+8,1.7) rectangle (3.7+4-0.9+8,-1.7);
\draw[myblue,fill=mybluefill,very thick,solid,rounded corners=2] (2.3+4+0.9+8,1.7) rectangle (3.7+4+8,-1.7);

\draw[] (1,2) node[rotate=0] {$\mathcal{T}^{sm}$};
\draw[] (5,2) node [rotate=0] {$\mathcal{T}^{sm}$};
\draw[] (9,2) node[rotate=0] {$\mathcal{T}^{sm}$};
\draw[] (13,2) node[rotate=0] {$\mathcal{T}^{sm}$};


\draw[] (7,5) node {\bf \large Model};

\draw[decoration={brace,mirror,raise=2pt},decorate]
  (-0.9,-2) -- node[below=3pt] {$\mathcal{A}^{sa}_{z_{t-1}}$} (2.9,-2);
\draw[decoration={brace,mirror,raise=2pt},decorate]
  (-0.9+4,-2) -- node[below=3pt] {$\mathcal{A}^{sa}_{z_{t}}$} (2.9+4,-2);
  \draw[decoration={brace,mirror,raise=2pt},decorate]
  (-0.9+8,-2) -- node[below=3pt] {$\mathcal{A}^{sa}_{z_{t+1}}$} (2.9+8,-2);
  \draw[decoration={brace,mirror,raise=2pt},decorate]
  (-0.9+12,-2) -- node[below=3pt] {$\mathcal{A}^{sa}_{z_{t+2}}$} (2.9+12,-2);


\draw[] (-0.35,3.5) node {\footnotesize $\rho^{m}_{t-1}$};
\draw[] (3.65,3.5) node {\footnotesize $\rho^{m}_{t}$};
\draw[] (7.65,3.5) node {\footnotesize $\rho^{m}_{t+1}$};
\draw[] (11.65,3.5) node {\footnotesize $\rho^{m}_{t+2}$};

\draw[white,fill=white] (-2,-2) rectangle (-1,4.2);

\draw[white,fill=white] (15,-2) rectangle (16,4.2);

\draw[] (-2,3) node {$m$};
\draw[] (-2,1) node {$s$};
\draw[] (-2,-1) node {$a$};

\end{tikzpicture}
}
\end{minipage}

\caption{{\bf The observed statistics of a multitime process versus a model of it.} The top frame corresponds to the information available to an experimenter: they know the policy $\mathbf{A}$ that they employ (blue comb). The combined action of the experimenter for a single step is labelled by $\mathcal{A}^{sa}_{z_{\tau}}=  \mathcal{M}^{sa}_{z_{\tau}} *_a \mathcal{P}^{sa}_{\tau}$, where $*_a$ is a shorthand for the link product~\cite{theoreticalframeworkforquantumnetworks} connecting $a$. These are applied in sequence, yielding an overall action-outcome sequence $\mathbf{A}_{\overleftrightarrow{z}}$, and the policy $\mathbf{A}$ is the collection of all such sequences. Physically, applying $\mathbf{A}_{\overleftrightarrow{z}}$ to a quantum process $\mathbf{T}$ (red comb) corresponds to an experiment where at a single episode, the experimenter prepares a state, lets it evolve freely with the environment, then measures the result. In the bottom frame, a hypothesised model is constructed to explain the observed statistics, given any policy $\mathbf{A}$.}
    \label{fig:process_vs_model}
\end{figure}
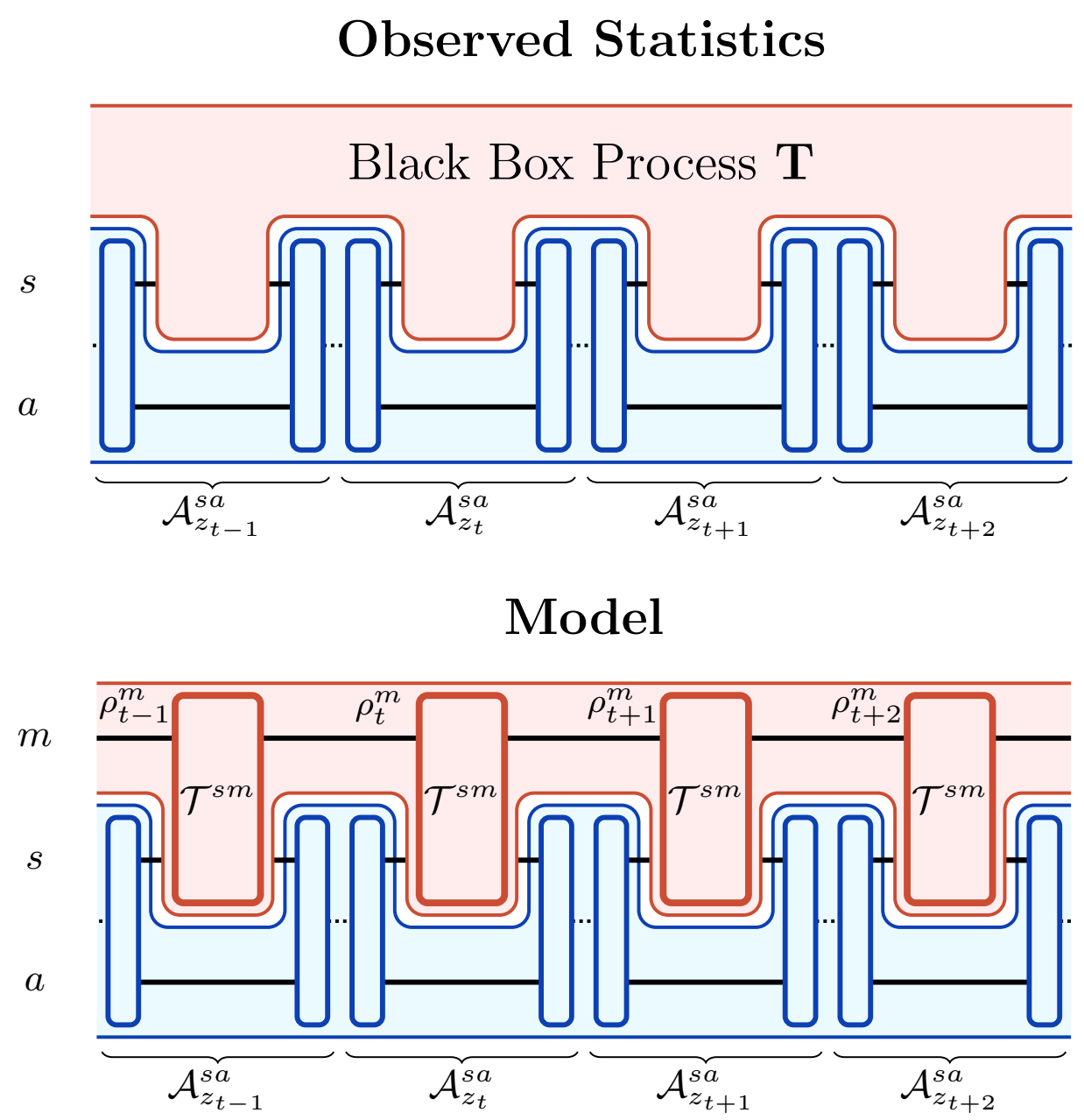

Policies mathematically correspond to testers~\cite{theoreticalframeworkforquantumnetworks,Chiribella_2012} (see App.~\ref{app:quantumpolicies} for technical details), which are the multitime generalisation of POVMs. Each $\mathbf{A}_{\overleftrightarrow{z}} \in \mathcal{L}\big( \mathcal{H}^{\text{in}_{t-n'}} \otimes \mathcal{H}^{\text{out}_{t-n'}} \otimes \dots \otimes \mathcal{H}^{\text{in}_{t+n-1}} \otimes \mathcal{H}^{\text{out}_{t+n-1}}\big)$ is represented by a positive semidefinite Hermitian operator via the Choi isomorphism. Policies satisfy a Born rule for any process $\mathbf{T}$, where both the process and policy elements are represented in Choi form,
\begin{equation} \label{eq:probabilityoftrajectory}
   p(\overleftrightarrow{z}|\mathbf{A}) 
   =  d_s^{2n}d_s^{2n'} \text{tr} \{ \mathbf{A}_{\overleftrightarrow{z}_{t-n':t+n-1}}^{\text{T}} \mathbf{T}_{{t-n':t+n-1}} \} .
\end{equation}
The prefactor of $d_s^{2n}d_s^{2n'}$ is caused by our convention of unit-normalised Choi states, where there is one input on $s$ for each time, for both the policy and the process. Note that testers satisfy a hierarchy of trace conditions similar to process tensors; see Fig.~\ref{fig:process_vs_model} for a graphical depiction of the process-policy interplay.

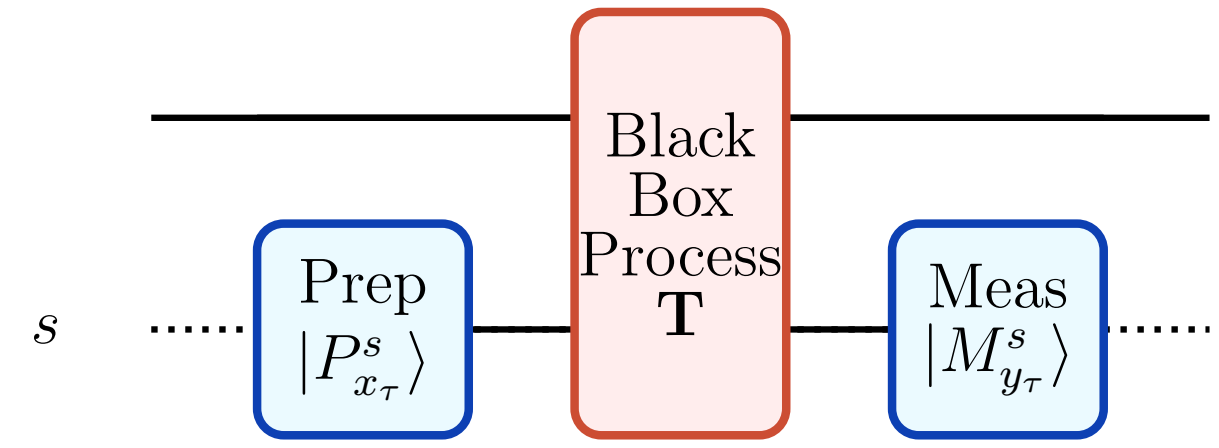
\begin{figure}[tbp]

\centering

\begin{tikzpicture}[scale=0.7]

\draw[black, very thick,solid] (0-2-1,7) -- (4+2+1,7);

\draw[black, very thick,dotted] (0-2-1,7-2) -- (4+2+1,7-2);


\draw[black, very thick,solid] (0-2,7) -- (4+2,7);

\draw[black, very thick,solid] (0,5) -- (4,5);


\draw[myred,fill=myredfill,ultra thick,solid,rounded corners=5] (1,4) rectangle (3,8);

\draw[myblue,fill=mybluefill,ultra thick,solid,rounded corners=5] (1-3,4) rectangle (3-3,8-2);

\draw[myblue,fill=mybluefill,ultra thick,solid,rounded corners=5] (1+3,4) rectangle (3+3,8-2);

\draw[] (2,6) node[rotate=0,black] {\large \shortstack{Black \\ Box \\ Process \\ $\mathbf{T}$}};


\draw[] (-1,5) node[rotate=0,black] {\large  \shortstack{Prep \\ $\ket{P^{s}_{x_{\tau}}}$}};

\draw[] (-1-3,5) node[rotate=0,black] {\large $s$};


\draw[] (-1+6,5) node[rotate=0,black] {\large \shortstack{Meas \\ $\ket{M^{s}_{y_{\tau}}}$}};

\end{tikzpicture}

\caption{{\bf Repeated interaction scenario.} An experimenter acts on a quantum system $s$ at a discrete number of episodes. The simplest scenario, and the direct analogue to classical input-output processes is a policy (see Eq.~\eqref{eq:prepareandmeasurepolicy}) consisting of a stochastically decided input state $\ket{P^{s}_{x_{\tau}}}$ to `black box' process $\mathbf{T}$ at each time $\tau$, followed by a measurement with elements $\ket{M^{s}_{y_{\tau}}}\bra{M^{s}_{y_{\tau}}}$. The unknown process $\mathbf{T}$ may possess a memory subsystem, but the experimenter cannot directly access it.}
    \label{fig:simplifiedscenario}
\end{figure}

The simplest class of policies in our framework are stochastic prepare-and-measure policies (see Fig.~\ref{fig:simplifiedscenario}). At each episode of evolution $\tau$, a random preparation of an input state is performed $P^{s}_{\tau}=\{ \ket{P^{s}_{x_{\tau}}}\bra{P^{s}_{x_{\tau}}} \}_{x_{\tau} \in \mathsf{X}}$ on $s$, with the preparation indexed by $x_{\tau}$, corresponding to random variable $X_{\tau}$. Subsequently within each episode of evolution, $s$ is acted on by the quantum process $\mathbf{T}$. The episode ends with a rank-1 measurement $M^{s}_{\tau}=\{ \ket{M^{s}_{y_{\tau}}} \bra{M^{s}_{y_{\tau}}} \}_{y_{\tau} \in \mathsf{Y}}$ being performed on $s$, with outcomes $\ket{M^{s}_{y_{\tau}}}$ indexed by a random variable $Y_{\tau}$. Applying the Choi isomorphism in the same manner as was done for $\mathbf{T}$ in Fig.~\ref{fig:process_choi_state} (plug in one half of a maximally entangled pair for each episode of evolution) for a single trajectory $(\overleftrightarrow{X},\overleftrightarrow{Y})=(\overleftrightarrow{x},\overleftrightarrow{y})$ of the quantum control described above yields a state 
\begin{equation} \label{eq:prepareandmeasurepolicy}
    \mathbf{A}_{\overleftrightarrow{z}} = 
    \bigotimes_{\tau=t-n'}^{t+n-1}   w_{z_{\tau}} \mathcal{A}^{s}_{z_{\tau}},
\end{equation}
where $\mathcal{A}^{s}_{z_{\tau}}= d_s^{-1} P^{s}_{x_{\tau}} \otimes M^{\text{T},s}_{y_{\tau}}$, and $\overleftrightarrow{z}:=(\overleftrightarrow{x},\overleftrightarrow{y})$. The transpose on $M^{\text{T},s}_{y_{\tau}}$ arises from the link product used to evaluate contractions between Choi states~\cite{theoreticalframeworkforquantumnetworks}, and the normalisation factors $w_{z_{\tau}}$ and $d_s^{-1}$ ensure that the overall policy conserves probability (see App.~\ref{app:quantumpolicies}). $w_{z_{\tau}}$ encodes the randomness associated with implementing the $d_s^2$-element input state distribution and measurement POVM, as is necessary to represent an informationally complete policy, while $d_s^{-1}$ arises from our policy normalisation convention.

Building from this simplest case, we introduce a memory subsystem $a$ held by the experimenter: $P^{s}_{\tau} \mapsto P^{sa}_{\tau}$ and $M^{s}_{\tau} \mapsto M^{sa}_{\tau}$, while retaining that $M^{sa}_{\tau}$ is a rank-1 measurement. The latter clause forbids the experimenter from maintaining quantum memory between episodes, which is a useful simplification because, if one were to retain quantum memory, $a$ could become entangled with $m$---and hence the future---as the experiment progresses. The result is that information about the future is stored jointly in $ma$, and the future is no longer a property of $\mathbf{T}$ alone. As interesting as this phenomenon might be, we exclude it for comparability to classical computational mechanics. We define the primary set of policies we consider in this work $\mathsf{A}^{*}$ as taking the form described above, allowing $P^{sa}_{\tau}$ and $M^{sa}_{\tau}$ to also depend on the classical measurement record for episodes $<\tau$, and allowing them to be stochastic as well.

A key dividing line for the model classes studied in this work is whether the model injects randomness into its memory subsystem to fulfil its operation. As such, allowing policies that introduce randomness themselves would mask this distinction. Thus, we also consider the set of policies $\mathsf{P} \subset \mathsf{A}^{*}$ where each episode of evolution begins with the preparation of a \emph{pure} state on $s$ and $a$, and ends with a measurement \emph{whose operators are proportional to rank-1 projectors} on the process' output on $s$ and on $a$.

Informationally complete policies are of particular importance to this work because they leave no operationally meaningful content of the quantum process hidden. They are policies $\mathbf{A}$ whose elements $\mathbf{A}_{\overleftrightarrow{z}}$ form a spanning set on the real vector space of Hermitian matrices in which $\mathbf{T}$ is embedded (see App.~\ref{app:quantumpolicies} for more details). Stochastic prepare-and-measure policies (contained in $\mathsf{A}^{*}$), as in Eq.~\eqref{eq:prepareandmeasurepolicy}, are informationally complete if the sets of all $w_{z_{\tau}}\mathcal{A}^{s}_{z_{\tau}}$ are individually spanning sets that define a valid single-step tester on $s$ for all times $\tau$. For example, for a qubit, $\{ P^{s}_{x_{\tau}} \}_{x_{\tau} \in \mathsf{X}}$ can be the four elements of the symmetric informationally complete (SIC) POVM qubit tetrahedron~\cite{Fuchs_2017}, each with weights of $1/4$, and $\{ M^{s}_{y_{\tau}} \}_{y_{\tau} \in \mathsf{Y}}$ can correspond to non-orthogonal projectors from the same SIC POVM. This means that the total weight is $w_{z_\tau}=1/8$ for all $z_\tau$. If these two informationally complete POVMs are applied to the input and output of $\mathbf{T}$ at each time $\tau$, the overall multitime POVM $\mathbf{A}$ enumerates all possible trajectories of outcomes, and thus will be an informationally complete POVM for all times.

Equipped with an informationally complete and linearly independent\footnote{Linear independence is not a necessary assumption, but simplifies our exposition because it guarantees that $\text{tr} \{ {\mathbf{A}}_{\overleftrightarrow{z}'} {\mathbf{D}}^{\dag}_{\overleftrightarrow{z}} \} =\delta_{\overleftrightarrow{z}\overleftrightarrow{z}'}$ always holds.} policy $\mathbf{A}$, and its dual set $\{ {\mathbf{D}}_{\overleftrightarrow{z}} \}_{\overleftrightarrow{z} \in \text{W}(\mathsf{Z}) }$ such that $\text{tr} \{ {\mathbf{A}}_{\overleftrightarrow{z}'} {\mathbf{D}}^{\dag}_{\overleftrightarrow{z}} \} =\delta_{\overleftrightarrow{z}\overleftrightarrow{z}'}$, $\mathbf{T}_{t-n':t+n-1}$ can be mathematically represented purely in terms of its measurement probabilities via the theory of linear inversion tomography~\cite{nonmarkovianquantumprocesses,anintroductiontooperationalquantumdynamics,nonmarkovianquantumprocesstomography,quantumstochasticprocesses} 
\begin{equation} \label{eq:infinitetomographic}
    \mathbf{T}_{t-n':t+n-1} =   \frac{1}{d_s^{2n}d_s^{2n'}}\sum_{\overleftrightarrow{z}}  p(\overleftrightarrow{z}|\mathbf{A}){\mathbf{D}}^{*}_{\overleftrightarrow{z}} .
\end{equation}
This state is the very same matrix as the Choi state. The use of dual matrices ensures that the expansion coefficients are measurement probabilities. Eq.~\eqref{eq:infinitetomographic} is the statement that---within a given time window---$\mathbf{T}$ is fully characterised by its classical probability distribution of measurement outcomes under any informationally complete policy.

\subsection{Conditional Futures and Stationarity} \label{sec:historiesandfutures}
\begin{figure}[tbp]

\centering
\resizebox{\linewidth}{!}{%
\begin{tikzpicture}[scale=0.49]


\draw[black, very thick,solid] (9.7+20-5+2-17,1-8-1) -- (10.3+20-5+2-17,1-8-1);
\draw[black, very thick,solid] (3.7+20-5+2-17,1-8-1) -- (6.3+20-5+2-17,1-8-1);
\draw[black, very thick,solid] (7.7+20-5+2-17,1-8-1) -- (8.3+20-5+2-17,1-8-1);

\draw[myred,fill=myredfill, thick,solid,rounded corners=4] (6.5+20-5+2-17,2.1-8-1) -- (5.9+20-5+2-17,2.1-8-1) -- (5.9+20-5+2-17,0.1-8-1)  -- (4.1+20-5+2-17,0.1-8-1) -- (4.1+20-5+2-17,2.1-8-1) -- (4.1+20-5+2-17,3.9-8-1)  -- (9.9+20-5+2-17,3.9-8-1) -- (9.9+20-5+2-17,0.1-8-1) -- (8.1+20-5+2-17,0.1-8-1) -- (8.1+20-5+2-17,2.1-8-1) -- (6.5+20-5+2-17,2.1-8-1) ;

\draw[decoration={brace,mirror,raise=2pt},decorate]
  (-0.9+20+1-17,-2-8+1) -- node[below=3pt] {$t$} (2.9+20+1-17,-2-8+1);
\draw[decoration={brace,mirror,raise=2pt},decorate]
  (-0.9+4+20+1-17,-2-8+1) -- node[below=3pt] {$t+1$} (2.9+4+20+1-17,-2-8+1);






\draw[] (7,3-8-1) node {\small $\mathbf{T}^{\mathbf{A}_{\overleftarrow{z}_{{t-1}}}}_{{t}:{t+1}}$};

\draw[black, very thick,solid] (9.7+20-5+2-17,1-8-1-7) -- (10.3+20-5+2-17,1-8-1-7);
\draw[black, very thick,solid] (3.7+20-5+2-17,1-8-1-7) -- (6.3+20-5+2-17,1-8-1-7);
\draw[black, very thick,solid] (7.7+20-5+2-17,1-8-1-7) -- (8.3+20-5+2-17,1-8-1-7);

\draw[black, very thick,solid] (9.7+20-5+2-17,1-8-1+2-7) -- (10.3+20-5+2-17,1-8-1+2-7);
\draw[black, very thick,solid] (3.7+20-5+2-17-0.5,1-8-1+2-7) -- (6.3+20-5+2-17,1-8-1+2-7);
\draw[black, very thick,solid] (7.7+20-5+2-17,1-8-1+2-7) -- (8.3+20-5+2-17,1-8-1+2-7);
\draw[black, very thick,solid] (14.3-0.1-4,1-0.2-7-7) -- (14.3+0.1-4,1+0.2-7-7);

\draw[myred,fill=myredfill, thick,solid,rounded corners=4] (6.5+20-5+2-17,2.1-8-1-7) -- (5.9+20-5+2-17,2.1-8-1-7) -- (5.9+20-5+2-17,0.1-8-1-7)  -- (4.1+20-5+2-17,0.1-8-1-7) -- (4.1+20-5+2-17,2.1-8-1-7) -- (4.1+20-5+2-17,3.9-8-1-7)  -- (9.9+20-5+2-17,3.9-8-1-7) -- (9.9+20-5+2-17,0.1-8-1-7) -- (8.1+20-5+2-17,0.1-8-1-7) -- (8.1+20-5+2-17,2.1-8-1-7) -- (6.5+20-5+2-17,2.1-8-1-7) ;

\draw[myred,fill=myredfill, thick,solid,rounded corners=2] (6.5+20-5+2-17-5+1-0.1,2.1-8-1+0.5-0.2-7) rectangle (6.5+20-5+2-17+1-5+1+0.1,2.1-8-1+1+0.5-7);
\draw[decoration={brace,mirror,raise=2pt},decorate]
  (-0.9+20+1-17,-2-8+1-7) -- node[below=3pt] {$t$} (2.9+20+1-17,-2-8+1-7);
\draw[decoration={brace,mirror,raise=2pt},decorate]
  (-0.9+4+20+1-17,-2-8+1-7) -- node[below=3pt] {$t+1$} (2.9+4+20+1-17,-2-8+1-7);






\draw[] (7,3-8-1-7) node {\small $\mathbf{T}^{\rho^m_t}_{t:t+1}$};

\draw[] (7-4,3-8-1-7) node {\tiny $\rho^m_t$};

\draw[black, very thick,solid] (-2,1) -- (0.3,1);
\draw[black, very thick,solid] (1.7,1) -- (2.3,1);
\draw[black, very thick,solid] (3.7,1) -- (4.3,1);
\draw[black, very thick,solid] (9.7,1) -- (10.3,1);
\draw[black, very thick,solid] (11.7,1) -- (14.3,1);
\draw[black, very thick,solid] (14.3-0.1,1-0.2) -- (14.3+0.1,1+0.2);
\draw[black, very thick,solid] (11.7-0.1,1-0.2) -- (11.7+0.1,1+0.2);
\draw[black, very thick,solid] (5.7,1) -- (6.3,1);
\draw[black, very thick,solid] (6.3,1) -- (7.7,1);
\draw[black, very thick,solid] (7.7,1) -- (8.3,1);



\draw[myred,fill=myredfill, thick,solid,rounded corners=4] (6.5,2.1) -- (5.9,2.1) -- (5.9,0.1)  -- (4.1,0.1) -- (4.1,2.1) -- (1.9,2.1) -- (1.9,0.1) --  (0.1,0.1) -- (0.1,2.1) --    (-2,4.1-4+2) --  (-2,3.9) -- (16,3.9) -- (16,2.1) -- (13.9,2.1) -- (13.9,0.1) -- (12.1,0.1) -- (12.1,2.1) -- (9.9,2.1) -- (9.9,0.1) -- (8.1,0.1) -- (8.1,2.1) -- (6.5,2.1) ;

\draw[myblue,fill=mybluefill,thick,solid,rounded corners=4]  (3,-0.1) -- (3,1.9) -- (2.1,1.9) --  (2.1,-0.1) --  (-0.1,-0.1) --  (-0.1,1.9) --  (-1.9,1.9) -- (-1.9,-1.9) --  (0.1,-1.9) -- (3,-1.9) -- (3,0.1) -- (3,-0.1) ;

\draw[black,thick, densely dotted] (6.8-8,0) -- (7.2-8,0);
\draw[black, very thick,solid] (-0.3,-1) -- (2.8,-1);
\draw[myblue,fill=mybluefill,very thick,solid,rounded corners=2] (2.3,1.7) rectangle (3.7-0.9,-1.7);
\draw[myblue,fill=mybluefill,very thick,solid,rounded corners=2] (2.3+0.9-4,1.7) rectangle (3.7-4,-1.7);

\draw[] (7,3) node {\small $\mathbf{T}$};
\draw[] (1,-1.5) node {\small $\mathbf{A}_{\overleftarrow{z}_{{t-1}}}$};

\draw[decoration={brace,mirror,raise=2pt},decorate]
  (-0.9,-2) -- node[below=3pt] {$t-1$} (2.9,-2);
\draw[decoration={brace,mirror,raise=2pt},decorate]
  (-0.9+4,-2) -- node[below=3pt] {$t$} (2.9+4,-2);
\draw[decoration={brace,mirror,raise=2pt},decorate]
  (-0.9+4+4,-2) -- node[below=3pt] {$t+1$} (2.9+4+4,-2);
  \draw[decoration={brace,mirror,raise=2pt},decorate]
  (-0.9+4+4+4,-2) -- node[below=3pt] {$t+2$} (2.9+4+4+4,-2);

\draw[white,fill=white] (-2.1,-2) rectangle (-1.1,4.2);

\draw[white,fill=white] (2.3+4,1.7) rectangle (3.7+4,-1.7);

\draw[white,fill=white] (15,-2) rectangle (17,4.2);


\draw[] (18-17,1-8) node {\huge $=$};

\draw[] (18-17,1-8-7) node {\huge $=$};

\end{tikzpicture}
}

    \caption{{\bf A finite conditional future process.} An action-outcome history $\mathbf{A}_{\overleftarrow{z}_{{t-1}}}$ prior to time $t$ is contracted (top) with the arbitrarily long process $\mathbf{T}$, as in Eq.~\eqref{eq:conditionalfinitefuture}, and all times subsequent to $t+1$ are ignored, yielding a conditional future process $\mathbf{T}^{\mathbf{A}_{\overleftarrow{z}_{{t-1}}}}_{t:t+1}$ (middle). The purpose of modelling is to find a memory state $\rho^m_t=e(\mathbf{A}_{\overleftarrow{z}_{{t-1}}})$ and evolution channel that produce the same observed statistics $\mathbf{T}^{\rho^m_t}_{t:t+1} = \mathbf{T}^{\mathbf{A}_{\overleftarrow{z}_{{t-1}}}}_{t:t+1}$ (bottom). We seek to store as little information as possible to encode a suitable $\rho^m_t$.}
    \label{fig:finite_conditional_future_process}
\end{figure}
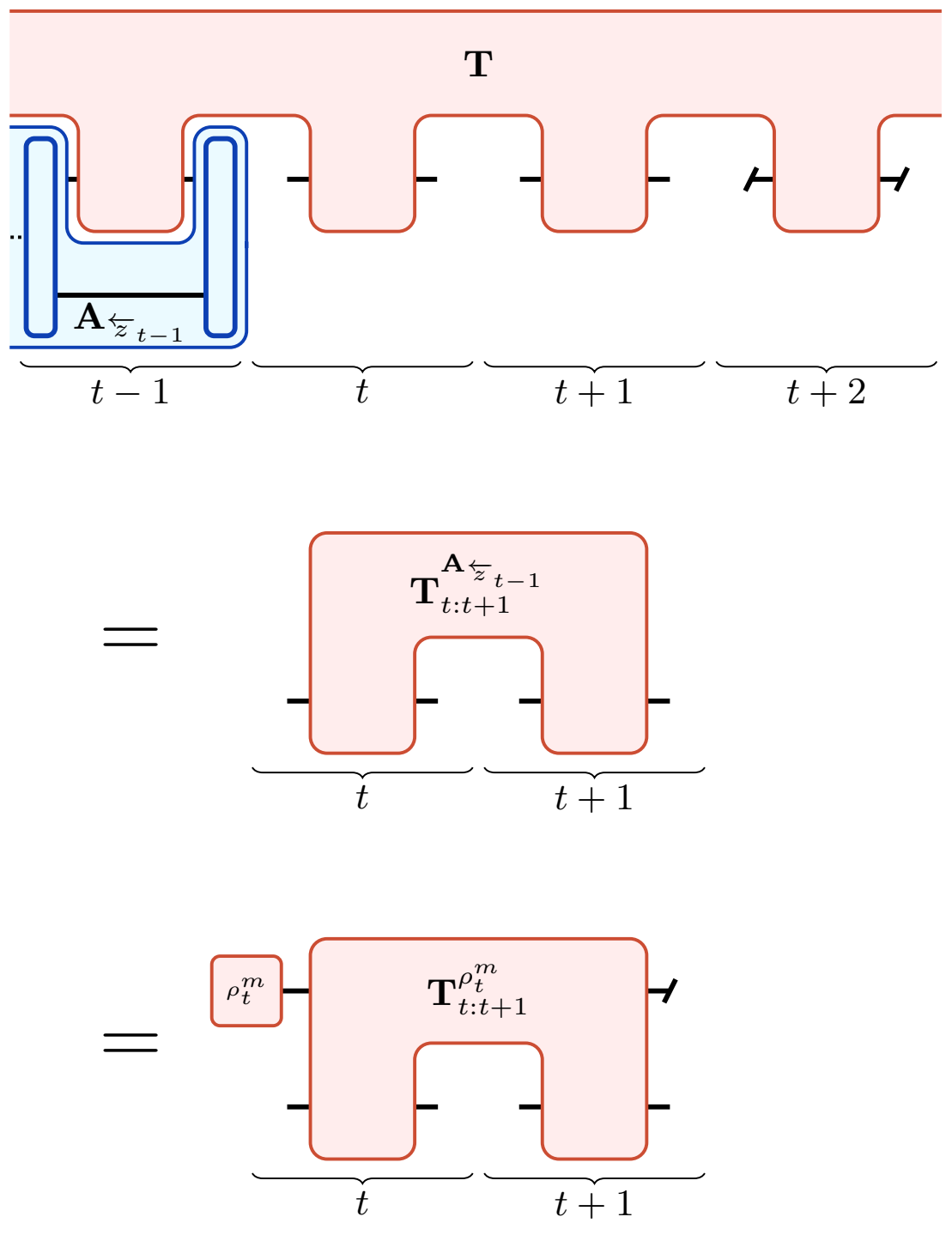

We can now construct the quantum generalisation of the central object of computational mechanics: the conditional future of a given past. Given policy $\mathbf{A}$, an action-outcome history $\mathbf{A}_{\overleftarrow{z}_{t-1}}$ is an outcome sequence terminating at time $t-1$. 
\begin{definition}[Conditional Future Process] \label{def:conditionalfuture}
Let $\mathbf{T}$ be a quantum process, and let $\mathbf{A}_{\overleftarrow{z}_{t-1}}$ be a history up to time $t-1$. Then the quantum conditional future process $\mathbf{T}^{\mathbf{A}_{\overleftarrow{z}_{t-1}}}_{t:t+n-1}$ is given by the contraction of the process $\mathbf{T}$ with the action-outcome history $\mathbf{A}_{\overleftarrow{z}_{t-1}}$,
\begin{equation} \label{eq:conditionalfinitefuture}
\mathbf{T}^{\mathbf{A}_{\overleftarrow{z}_{t-1}}}_{t:t+n-1} 
   :=  \eta_{n'}   \text{\emph{tr}}_{\text{\emph{past}}} \{\mathbf{A}^{\text{\emph{T}}}_{\overleftarrow{z}_{t-n':t-1}} \mathbf{T} \} ,
\end{equation}
where $\text{\emph{tr}}_{\text{\emph{past}}}$ denotes the partial trace over all subsystems corresponding to past times, $\eta_{n'}:=d_s^{2n'}/p(\overleftarrow{z}_{t-n':t-1}|\mathbf{A})$ ensures that the future is unit-normalised for all $n'$, and $p(\overleftarrow{z}_{t-n':t-1}|\mathbf{A})= d_s^{2n'} \text{\emph{tr}} \{\mathbf{A}^{\text{\emph{T}}}_{\overleftarrow{z}_{t-n':t-1}} \mathbf{T} \} $ is the probability of the history occurring.
\end{definition} 
 Note that the conditional future process of a zero-probability history is meaningless, and we exclude these cases from our discussion. See Fig.~\ref{fig:finite_conditional_future_process} for a graphical depiction of conditional future processes.

 The conditional future is a valid unit-normalised process tensor, maintaining the Born rule of Eq.~\eqref{eq:probabilityoftrajectory} for the future episodes. This is the crucial difference from models of classical processes: conditional future process tensors encode future outcome probabilities \emph{as a linear functional on policies that can be performed}, while conditional future probability distributions encode future outcome probabilities full stop.

We desire to only consider quantum processes satisfying the consistency condition that the outcomes of an experiment do not depend on when the experiment is performed. However, unlike classical stochastic processes---which are well-defined without reference to any experimenter---quantum processes require an assumption about what was done to the quantum system before the experimenter began looking. Probably the most natural assumption is that nobody did anything, corresponding to slotting in identity channels at each time, consistent with the generalised Kolmogorov extension theorem~\cite{Milz_2020}. However, another non-equivalent assumption is to trace out all past subsystems of the Choi state~\cite{chang2026randomnessquantumprocessexplained}. Here, we provide a framework general enough to handle either perspective.\footnote{Using a background corresponding to slotting in identity channels requires a quantum comb convention where an episode begins with an output, not an input. Our definitions still work in this case, provided the relevant adjustments are made.} We define a background $\mathbf{K}$ to be a single-element policy obtained by performing some other policy derived from the same recurrent action taken at every time and discarding the outcomes. Stationarity is time-translation invariance with respect to the background.
\begin{definition}[Quantum Stationarity] \label{def:quantumstationarity}
Let $\mathbf{T}$ be a quantum process, let $\mathbf{K}$ be a background, and let $\mathbf{K}_{t:t-1+L}$ be the $L$-time marginal of $\mathbf{K}$ starting at time $t$. $\mathbf{T}$ is stationary with respect to $\mathbf{K}$ if for all integers $t$ and all natural numbers $n,L$,
\begin{equation} \label{eq:stationarity}
    \mathbf{T}^{\mathbf{K}_{t:t-1+L}}_{t+L:t+n-1+L} = \mathbf{T}_{t:t+n-1}.
\end{equation}
\end{definition}
When we refer to a quantum process as stationary, we mean it is stationary with respect to some background $\mathbf{K}$. $\mathbf{T}_{t:t+n-1}$ is the process seen initially at time $t$, having interacted with the background $\mathbf{K}$ for all times prior, and explicitly depends on the choice of background. We view the choice of background as a physical assumption about the family of accessible windows $\mathbf{T}$.

Given Def.~\ref{def:quantumstationarity}, it naturally follows that a process only depends on what has been observed, and not on when those observations were made.
\begin{proposition}[Operational Meaning of Quantum Stationarity] \label{prop:quantumstationarity}
Let $\mathbf{T}$ be a quantum process. It is stationary with respect to some background $\mathbf{K}$ if and only if for all policies $\mathbf{A}$, all action-outcome histories of non-zero probability $\mathbf{A}_{\overleftarrow{z}}$, all integers $t$, all natural numbers $L$, and all natural numbers $n,n'$ it holds that
\begin{equation} \label{eq:stationarity_prop}
     \mathbf{T}^{\mathbf{A}_{\overleftarrow{z}_{t-n'+L:t-1+L}}}_{t+L:t+n-1+L}=\mathbf{T}^{\mathbf{A}_{\overleftarrow{z}_{t-n':t-1}}}_{t:t+n-1} .
\end{equation}
\end{proposition}
\begin{proof}
    Starting at time $t-n'$ instead of $t$, apply Eq.~\eqref{eq:stationarity} and then apply the histories $\mathbf{A}_{\overleftarrow{z}}$ to each side:
    \begin{equation} \label{eq:stationarity_proof}
      \mathbf{T}^{{\mathbf{K}_{t-n':t-n'+L-1}},\mathbf{A}_{\overleftarrow{z}_{t-n'+L:t-1+L}}}_{t+L:t+n-1+L} = \mathbf{T}^{\mathbf{A}_{\overleftarrow{z}_{t-n':t-1}}}_{t:t+n-1}.
\end{equation}
Eq.~\eqref{eq:stationarity_proof} is the explicit statement of what was happening while the experimenter was not looking during the generation of the time translation. Additionally, observe that the case of length $n'=0$ histories reduces precisely to Def.~\ref{def:quantumstationarity}, so the equivalence goes in both directions.
\end{proof}

We stress that there is no requirement that the policy $\mathbf{A}$ in Prop.~\ref{prop:quantumstationarity} satisfies any notion of stationarity itself, because it is being time-translated too. If a quantum process is stationary, it is safe to omit the `present' time $t$ from its mathematical description, and we also omit $n,n'$ for succinctness.

\section{Quantum Models} \label{sec:quantummodels}
\subsection{Core Definition}
Thus far, we have obtained a characterisation of a quantum process purely as its statistical reaction to a chosen policy. However, making use of this description formally requires knowledge of the arbitrarily long past, which requires an arbitrarily large amount of memory in general. The programme of computational mechanics---and our goal---is to retain as little information about the past as possible, while still storing and processing all features of the past that are \emph{relevant} to the future. As discussed in App.~\ref{app:classicalcomputationalmechanics}, HMMs are the models most often considered in classical computational mechanics. The models we consider can be understood as a direct quantum generalisation of HMM.
\begin{definition}[Quantum Model] \label{def:quantummodel}
    A quantum model of quantum process $\mathbf{T}$ is $\mathscr{M}(\mathbf{T})=(\mathcal{H}^{s},\mathcal{H}^{m},\mathcal{T}^{sm},\omega^{m})$, where $\mathcal{H}^{s}$ is the accessible subsystem's Hilbert space specified by $\mathbf{T}$, and $\mathcal{H}^{m}$ is an inaccessible memory space of dimension $d_m$ hypothesised to explain the observed statistics of $\mathbf{T}$ on $s$ via the bipartite channels $\mathcal{T}^{sm}$. $\omega^{m}$ is the memory state initialised at the beginning of every experiment.
\end{definition}
See Fig.~\ref{fig:process_vs_model} for a depiction of a quantum model. This is a generalisation of a classical HMM $(\mathsf{X},\mathsf{M}, T_{\mathsf{X}\mathsf{M}})$, where the outcome alphabet of the HMM is analogous to the joint input-output spaces of the Choi state $\mathsf{X} \leftrightarrow \mathcal{H}^{\text{in}_{\tau}} \otimes \mathcal{H}^{\text{out}_{\tau}}$ (which we assume to be invariant for all episodes $\tau$), the memory alphabet is analogous to the memory Hilbert space $\mathsf{M} \leftrightarrow \mathcal{H}^{m}$, and the transition matrices are analogous to the evolution channel $T_{\mathsf{X}\mathsf{M}} \leftrightarrow \mathcal{T}^{sm}$. For comparison, the more often considered quantum hidden Markov models (QHMM)~\cite{implementationandlearningofquantum,hiddenquantummarkovmodels} are an intermediate object between our models and HMM, where the accessible space remains as $\mathsf{X}$ rather than $\mathcal{H}^{\text{in}_{\tau}} \otimes \mathcal{H}^{\text{out}_{\tau}}$, and the analogue of $T_{\mathsf{X}\mathsf{M}}$ is a collection of Kraus operators acting on $\mathcal{H}^{m}$, indexed by the classical random variable $X_{\tau}$. As such, what distinguishes the models we consider is that they assume nothing about how an external experimenter intends to probe $s$. 

Additionally, ergodicity is usually assumed in computational mechanics, implying that the stationary distribution $\pi^m$ that each experiment begins with is fixed by $T_{\mathsf{X}\mathsf{M}}$, and hence does not need to be specified in addition to $T_{\mathsf{X}\mathsf{M}}$. We have concern for the applicability of this assumption for important quantum processes. Consider the case where $\mathcal{T}^{sm}$ is a CNOT gate targeting $s$, and let either $\omega^{m}=\ket{0^m}\bra{0^m}$ or $\omega^{m}=\ket{1^m}\bra{1^m}$. Both initial states remain invariant regardless of what the experimenter does, so both are perfectly good stationary states. However, starting in a statistical mixture of the two, inputting $\ket{0^s}$, and performing a $z$-basis measurement fixes the model to occupy only one of these states forever more. This is non-ergodicity, and it is exhibited by one of the most ubiquitous entangling gates. Thus, we do not assume $\omega^{m}$ to be fixed by $\mathcal{T}^{sm}$ alone, in congruence with non-ergodic computational mechanics~\cite{e28010111}. Instead, the value of $\omega^{m}$ can be interpreted as a parameter that is constrained by forcing the model to faithfully reproduce the observed process.

\subsection{Encoding Functions}
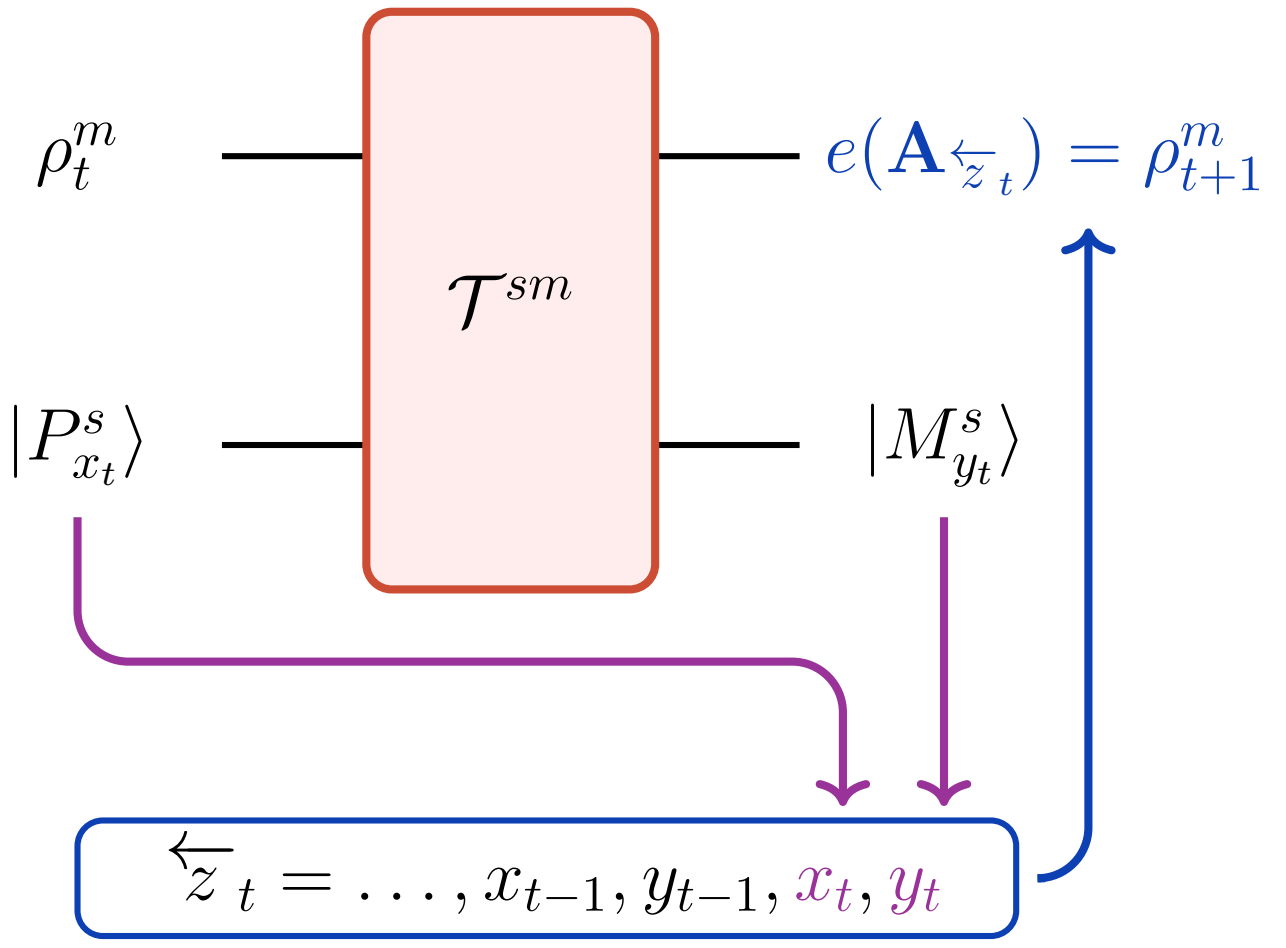
\begin{figure}[tbp]

\centering

\resizebox{\linewidth}{!}{%
\begin{tikzpicture}[scale=1]

\draw[black, very thick,solid] (0,7) -- (4,7);

\draw[black, very thick,solid] (0,5) -- (4,5);

\draw[myred,fill=myredfill,ultra thick,solid,rounded corners=5] (1,4) rectangle (3,8);

\draw[] (2,6) node[rotate=0,black] {\Large $\mathcal{T}^{sm}$};

\draw[] (-1,7) node[rotate=0] {\Large $\rho^{m}_{t}$};

\draw[] (-1,5) node[rotate=0,black] {\Large $\ket{P^{s}_{x_{t}}}$};

\draw[] (5,5) node[rotate=0,black] {\Large $\ket{M^{s}_{y_{t}}}$};

\draw[] (5.7,7) node[rotate=0] {\Large ${\color{myblue}e(\mathbf{A}_{\overleftarrow{z}_{t}})=\rho^{m}_{t+1}}$};



\draw[myblue,very thick,solid,rounded corners=5] (-1,1.6) rectangle (5.5,2.4);


\draw[] (2.3,2) node[rotate=0] {\Large $\overleftarrow{z}_{t}=\dots,x_{t-1},y_{t-1},{\color{mypurple}x_{t}},{\color{mypurple}y_{t}}$};


\draw[mypurple,ultra thick, ->,rounded corners=10] (5,4.5) -- (5,2.5);

\draw[mypurple,ultra thick, ->,rounded corners=10] (-1,4.5) -- (-1,3.5) -- (4.3,3.5) -- (4.3,2.5);

\draw[myblue,ultra thick, ->,rounded corners=10] (5.7,2) -- (6,2) -- (6,6.5) ;


\end{tikzpicture}
}

\caption{{\bf Encoding functions.} A model of a quantum process is defined by an encoding $e$ from histories to memory states, consistent with a system-memory ($sm$) evolution channel. Given a policy $\mathbf{A}$ of the form in Eq.~\eqref{eq:prepareandmeasurepolicy}, at each time $t$, a pure state $\ket{P^{s}_{x_{t}}}$ is determined by $x_{t}$. This input state is then sent through the process being modelled. Finally, the output is measured, with measurement operator $\ket{M^{s}_{y_{t}}} \bra{M^{s}_{y_{t}}}$ corresponding to outcome $y_t$. The input-output pair $z_t=(x_t,y_t)$ is then appended to the existing history $\overleftarrow{z}_{t}=\overleftarrow{z}_{t-1},z_t$. For prediction, $\mathbf{A}_{\overleftarrow{z}_{t}}$ is fed into the encoding function $e$ of the model, which tells the experimenter that the next memory state will be $\rho^{m}_{t+1}$ (blue text). In general, an encoding $e$ may produce a high-dimensional memory Hilbert space $\mathcal{H}^{m}$, whose states are difficult to store and manipulate.}
    \label{fig:modelling}
\end{figure}

Given a model $\mathscr{M}(\mathbf{T})=(\mathcal{H}^{s},\mathcal{H}^{m},\mathcal{T}^{sm},\omega^{m})$ of a stationary process, and some policy $\mathbf{A}$, one can compute how $\omega^{m}$ evolves to some other state after seeing some history $\mathbf{A}_{\overleftarrow{z}}$ of length $n'$; this provides the memory mixed-state description at time $t$ in the absence of any information about what happened before time $t-n'$. The general method to do this is to construct a process tensor up to time $t-1$ from $\mathscr{M}(\mathbf{T})$ by iterating the link product~\cite{theoreticalframeworkforquantumnetworks}, forgoing tracing out $m$ at the end, and then contracting the accessible subsystems with a policy element and tracing out $a$. A concrete formula for evaluating this in the case of $\mathbf{A} \in \mathsf{A}^{*}$ is given by 
\begin{equation} \label{eq:dilationencodingfunction}
\begin{aligned}
   e_{n'}(\mathbf{A}_{\overleftarrow{z}_{t-1}}) := & \frac{1}{p_{\mathscr{M}}(\overleftarrow{z}_{t-n':t-1}|\mathbf{A})} \\ 
   &     \Big(\bigcirc_{\tau=t-n'}^{t-1} w_{z_{\tau}|\overleftarrow{z}_{t-n':\tau-1}} \mathcal{M}^{sa}_{y_{\tau}|\overleftarrow{z}_{t-n':\tau-1},x_{\tau}} \\
   \circ &  \mathcal{T}^{sm} \circ \mathcal{P}^{sa}_{x_{\tau}|\overleftarrow{z}_{t-n':\tau-1}}  \Big) \circ \omega^{m}  ,
   \end{aligned}
\end{equation}
where the adaptive state preparation at any episode $\tau$ is $\mathcal{P}^{sa}_{x_{\tau}|\overleftarrow{z}_{t-n':\tau-1}}$, the subsequent adaptive measurement is $\mathcal{M}^{sa}_{y_{\tau}|\overleftarrow{z}_{t-n':\tau-1},x_{\tau}}$, and the corresponding weight is $w_{z_{\tau}|\overleftarrow{z}_{t-n':\tau-1}}$. Here, we have specified that $z_\tau=(x_\tau,y_\tau)$, where $y_\tau$ decides the measurement outcome and $x_\tau$ decides the preparation. $y_\tau$ can depend on $x_\tau$ but not the converse. Due to the policy being in $\mathsf{A}^{*}$, it is implicit in Eq.~\eqref{eq:dilationencodingfunction} that $\mathcal{P}^{sa}_{x_{\tau}|\overleftarrow{z}_{t-n':\tau-1}}$ prepares a fresh state on $sa$ and $\mathcal{M}^{sa}_{y_{\tau}|\overleftarrow{z}_{t-n':\tau-1},x_{\tau}}$ traces out $sa$ at each episode. See Fig.~\ref{fig:modelling} for a schematic depiction. Note that the subscript on $p_{\mathscr{M}}(\overleftarrow{z}_{t-n':t-1}|\mathbf{A})$ denotes that it is the probability of the history $\mathbf{A}_{\overleftarrow{z}_{t-n':t-1}}$ under the \emph{model} $\mathscr{M}(\mathbf{T})$. Whether that corresponds to the true probability for the process will be addressed in Sec.~\ref{sec:prediction_functions}. We have cast the encoding function as a mapping from history positive semidefinite Hermitian matrices to memory density matrices $e_{n'}(\mathbf{A}_{\overleftarrow{z}_{t-1}}): \mathcal{L}\big( \mathcal{H}^{\text{in}_{t-n'}} \otimes \mathcal{H}^{\text{out}_{t-n'}} \otimes \dots \otimes \mathcal{H}^{\text{in}_{t-1}} \otimes \mathcal{H}^{\text{out}_{t-1}}\big) \rightarrow \mathcal{D}(\mathcal{H}^m)$ that we refer to as the \emph{encoding function} of the model, cf. the classical encoding function of Eq.~\eqref{eq:classicalencoding}.

Recall that classically we defined both an encoding function $e_{\text{cl}}$ taking histories to memory state probability distributions, and an encoding map $\tilde{e}_{\text{cl}}$ taking histories to memory state random variables governed by the aforementioned distributions. In the quantum case, only the former definition remains obviously viable. Classically, a memory must occupy a definite state at all times; without disturbance, one can always look at the computer running their model and read off the state it is in.\footnote{Classical belief states pertaining to an observed system that is physically distinct from the model can be mixed. However, implementing those belief states on a model running on a classical computer still requires specifying the belief state as a concrete string, which is consistent with the description we have provided here.} Quantum mechanically, on the other hand, a density matrix does not uniquely specify a random variable over quantum states, and nor can one be meaningfully distinguished from the mixed state of its average---unless one has access to its purifying subsystem. As we will see in Sec.~\ref{sec:quantumprediction}, this has profound impact on how our model classes apply to quantum processes. Note that $e_{n'}$ is a member of a family of encoding functions, each defined by its value for the integer parameter $n'$. We write $e$ with no subscripts as a shorthand for all finite members of this family. We comment that $e_{n'}$ are not linear due to the re-normalisation in Eq.~\eqref{eq:dilationencodingfunction}.

\subsection{Prediction Functions} \label{sec:prediction_functions}
$e$ captures half of the operation of a quantum model: the encoding of relevant features from the past into a compressed memory representation. The other half of modelling is to use the encoded memory state to produce a forecast of the future. We define a family of CPTP maps $\Phi$---one $\Phi_n$ for each future length $n$---and refer to each member as a \emph{prediction function} of the model, taking model memory states to length $n$ future process tensors by enacting the Choi isomorphism,
\begin{equation} \label{eq:quantumprocessdef}
  \Phi_n(\rho_t) := {\text{tr}}_{m} \Bigg\{   \bigcirc_{\tau=t}^{t+n-1}     \mathcal{T}^{{\text{out}}_\tau,m} \Bigg[ \bigotimes_{\tau=t}^{t+n-1}  \Psi^{{\text{in}}_{\tau},{\text{out}}_\tau}   \otimes  \rho_{t}^{m}  \Bigg] \Bigg\},
\end{equation}
where $\text{in}_{\tau}$ and $\text{out}_{\tau}$ are the $\tau$th input and output on $s$, and $\Psi^{\text{in}_{\tau},\text{out}_{\tau}}$ is a maximally entangled pair on those subsystems. Each $\mathcal{T}^{{\text{out}}_\tau,m}$ acts on one half of its corresponding $\Psi^{\text{in}_{\tau},\text{out}_{\tau}}$, and on the memory. 

Equipped with encoding functions and prediction functions, we can write out the defining separation that distinguishes a meaningful model \emph{of a process} from an arbitrary recurrent quantum circuit. 
\begin{definition}[Faithfulness] \label{def:faithfulness}
A model $\mathscr{M}(\mathbf{T})$ of a stationary quantum process $\mathbf{T}$ is \emph{faithful} ($\texttt{F}$) if 
\begin{equation} \label{eq:faithfulness1}
    \Phi (\omega^m) = \mathbf{T}.
\end{equation}
\end{definition}
Recall that $\Phi$ refers to the family of maps from memory states to length $n$ futures for all values of $n>0$. Thus, it produces the family of all length $n$ windows of the stationary process, which is $\mathbf{T}$ itself. Def.~\ref{def:faithfulness} directly implies that correct statistics are obtained for any policy.
\begin{proposition}[Operational Meaning of Faithfulness] \label{prop:faithfulness}
If a model $\mathscr{M}(\mathbf{T})$ of a stationary quantum process $\mathbf{T}$ is faithful, then for all policies $\mathbf{A}$, and all non-zero probability histories $\mathbf{A}_{\overleftarrow{z}} \in \mathbf{A}$,
\begin{equation} \label{eq:faithfulness2}
    \Phi \circ e (\mathbf{A}_{\overleftarrow{z}}) = \mathbf{T}^{\mathbf{A}_{\overleftarrow{z}}}.
\end{equation}
\end{proposition}
\begin{proof}
    Faithfulness implies that for all $l>0$, $\Phi_l (\omega^m) = \mathbf{T}_{t:t+l-1}$. Fixing some value of $l$, choose any pair $n,n'$ such that $n+n'=l$. Then for any non-zero probability length $n'$ history $\mathbf{A}_{\overleftarrow{z}}$ of any policy $\mathbf{A}$ we can apply Eq.~\eqref{eq:conditionalfinitefuture} to $\Phi_l (\omega^m)$ to compute the conditional future of the model $\tilde{\mathbf{T}}^{\mathbf{A}_{\overleftarrow{z}}}_{t+n':t+l-1}$. Due to the fact that any model's operation correctly yields the process it generates, it must hold that $\Phi_n \circ e_{n'} (\mathbf{A}_{\overleftarrow{z}})=\tilde{\mathbf{T}}^{\mathbf{A}_{\overleftarrow{z}}}_{t+n':t+l-1}$. Since the model is faithful $\tilde{\mathbf{T}}^{\mathbf{A}_{\overleftarrow{z}}}_{t+n':t+l-1}=\mathbf{T}^{\mathbf{A}_{\overleftarrow{z}}}_{t+n':t+l-1}$, and Eq.~\eqref{eq:faithfulness2} holds.
\end{proof}
The RHS of Eq.~\eqref{eq:faithfulness2} represents the future statistics of the process being modeled for the given past, as in Eq.~\eqref{eq:conditionalfinitefuture}, while the LHS corresponds to the output of the model, provided with the same past. Thus faithfulness specifies that our model of the quantum process and the quantum process itself have statistically indistinguishable conditional future statistics.

\section{Quantum Model Classes} \label{sec:quantumprediction}
Memory cost depends on what is required from a model. Indeed, there exist classical processes for which a classical $\texttt{F}$ model requires less memory than the $\epsilon$-machine~\cite{predictionandgeneration,onthegenerativenatureofprediction}, which is the minimal $\texttt{nOF}$ model of a classical stochastic process. The most rudimentary such example is the simple non-unifilar source, for which a two state generator exists, but whose $\epsilon$-machine has a countable infinity of memory states~\cite{thecalculiofemergence}. In this section, we produce the hierarchy of quantum model classes $\texttt{SF}$ $\subset$ $\texttt{nOF}$ $\subset$ $\texttt{F}$, and analyse both the memory constraints and the operational consequences of each. Due to the non-passive nature of quantum observation, we define these properties relative to a given experimental policy, and add the prefix `uniformly' to denote when the property holds for a relevant set of policies. Our primary interest, however, is in $\texttt{nOF}$ models, due to their special position in classical computational mechanics.

\subsection{Quantum Quantum \texorpdfstring{$\texttt{nOF}$}{nOF}}
Classically, we expect that $\texttt{nOF}$ and $\texttt{pRF}$ are the same property, despite the former being defined entropically while the latter is defined operationally. We provide a sketch of why. 

Any classical HMM generates its future from its memory states $\mu_i$ alone. A faithful model as in Eq.~\eqref{eq:classicalfaithfulness} generates the \emph{correct} future on average: $\sum_{i} p(\mu_i|\overleftarrow{x})p(\overrightarrow{X}|\mu_i)=p(\overrightarrow{X}|\overleftarrow{x})$. Oracular information~\cite{synchronizationandcontrolinintrinsic} is the information a model's memory shares with its own future beyond that contained in the past, 
\begin{equation} \label{eq:classicaloracular}
    \zeta_{\text{cl}} := I(M:\overrightarrow{X}|\overleftarrow{X}) ,
\end{equation}
see Fig.~\ref{fig:three_model_classes}. Provided a faithful HMM, we can express oracular information as the history average of the state average of all deviations from what the history provides,
\begin{equation} \label{eq:classicaloracular2}
    \zeta_{\text{cl}} = \sum_{\overleftarrow{x}} p(\overleftarrow{x})\sum_{i} p(\mu_i|\overleftarrow{x})D_{\text{KL}}\big( p(\overrightarrow{X}|\mu_i) \big\Vert p(\overrightarrow{X}|\overleftarrow{x}) \big),
\end{equation}
where $D_{\text{KL}}(\cdot \Vert \cdot)$ is the Kullback-Leibler divergence, and $\mu_i$ are the memory states of a classical HMM. Using the property that $D_{\text{KL}}=0$ if and only if its arguments are equal, and limiting our consideration to the case of finite $n,n'$, it holds that $\zeta_{\text{cl}}=0$ if and only if for every non-zero probability history and for all $\mu_i$ in the support of the encoded memory probability distribution of $\overleftarrow{x}$, $p(\overrightarrow{X}|\mu_i)=p(\overrightarrow{X}|\overleftarrow{x})$, which is equivalent to Eq.~\eqref{eq:classicalprfaithfulness} holding.

A direct quantum analogue of Eq.~\eqref{eq:classicaloracular} naively seems impossible because the memory state must be consumed by the model to produce the future. We circumvent this problem by a) purifying the memory state on some arbitrary subsystem $m'$, and b) using policies in $\mathsf{A}^{*}$ whereby the past-future bipartition is a classical-quantum state. Jointly, this allows quantum conditional mutual information to be expressed as a history average over quantum mutual information terms between the purifying $m'$ subsystem of the consumed memory and the future.
\begin{definition}[Oracular Information] \label{def:oracular_information}
The oracular information of a model $\mathscr{M}(\mathbf{T})$ of a stationary process $\mathbf{T}$ is 
\begin{equation} \label{eq:oracular_information}
    \zeta\big(\mathbf{A},\mathscr{M}(\mathbf{T})\big):= \lim_{n=n' \rightarrow \infty}\sum_{\overleftarrow{z}} p(\overleftarrow{z}|\mathbf{A})I_{\mathbf{A}_{\overleftarrow{z}}}(f:m'),
\end{equation}
where $I_{\mathbf{A}_{\overleftarrow{z}}}$ pertains to the state
\begin{equation} \label{eq:upsilon_oracular}
    \Upsilon_{\mathbf{A}_{\overleftarrow{z}}} = \Phi_n^{m \rightarrow f} \otimes \mathcal{I}^{m'} \big( \ket{\psi^{mm'}_{\rho}} \bra{\psi^{mm'}_{\rho}} \big),
\end{equation}
and $\ket{\psi^{mm'}_{\rho}}$ is a purification of $\rho^{m}:=e(\mathbf{A}_{\overleftarrow{z}})$. $\mathcal{I}^{m'}$ is the identity channel on the purifying subsystem $m'$.
\end{definition}
A model $\mathscr{M}(\mathbf{T})$ is $(\mathbf{A},\mathbf{T})$-\emph{non-oracular} if $\zeta\big(\mathbf{A},\mathscr{M}(\mathbf{T})\big)=0$. We note that oracular information is purification-independent, and upper bounded by twice the average entropy of the encoded memory state via the Araki-Lieb inequality. The potential for the oracular information to exceed the average memory entropy is a signature of a stronger-than-classical dependence of the future on side-information.

An example where this occurs is a model of the fully depolarising channel where $\mathcal{T}^{sm}$ is a SWAP gate followed by replacing the $m$ state with a maximally mixed one. Here, $\zeta\big(\mathbf{A},\mathscr{M}(\mathbf{T})\big)=2\log(d_m)$, because the uncertainty in the process can be interpreted as the next output becoming maximally entangled with a hidden half of a maximally entangled state on some external subsystem $m'$. On the other hand, the encoded memory state is always the maximally mixed state, which has an entropy of $\log(d_m)$. In this example, the purifying `oracular' subsystem $m'$ is maximally entangled with the future. Crucially, however, this oracular information only pertains to \emph{the model's own future}, not the future of some external physical process, whose own memory we assume to be fundamentally inaccessible.

We characterise quantum $(\mathbf{A},\mathbf{T})$-$\texttt{nOF}$ as the combination of $(\mathbf{A},\mathbf{T})$-non-oracularity and faithfulness as in Eq.~\eqref{eq:faithfulness1}. 
\begin{definition}[$\texttt{nOF}$] \label{def:prescience}
    A faithful model $\mathscr{M}(\mathbf{T})$ of a stationary process $\mathbf{T}$ is $(\mathbf{A},\mathbf{T})$-$\texttt{nOF}$ if it is $(\mathbf{A},\mathbf{T})$-non-oracular.
\end{definition}

\subsection{Quantum Per-Realisation Faithfulness}
Following the classical intuition in App.~\ref{app:classicalcomputationalmechanics}, one should suspect that quantum $\texttt{nOF}$ is equivalent to some notion of providing a faithful conditional distribution `every time'. However, in quantum mechanics it is not immediately clear that such a notion is even meaningful at all. Consider the classical case where a history $\overleftarrow{x}$ has occurred, leading to a memory probability distribution $e_{\text{cl}}(\overleftarrow{x})$, from which the memory state random variable $M^m_t$ takes some definite value $\mu^m_i$. This results in a forecast of the future fully determined by $\mu^m_i$, albeit not necessarily fully determined by $\overleftarrow{x}$. 

Now looking at the quantum analogue, let $\mathscr{M}(\mathbf{T})$ be a model of a stationary process $\mathbf{T}$ and let $\mathbf{A}$ be a policy. Given a history $\mathbf{A}_{\overleftarrow{z}}$, consider the encoded state $\rho^{m}=e(\mathbf{A}_{\overleftarrow{z}})$. Unlike in the classical case, $\rho^m$ fundamentally does not contain information about which ensemble of pure states (or what entanglement) produced it, meaning that the quantum analogue of $\tilde{e}_{\text{cl}}$ is non-unique. Since the memory state cannot be uniquely pinned down to one set of options, neither can the prediction of the model in general, meaning that the quantum notion of providing a faithful conditional distribution `every time' becomes inherently ambiguous. For precisely the same dynamics, some assignments of memory density matrices to pure-state ensembles may yield the same future for all members, while other ensembles may fail. An example of this is a dephase-then-SWAP model, in which a qubit full dephasing channel along the $z$ axis is performed on $m$, followed by an $sm$ qubit SWAP gate. Here, the experimenter inserts a maximally mixed state $\mathbbm{1}/d_s$ at each time. In this case, the memory state will be the maximally mixed state, but an ensemble $\big\{(1/2, \ket{0^m} ), (1/2, \ket{1^m}) \big\}$ will yield distinct futures, while the ensemble $\big\{(1/2, \ket{+^m} ), (1/2, \ket{-^m}) \big\}$ will not. Here, the answer to whether `the correct output is obtained each time' depends on information not contained in the memory density matrix, rendering the question operationally meaningless when only provided with that density matrix. 

The resolution to this ambiguity is to strengthen per-realisation faithfulness in the quantum setting. We demand ontological neutrality: our definition should not depend on underlying facts about a quantum system that quantum mechanics forbids an observer from learning directly. In our case this forbidden information is the ensemble that produced a density matrix. Thus, informally, a model is $(\mathbf{A},\mathbf{T})$-$\texttt{pRF}$ if all elements of all pure-state ensembles of the encoded state $\rho^m$ yield the correct future, except for cases with vanishing probability. 
\begin{definition}[Per-Realisation Faithfulness] \label{def:per-realisationfaithfulness}
    A faithful model $\mathscr{M}(\mathbf{T})$ is $(\mathbf{A},\mathbf{T})$-per-realisation faithful ($\texttt{pRF}$) if for all assignments that map each history $\mathbf{A}_{\overleftarrow{z}} \in \mathbf{A}$ to a finite pure-state ensemble $\big\{ p(k|\overleftarrow{z}), \ket{\sigma^m_{k,\mathbf{A}_{\overleftarrow{z}}}} \big\}$ of its encoded state $\rho^{m}$, and for all $\delta>0$, the probability that
    \begin{equation} \label{eq:PAC}
       E_{n,n'}(k,\overleftarrow{z},\mathbf{A}):= \left\Vert \Phi_n\big(\ket{\sigma^m_{k,\mathbf{A}_{\overleftarrow{z}}}} \bra{\sigma^m_{k,\mathbf{A}_{\overleftarrow{z}}}}  \big) - \mathbf{T}^{\mathbf{A}_{\overleftarrow{z}}}_{t:t+n-1} \right\Vert 
    \end{equation}
    exceeds $\delta$, given by
    \begin{equation}
        q_{n,n'}(\delta):=\sum_{k,\overleftarrow{z}}p(k|\overleftarrow{z})p(\overleftarrow{z}|\mathbf{A})\texttt{I}\big[  E_{n,n'}(k,\overleftarrow{z},\mathbf{A}) > \delta  \big],
    \end{equation}
approaches zero as $n=n' \rightarrow \infty$. $\texttt{I}[\cdot ]$ is the indicator function, which is one if the argument holds and zero otherwise.
\end{definition}
$\Vert \cdot \Vert$ is the trace norm. A $\texttt{pRF}$ quantum model does not necessarily operate using pure states, yet one can still describe it as being in any pure state consistent with the past without breaking its output. While mathematically simpler, an exact analogue of Def.~\ref{def:per-realisationfaithfulness} would be too strict; it would rule out the case where it is only met approximately after seeing any finite time window. We expect the probabilistic, approximate, and asymptotic case to be generic in our setting, because classical $\epsilon$-machines are only known to be guaranteed to satisfy Eq.~\eqref{eq:classicalprfaithfulness} asymptotically~\cite{exactsynchronization,asymptoticsynchronization}---with a simple counterexample being the classical $\epsilon$-machine of the Even Process~\cite{crutchfield2001regularitiesunseenrandomnessobserved}---and because there do not exist non-Markovian quantum processes that have finite Markov order with respect to all possible instruments~\cite{quantummarkovorder}. We say that a model is uniformly $\texttt{pRF}$ if it is $(\mathbf{A},\mathbf{T})$-$\texttt{pRF}$ for all informationally complete $\mathbf{A} \in \mathsf{P}$. 

Since small trace distance between high-dimensional Choi states does not necessarily constrain operational distinguishability~\cite{rosenthal2024quantumchanneltestingaveragecase}, one may be led to question the meaningfulness of $E_{n,n'}(k,\overleftarrow{z},\mathbf{A})$ as a quantifier of error. In App.~\ref{app:operational_distinguishability}, we show that this does not obstruct our use: for finite-dimensional faithful models of stationary processes, $\texttt{pRF}$ with respect to trace distance on Choi states is equivalent to $\texttt{pRF}$ with respect to differences in statistics under any future policy. In order to complete this proof, we derive an upper bound on measurement error in terms of trace norm error, which may be useful in its own right. What makes this bound serve our purposes is that it is uniform for all $n$ at least as large as some finite $n_0$, meaning that trace distance controls measured error for arbitrarily long processes.

\subsection{Memory Correspondence} \label{sec:presciencecorrespondence}
We establish a one-to-one correspondence between $\texttt{nOF}$ and $\texttt{pRF}$. This theorem characterises the quantum generalisation of the classical model class that $\epsilon$-machines are minimal within, revealing the analogous relationship between memory and behaviour.
\begin{theorem}[Memory Correspondence] \label{thm:predictioncorrespondence}
    Let $\mathscr{M}(\mathbf{T})$ be a faithful model of a stationary process $\mathbf{T}$ with $d_m<\infty$, and let $\mathbf{A} \in \mathsf{A}^{*}$. Then, $\mathscr{M}(\mathbf{T})$ is $(\mathbf{A},\mathbf{T})$-$\texttt{pRF}$ if and only if it is $(\mathbf{A},\mathbf{T})$-$\texttt{nOF}$.
\end{theorem}
The formal proof is intricate due to the need to biconditionally tie an entropic quantity to a notion of asymptotic approximate probabilistic convergence. Particular difficulties arise in the `only if' direction, from the need to control off-diagonal contributions to the prediction error, and the need to separately treat terms whose eigenvalue weights are small enough to be poorly constrained by per-realisation faithfulness alone. However, the core physics that underlies Thm.~\ref{thm:predictioncorrespondence} can still be gleaned from a finite exact deterministic simplification. In App.~\ref{app:finite_case} we provide finite history versions of per-realisation faithfulness and non-oracularity, and prove a corresponding simplification of Thm.~\ref{thm:predictioncorrespondence}. This provides a flavour of the proof method and underlying physics that go into the asymptotic case, which is covered in full detail in App.~\ref{app:asymptotic_case}. 

We also point out that what is proved in App.~\ref{app:asymptotic_case} is significantly stronger than what is claimed in Thm.~\ref{thm:predictioncorrespondence}: the proof only uses that for all $n$, $\Phi_n$ is invoked in Eq.~\eqref{eq:upsilon_oracular} and Eq.~\eqref{eq:PAC} as a quantum channel that maps encoded memory states $e(\mathbf{A}_{\overleftarrow{z}})$ to target states $\mathbf{T}^{\mathbf{A}_{\overleftarrow{z}}}_{t:t+n-1}$, as in Eq.~\eqref{eq:faithfulness2}. No relation between $\Phi_n$ for different values of $n$ is used. Thus, alternate versions of Thm.~\ref{thm:predictioncorrespondence} hold when each $\Phi_n$ is replaced by some other channel $\Theta_n$ such that for the target states $\mathbf{S}^{\mathbf{A}_{\overleftarrow{z}}}_{t:t+n-1}$ of all histories $\mathbf{A}_{\overleftarrow{z}}$ with $p(\overleftarrow{z}|\mathbf{A})>0$, $\mathbf{S}^{\mathbf{A}_{\overleftarrow{z}}}_{t:t+n-1}:=\Theta_n \circ e(\mathbf{A}_{\overleftarrow{z}})$. $\mathbf{S}^{\mathbf{A}_{\overleftarrow{z}}}_{t:t+n-1}$ could be a quantum space series~\cite{elliott2026causalmemorystructuralcomplexity}, a non-Choi state representation of the future, or the classical probability distribution of the outcomes of some specific experiment performed in the future. Notably, $\mathbf{S}^{\mathbf{A}_{\overleftarrow{z}}}_{t:t+n-1}$ can be the target of a reservoir computing-like task for input data $\mathbf{A}_{\overleftarrow{z}}$.

Observe that because Thm.~\ref{thm:predictioncorrespondence} is an `if and only if' statement, our stipulation of ontological neutrality---insensitivity to underlying facts that cannot be directly observed---was essential; if a model just satisfied per-realisation faithfulness for one ensemble decomposition but not another, the model need not be non-oracular. 

As an intermediate step in the proof of Thm.~\ref{thm:predictioncorrespondence}, we identify that $\texttt{nOF}$/$\texttt{pRF}$ is also equivalent to the property that applying a POVM to a purification of the memory state cannot steer (in the sense of the HJW Theorem~\cite{Hughston1993}) the prediction of the model away from the correct one. This implies that $\texttt{nOF}$ models possess a security guarantee: no party with access to a purification of the model's memory state can obtain `insider information' that enables more accurate prediction than the history itself allows.

\subsection{Synchronising Faithfulness}
Classically, a particularly elegant subset of $\texttt{pRF}/\texttt{nOF}$ models is those where upon seeing almost any asymptotically long history, the uncertainty in the memory state approaches zero. This is called asymptotic synchronisation~\cite{asymptoticsynchronization}. This guarantees that in the asymptotic regime, one only ever needs to access definite states, allowing Eq.~\eqref{eq:classicalprfaithfulness} to be satisfied without imposing a stochastic memory state selection rule $\mathsf{select}$. The quantum generalisation of this is that the memory state asymptotically purifies, which neatly falls out as a strengthening of our notion of non-oracularity. Not only does the memory not contain information relevant to the future given the past, but it contains no information at all given the past.
\begin{definition}[Quantum Synchronising Faithfulness] \label{def:sync}
   A faithful model $\mathscr{M}(\mathbf{T})$ of a stationary process $\mathbf{T}$ is $(\mathbf{A},\mathbf{T})$-synchronising faithful ($\texttt{SF}$) if the policy $\mathbf{A}$ yields that the average state entropy obtained by the encoding function is zero in the long history limit,
\begin{equation}
    \lim_{n' \rightarrow \infty} \sum_{\overleftarrow{z}} p(\overleftarrow{z}|\mathbf{A}) S\big(e(\mathbf{A}_{\overleftarrow{z}})\big) = 0.
\end{equation}
\end{definition}
It is an immediate consequence of Thm.~\ref{thm:predictioncorrespondence} that $\texttt{SF} \subseteq \texttt{pRF}$.
\begin{corollary}[$\texttt{SF}$ Implies $\texttt{pRF}$]
    If a model $\mathscr{M}(\mathbf{T})$ with $d_m<\infty$ is $(\mathbf{A},\mathbf{T})$-$\texttt{SF}$ for $\mathbf{A} \in \mathsf{A}^{*}$, it is also $(\mathbf{A},\mathbf{T})$-$\texttt{pRF}$.
\end{corollary}
\begin{proof}
    By the Araki-Lieb inequality we have that
    \begin{equation}
    \begin{aligned}
        \zeta\big(\mathbf{A},\mathscr{M}(\mathbf{T})\big):=& \lim_{n=n' \rightarrow \infty}\sum_{\overleftarrow{z}} p(\overleftarrow{z}|\mathbf{A})I_{\mathbf{A}_{\overleftarrow{z}}}(f:m') \\ 
        \leq & 2 \lim_{n' \rightarrow \infty} \sum_{\overleftarrow{z}} p(\overleftarrow{z}|\mathbf{A}) S\big(e(\mathbf{A}_{\overleftarrow{z}})\big)\\ 
        =&0.
    \end{aligned}
    \end{equation}
    This establishes that $\mathscr{M}(\mathbf{T})$ is $(\mathbf{A},\mathbf{T})$-non-oracular. To complete the proof we invoke Thm.~\ref{thm:predictioncorrespondence}, implying that it is $(\mathbf{A},\mathbf{T})$-$\texttt{pRF}$. 
\end{proof}
The inclusion $\texttt{SF} \subset \texttt{pRF}$ is in fact strict. The simplest example is a process consisting of a sequence of identity channels on $s$ alone, which has a sequence of identity channels on $s$ alone as an $(\mathbf{A},\mathbf{T})$-$\texttt{SF}$ model for any policy (the memory has dimension 1 so it is synchronised always). If we append an irrelevant evolving memory subsystem that never purifies to this model (e.g. an initial mixed memory state undergoing local unitary evolution), it remains $(\mathbf{A},\mathbf{T})$-$\texttt{pRF}$ for all policies, but now fails to be $(\mathbf{A},\mathbf{T})$-$\texttt{SF}$.

We say that the model is uniformly $\texttt{SF}$ if it is $(\mathbf{A},\mathbf{T})$-$\texttt{SF}$ for all informationally complete $\mathbf{A} \in \mathsf{P}$. The informational completeness clause is useful here because failing to fully probe the process may sabotage synchronisation by no fault of the model itself. In summary, $\texttt{SF}$ $\subset$ $\texttt{pRF}$ $\subset$ $\texttt{F}$, and:
\begin{enumerate}
    \item An $\texttt{SF}$ model's memory almost always occupies an arbitrarily pure state, as history length tends to infinity, and that state is uniquely determined by a historical record. Since the memory state is pure, there is no non-trivial purification to use for steering.
    \item An $\texttt{nOF}$ model's output is almost always arbitrarily indistinguishable from that produced by all members of a set of pure states, as history length tends to infinity, and those states may not be uniquely determined by a historical record. Despite the memory state being mixed, no party accessing the memory state's purification can steer the predictions of the model.
    \item An $\texttt{F}$ model's memory is only fully characterised by its full density matrix description. The predictions of the model can be steered by a recorded measurement on the memory state's purification.
\end{enumerate}

\section{Fundamental Limits on Quantum Models} \label{sec:fundamentallimitations}
While these model classes differ in what types of information get stored in memory, faithfulness to $\mathbf{T}$ requires them all to be bound by a universal constraint: the memory space must be sufficiently large to produce all observable future behaviour. We now address how the behaviour of $\mathbf{T}$ constrains all quantum models of it. 

\subsection{Linear Dependence of Futures} \label{sec:LDFM} 
Classically, we saw that the causal structure of stochastic processes could be characterised with an equivalence relation $\sim_{\epsilon}$, whereby histories $\overleftarrow{x},\overleftarrow{x}'$ satisfy $\overleftarrow{x} \sim_{\epsilon} \overleftarrow{x}'$ if and only if they lead to indistinguishable future statistics $p(\overrightarrow{X}|\overleftarrow{x})=p(\overrightarrow{X}|\overleftarrow{x}')$. Causal equivalence dictates that in such cases $\overleftarrow{x}$ and $\overleftarrow{x}'$ should be assigned to the same memory state. 

It turns out that applying this pairwise test of equivalence is neither necessary nor sufficient as an `Occam's razor' for quantum processes. The necessity direction is straightforward to see. Quantum histories, memory states, and futures are all represented by quantum states, which exist on vector spaces that have a notion of linear dependence. As such, having a pair of pasts yield two distinct futures does not imply that the model needs to expand its memory space to account for them; there may be some superposition or statistical mixture of existing pasts that can produce the desired future instead. Needing a new memory state does not imply needing a new memory dimension. This is a property of models whose memory states are elements of a vector space, which is not only satisfied by quantum models. For example, observable operator models have precisely this property~\cite{observableoperatormodels}. 

What is more interesting is that quantum superpositions can hide inequivalences. Consider again our recurring dephase-then-SWAP model. Here, the memory states $\ket{+^{m}}$ and $\ket{-^{m}}$ can be produced by any history that sends in $\ket{+^{s}}$ and $\ket{-^{s}}$ respectively at its final step. The dephasing on the $s$ output implies that all such histories result in identical conditional futures. However, if we put these memory states in superposition, we find that they did not contain identical information about the future: $\frac{1}{\sqrt{2}}(\ket{+^{m}} + \ket{-^{m}})=\ket{0^{m}}$ and $\frac{1}{\sqrt{2}}(\ket{+^{m}} - \ket{-^{m}})=\ket{1^{m}}$. Putting two memory states or pasts in superposition can yield statistics that are distinct from either in isolation.

In light of these failures at both necessity and sufficiency, instead of iteratively asking whether each pair of histories leads to the same future, we perform a single global comparison of whether a whole set of histories leads to a linearly dependent set of futures. This can be achieved via a Gram-Schmidt orthonormalisation procedure on the real vector space of Hermitian matrices that quantum density matrices representing futures live within. This is written as Alg.~\ref{alg:NLDFM} in pseudocode in App.~\ref{app:algorithm}, which we numerically demonstrate in Sec.~\ref{sec:numerics}. This algorithm takes the set of all possible length $n'$ histories that can be produced with non-zero probability, computes the set of resultant futures of length $n$, and finds a minimal spanning set. The key number---the size of this basis as $n=n'\rightarrow \infty$--- is denoted by $D$, and represents the number of linearly independent futures a process $\mathbf{T}$ can exhibit, provided with any possible history. $D$ is a property of $\mathbf{T}$ alone, and is independent of any model, or what any experimenter might choose to do. 

We show that our computational mechanics approach to describing the past-future causal relations of quantum processes results in a quantity that can be understood as the bond dimension~\cite{Pirvu_2010} of the matrix product operator representation of a process tensor~\cite{J_rgensen_2019}. We stress that this is distinct from the bond dimension of the matrix product \emph{state} representation of a process tensor.
\begin{proposition}[$D$ Equals Operator Schmidt Rank] \label{prop:schmidt}
    Let $\mathbf{T}$ be a stationary quantum process. Its number of linearly independent futures $D$ is equal to the operator Schmidt rank $R$ of its Choi state as $n=n' \rightarrow \infty$.
\end{proposition}
The proof is in App.~\ref{app:choi_schmidt}, and leverages a particular form of the tomographic representation of a process that can be likened to an operator Schmidt decomposition.

\subsection{Fundamental Limit from Observed Statistics} \label{sec:fundamentallimits}
In classical computational mechanics, the number of distinct futures a stochastic process can exhibit fixes the number of memory states its $\epsilon$-machine requires. Here, the number of linearly independent futures $D$ defines a \emph{lower bound} on the memory dimension of a quantum process' minimal model.
\begin{theorem}[Memory Dimension Lower Bound] \label{thm:lbfm}
Let $\mathbf{T}$ be a stationary quantum process, and let $D$ be the number of linearly independent futures of process $\mathbf{T}$. Let $d$ be the minimal memory dimension of any faithful quantum model of $\mathbf{T}$. Then,
\begin{equation}
    \sqrt{D} \leq d.
\end{equation}
\end{theorem}
\begin{proof}

Since we assumed $\mathscr{M}_{\text{min}}(\mathbf{T})$ to be a faithful model, $\mathbf{T}$ can be expressed using $\mathscr{M}_{\text{min}}(\mathbf{T})$, via Eq.~\eqref{eq:quantumprocessdef}. Observe that the channel $\Phi_n$ that maps a memory state to its corresponding Choi state is linear on $\mathcal{L}(\mathcal{H}^{m})$. Due to this linearity, the image $\Phi\big( \mathcal{L}(\mathcal{H}^{m}) \big)$ of $\Phi$ has real dimension at most $d^2$. In general, the set of achievable futures (given any history rather than any memory state) is contained within this image, leading to the inequality $\sqrt{D} \leq d$. 
\end{proof}
$D$ can be interpreted as the memory \emph{exhibited} by a process, and Thm.~\ref{thm:lbfm} implies that it defines a lower bound on the number of memory basis states that must be \emph{used} to model it. What is uniquely quantum about Thm.~\ref{thm:lbfm} is that the bound is quadratic rather than linear. The substance here lies in what this quadraticity says about each of our model classes.

We believe that the number of linearly independent futures $D$ can be understood as an extension of the notion of stochastic language rank that is independent of how one chooses to interact with the process. As such, quadratic lower bounds similar to those in the existing literature~\cite{implementationandlearningofquantum,riechers2025identifiabilityminimalityboundsquantum,Zonnios_2025} can be obtained from Thm.~\ref{thm:lbfm}. 

Consider the case of $\texttt{F}$ models, where the minimal model has a $d$-dimensional Hilbert space, and can produce up to $d^2$ linearly independent futures. If the desire is to simply generate some number of linearly independent futures on a quantum computer with a fixed number of qubits, this gap between memory and behaviour genuinely is quadratic. However, for such a model, there is no guarantee that a given future can be generated by a pure memory state. In other words, the causal explanation of the model's behaviour is irreducibly a mixed state description, where the relevant memory dimension is that of Hermitian matrices, with dimension $\tilde{D} =  d^2$. From this perspective, $D \leq \tilde{D}$. $\texttt{nOF}$ models are the opposite of this case: for the futures of sufficiently long pasts, there is almost always a pure memory state that approximately produces them. $\texttt{SF}$ models simplify this picture: not only can their operation be \emph{explained} using only pure states, but their operation can actually be \emph{implemented} using only pure states. Recalling the non-steering interpretation of $\texttt{pRF}$, such models can not only generate statistics with a quadratic gap between memory and behaviour, but they do so \emph{securely}, where there exists no external information that can impact the model's output.

\subsection{Probabilistic Distinctness} \label{sec:probabilisticdistinctness}
A natural question one might ask about Thm.~\ref{thm:lbfm} is which types of processes saturate the bound, and consequently maximise the quadratic separation between memory and cause. To understand this, we consider a key property possessed by classical $\epsilon$-machines~\cite{equivalenceofhistoryandgenerator}---probabilistic distinctness. In this context, probabilistic distinctness is that each classical memory state yields a distinct conditional future probability distribution:
\begin{equation} \label{eq:classicalprobabilisticdistinctness}
    \mu_i = \mu_j \quad \text{iff} \quad p(\overrightarrow{X}|\mu_i) = p(\overrightarrow{X}|\mu_j).
\end{equation}
In analogy, define probabilistic distinctness for quantum models as: for all $\ket{\psi^m},\ket{\phi^m} \in \mathcal{H}^{m}$,
\begin{equation} \label{eq:quantumprobabilisticdistinctness}
    \begin{gathered}
    \ket{\psi^m}\bra{\psi^m} = \ket{\phi^m}\bra{\phi^m} \\
    \text{iff} \quad \Phi\big(\ket{\psi^m} \bra{\psi^m}\big) = \Phi\big(\ket{\phi^m} \bra{\phi^m}\big).
    \end{gathered}
\end{equation}
As emphasised by our dephase-then-SWAP model, not all minimal quantum models exhibit this property; states differing by only a phase are not probabilistically distinct, despite there being no simpler quantum model producing the same process. In quantum mechanics, there exist memory states that are structurally necessary, yet causally redundant. Interestingly, if we shift the order of operations to define a SWAP-then-dephase model of the same process, probabilistic distinctness is recovered, but at the cost of all $z$-basis coherent density matrices being unreachable under any history. As such, we expect that this gap between causal structure and memory is very much model-dependent.

We strengthen this notion: a model satisfies strong probabilistic distinctness if for all memory states (pure or mixed) $\rho^m,\sigma^m \in \mathcal{D}(\mathcal{H}^{m})$,
\begin{equation} \label{eq:quantumstrongprobabilisticdistinctness}
    \rho^m = \sigma^m \quad \text{iff} \quad \Phi(\rho^m) = \Phi(\sigma^m).
\end{equation}
This puts the memory and future spaces on equal footing, allowing us to prove a minimality result. 
\begin{theorem}[Minimality]  \label{thm:probdistinct}
If a faithful model $\mathscr{M}(\mathbf{T})=(\mathcal{H}^{s},\mathcal{H}^{m},\mathcal{T}^{sm},\omega^{m})$ of a stationary process $\mathbf{T}$ with $d_m < \infty$ satisfies strong probabilistic distinctness and all memory density matrices are in the span of those reached under some policy, then it has minimal memory dimension.
\end{theorem}
The proof is in App.~\ref{supp:SPDproof}. A simple example of a model satisfying these conditions is the system-memory SWAP gate. In this case, all information about the memory is transmitted directly to the accessible system. This means that the number of linearly independent futures $D$ is the size of an informationally complete set on the memory, which is $d^2$. In turn, this implies that our lower bound of Thm.~\ref{thm:lbfm} is saturated, and the model must be minimal.

\subsection{Numerical Experiment} \label{sec:numerics}
In this section, we deploy our formalism to demonstrate a feature of quantum dynamics which may be leveraged to save memory in quantum modelling. Suppose one is provided with a unitary model $\mathscr{M}(\mathbf{T})=(\mathcal{H}^{s},\mathcal{H}^{m},\mathcal{U}^{sm},\omega^{m})$, and asked whether it is the minimal model of the process it generates, or if it has redundant memory. Further, assume that the unitary $\mathcal{U}^{sm}(\cdot)=U\cdot U^{\dag}$ is derived from a Hamiltonian evolution $U=\text{exp}(-i\Delta t H)$, where $H$ is the Hamiltonian and $\Delta t$ is the duration of the evolution. Further, assume some finite precision $\xi$ in being able to distinguish futures. Intuitively, one might expect that the fundamental amount of memory $D$ \emph{exhibited} by the process depends on $H$ alone, because it defines the way $s$ and $m$ interact. However, we demonstrate that this is false at finite precision---except in the large $\Delta t $ limit. Furthermore, due to Thm.~\ref{thm:lbfm}, the minimum memory \emph{used} to approximately model $\mathbf{T}$ is timescale dependent. This implies that if one seeks to model Hamiltonian evolutions for small timescales, one may be able to get away with less memory than the structure of the Hamiltonian would indicate.

Our numerical experiment employs a bipartite $sm$ unitary model $\mathscr{M}(\mathbf{T})=(\mathcal{H}^{s},\mathcal{H}^{m},\mathcal{U}^{sm},\omega^m)$, where $d_s=2$, $d_m=6$, and $\omega^m=\mathbbm{1}/d_m$. Using the Python package QuTiP, we randomly generate an ensemble of 100 dense random Hamiltonians $H^{sm}$, and apply them using values of $\Delta t$ within the interval $[0.1, 10 ]$. Then we use Alg.~\ref{alg:NLDFM} for $n=n'=1$ (total of 2 steps), and $n=n'=2$ (total of 4 steps), and a tolerance of $\xi = 10^{-5}$ for whether to reject almost linearly dependent futures.

\begin{figure}[tbp] 
\centering
        \centering
        \includegraphics[width=\linewidth]{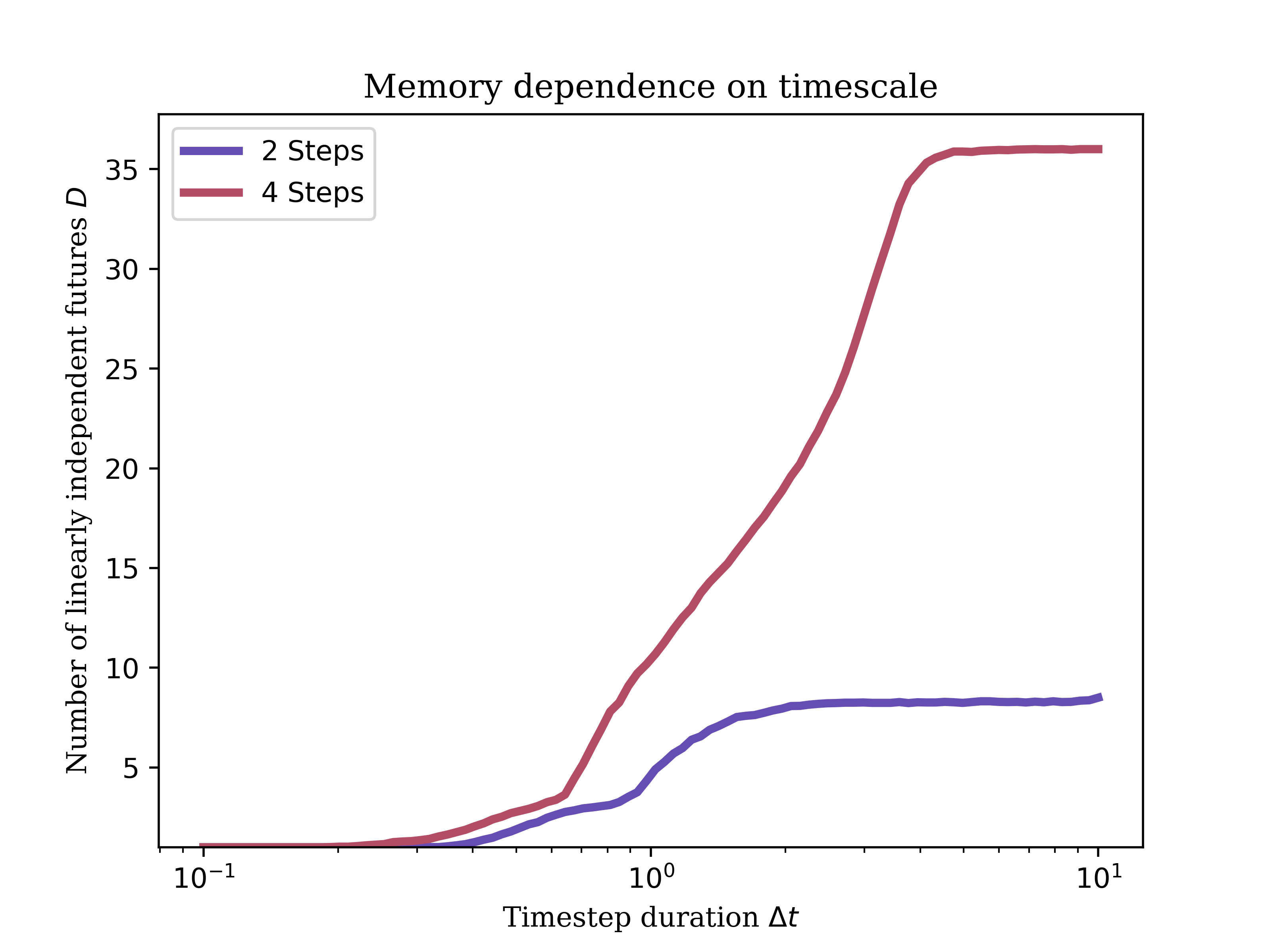}

\caption{{\bf Memory dependence on timescale.} The blue line represents the ensemble average of $D$ over all $n=n'=1$ (total of 2 steps) simulations, and the red line represents $n=n'=2$ (total of 4 steps). Both scenarios begin with $\mathcal{U}^{sm}$ being close to the identity channel for small $\Delta t$, which correspondingly yields $D=1$ (no memory). As $\Delta t$ grows, the ensemble average of $D$ grows, with a kink in both graphs. The $n=n'=1$ line saturates early because there are not enough linearly independent channels to express the structure of the underlying memory of the process. On the other hand, the $n=n'=2$ line saturates at $D=36$ because $\sqrt{D}\leq d \leq d_m=6$ as in Thm.~\ref{thm:lbfm}. In this case, there are not enough memory dimensions in the underlying process to allow each future process tensor to be distinct.}\label{fig:numerics}
\end{figure}

We observe in Fig.~\ref{fig:numerics} that both the $n=n'=1$ line and the $n=n'=2$ line saturate for large $\Delta t$, but for distinct reasons. In the case of $n=1$, apparent saturation within the studied $\Delta t$ range, around $D \approx 9$ occurs because there is a limited number of linearly independent channels on $s$, somewhat undershooting the dimension-counting upper bound of $D \leq 13$. Here, a channel-valued future simply cannot convey the information contained in the memory. The opposite bottleneck occurs for $n=2$ futures, where the memory of the unitary itself becomes saturated at $D = d_m^2=36$, consistent with Thm.~\ref{thm:lbfm}. 

Our observations suggest that $D$ is timescale-dependent at finite precision. The implication is that since $\sqrt{D} \leq d$, modelling a quantum process defined by a Hamiltonian $H$ for small $\Delta t$ may be possible with a model that does not contain all the information that $H$ does. This points to the idea that brief/weak evolutions fundamentally require lower memory resources to faithfully predict, irrespective of the underlying structure that generates them. 

\section{Discussion}
In complexity science, computational mechanics provides a comprehensive framework for inferring the underlying causal relations in stochastic time-series data. Here, our work provides theoretical foundations for generalising the key concepts of computational mechanics to settings where the process we wish to model is quantum mechanical. Instead of a probability distribution over strings, our quantum world is described by a process tensor that contains all information about how the world will behave depending on how it is measured (or not measured) at each time-step. In place of a machine that takes past data strings to sample appropriate conditional futures, we consider machines that can synthesise conditional future process tensors---quantum states that can completely describe a process's conditional future behaviour regardless of how we choose to interact with it during future time-steps. We make progress in three key areas. Thm.~\ref{thm:predictioncorrespondence} characterises the class of models within which quantum $\epsilon$-machine analogues should be sought. Thm.~\ref{thm:lbfm} and Prop.~\ref{prop:schmidt} provide two formulations of a policy-independent lower bound on model memory requirements. Finally, Thm.~\ref{thm:probdistinct} highlights an $\epsilon$-machine-like special case where model minimality is guaranteed. These developments provide the first steps towards understanding how to isolate causes within quantum processes.

One interesting direction for further research is to understand the relation between our results and recent work in extending computational mechanics to the modelling of a sequence of input-free correlated quantum states~\cite{elliott2026causalmemorystructuralcomplexity}. Indeed, any such sequence of correlated quantum states has a quantum process tensor description and thus corresponds to a special case of causal modelling of general quantum processes, mirroring the classical distinction between stochastic processes and input-output processes~\cite{thecomputationalmechanicsofinputoutput}. One interesting result from Ref.~\cite{elliott2026causalmemorystructuralcomplexity} in this context is that for memoryful stationary quantum sequences that admit a finite-dimensional primitive causal model, there exists a memory-minimal model that can be realised without an iteratively discarded `junk' environment. It would certainly be interesting to see if such ideas generalise to full quantum processes with inputs and outputs.

Another direction is to better understand the qualitative differences between classical and quantum worlds. Here, we have already seen that certain intuitions regarding the principle of causal equivalence require care in their application. Classically, it is a given that if two different pasts have identical futures, producing statistically faithful future predictions should not require us to commit memory to discerning between those pasts. In the quantum setting, we see that a similar principle may not apply. Two quantum states $\ket{\phi}$ and $\ket{\phi'}$ may individually have identical futures while their superpositions do not, implying that pairwise equality of futures does not certify that memory states are mutually redundant. Our dephase-then-SWAP model exemplifies this phenomenon. Such observations may have fascinating consequences. For example, they call into question the transferability of certain classical methods for inferring causal models, such as Causal-State Splitting Reconstruction (CSSR)~\cite{analgorithmforpatterndiscovery}, whose basis is the assumption that similarly behaving states can be merged together. Meanwhile, they may also indicate that, for a quantum agent to properly learn minimal models of a quantum environment, probing in superposition is essential.


\begin{acknowledgments}
This work is supported by the National Research Foundation of Singapore through the NRF Investigatorship Program (Award No. NRF-NRFI09-0010), the National Quantum Office, hosted in A*STAR, under its Centre for Quantum Technologies Funding Initiative (S24Q2d0009), the RIE 2025 AQAS projects S25Q9DA001 and S25Q9DA002, the Singapore Ministry of Education Tier 1 Grant RT4/23 and RG91/25, the RIE25 Japan-Singapore Joint Call on Quantum (Project ID H25-MRO3490) and the NRF AI-for-Science (AI4S) Catalytic Grant NRF-AI4SCT-2025-0005.
\end{acknowledgments}




\appendix
\onecolumngrid

\section{Classical Computational Mechanics}
\label{app:classicalcomputationalmechanics}
Before delving into our quantum formalism, we present the classical analogue for a concrete introduction to the relevant concepts. A classical stochastic process is a temporally ordered sequence of random variables $X_\tau$, chosen from an alphabet $\mathsf{X}$. Specifying a `present' time $t$, plus future and past lengths $n$ and $n'$ respectively, we concern ourselves with the empirically accessible finite block of this classical stochastic process $\overleftrightarrow{X}_{t-n':t+n-1}:=\{ X_{t-n'}, \dots, X_t,\dots, X_{t+n-1} \}$. This finite window of observation is fully characterised by a joint probability distribution $p(\overleftrightarrow{X}_{t-n':t+n-1})$, which we physically assume to exist and be mutually consistent for all combinations of $t,n,n'$, as required by the Kolmogorov extension theorem~\cite{Milz_2020}. 

Employing the future-past bipartition made explicit in the definition of a classical stochastic process, we are primarily concerned with the relationship between any past that has occurred $\overleftarrow{x}_{t-n':t-1}$ and the future $\overrightarrow{X}_{t:t+n-1}$, fully characterised by the conditional probability distribution $p(\overrightarrow{X}_{t:t+n-1}|\overleftarrow{x}_{t-n':t-1})$.

A key consistency property for realistic experiments is that future statistics depend only on observed outcomes $\overleftarrow{x}_{t-n':t-1}$ in the past, rather than when that past occurred, i.e. the value of $t$. This is the property that $p(\overrightarrow{X}_{t:t+n-1}|\overleftarrow{x}_{t-n':t-1})=p(\overrightarrow{X}_{t+L:t+n-1+L}|\overleftarrow{x}_{t-n'+L:t-1+L})$, for any integer $t$, any natural number $L$ and any natural numbers $n,n'$, which is known as stationarity~\cite{computationalmechanics}. Note that we allow $n'=0$ `empty' pasts in this definition. When a process is stationary, the explicit time $t$ for the future-past bipartition has no physical impact, and will be omitted except where explicitly relevant. We also typically omit $n,n'$ and use the bare over arrow notation to signify `any arbitrarily long finite futures and pasts'. We will explicitly take limits where it is relevant. 

The key approach to saving memory in modelling stationary processes is to assign histories $\overleftarrow{x} \in \overleftarrow{\mathsf{X}}$ to a discrete set of memory states $\mu^m_i \in \mathsf{M}$. These memory states evolve according to a conditional update rule $T_{\mathsf{X}\mathsf{M}}=\{ T^{x_t}_{\mathsf{X}\mathsf{M}} \}_{x_t \in \mathsf{X}}$ that dictates both the probability of an outcome $x_t$ and how each memory state updates after $x_t$ has been observed. The tuple $(\mathsf{X},\mathsf{M}, T_{\mathsf{X}\mathsf{M}})$ is a hidden Markov model (HMM) of process $\overleftrightarrow{X}$. Given an HMM of a stationary process, the stationary distribution $\pi^m$ is the probability distribution describing the random variable $M^m_{t-n'}$ for the memory state at the beginning of any finite time window, and is characterised by the property of being invariant under forgetting the past. 

An encoding function $e_{\text{cl}}: \overleftarrow{\mathsf{X}} \rightarrow D(\mathsf{M})$ is a mapping from histories of non-zero probability to probability distributions $D(\mathsf{M})$ over memory states. For length $n'$ histories, this can be written in terms of a model as 
\begin{equation} \label{eq:classicalencoding}
   e_{\text{cl}}(\overleftarrow{x}_{t-n':t-1}) = \frac{1}{p(\overleftarrow{x}_{t-n':t-1})} T_{\mathsf{X}\mathsf{M}}^{x_{t-1}}  \dots  T_{\mathsf{X}\mathsf{M}}^{x_{t-n'}} \pi^m,
\end{equation}
where $p(\overleftarrow{x}_{t-n':t-1})$ is the probability of the past sequence $\overleftarrow{x}_{t-n':t-1}$, obtained by computing $T_{\mathsf{X}\mathsf{M}}^{x_{t-1}}  \dots  T_{\mathsf{X}\mathsf{M}}^{x_{t-n'}} \pi^m$ and taking the sum of entries. At each time $\tau$, the memory state is updated according to the observed outcome $x_{\tau-1}$ at the previous time, leading to a new random variable $M^m_{\tau}$. The random variable at the end of this chain is $M^m_{t}$, which is governed by the probability distribution $e_{\text{cl}}(\overleftarrow{x}_{t-n':t-1})$. 

We can see then that while the probability distribution governing the random variable $M^m_{t}$ is a fixed function of the past, the memory state $\mu^m_{i}$ achieved in a given run of the process at time $t$ can be a \emph{stochastic} function of the history $\overleftarrow{x}_{t-n':t-1}$: the same history can yield different memory states in different runs. We capture this evolution using the encoding \emph{map} $\tilde{e}_{\text{cl}}: \overleftarrow{\mathsf{X}} \rightarrow R(\mathsf{M})$, where $M^m \in R(\mathsf{M})$ are random variables over memory states $\mu^m_i$ described by probability distributions $p(\mu^m_i) \in D(\mathsf{M})$. We define $\tilde{e}_{\text{cl}}(\overleftarrow{x}):=\mathsf{select} \circ e_{\text{cl}}(\overleftarrow{x})$, where $\mathsf{select}$ draws a single memory state $\mu^m_i$ from the probability distribution $e_{\text{cl}}(\overleftarrow{x})$ according to its probability.

We define a classical prediction function as a map $\Phi_{\text{cl}} : R(\mathsf{M}) \rightarrow R\big(D(\mathsf{X}^{n})\big)$ that takes random variables over memory states to random variables over probability distributions describing conditional futures. Analogously to Eq.~\eqref{eq:classicalencoding}, the action of $\Phi_{\text{cl}}$ can be defined from an HMM as 
\begin{equation} \label{eq:classicalpredictionfunction}
    \Phi_{\text{cl}}\big(q(\mu^m_i) \big)= \Big\{ p\big( \overrightarrow{x}_{t:t+n-1} \big\vert q(\mu^m_i)  \big)  \Big\}_{ \overrightarrow{x}_{t:t+n-1} \in \mathsf{X}^n},
\end{equation}
where all $p\big( \overrightarrow{x}_{t:t+n-1} \big\vert q(\mu^m_i)  \big) $ are computed by applying the transition matrices $T_{\mathsf{X}\mathsf{M}}^{x_{t+n-1}}  \dots  T_{\mathsf{X}\mathsf{M}}^{x_{t}}$ to the probability distribution $q(\mu^m_i)$, then taking the sum of the entries. 

In this work, we study quantum generalisations of a hierarchy of three classes of models, all of which we believe to be closely related to definitions that have previously appeared in the classical literature~\cite{onthegenerativenatureofprediction,predictionandgeneration}, although we make no claims about formal equivalence. For illustrative purposes, consider the thought experiment of a betting game where Alice must wager some chosen quantity of money on the outcomes of a stochastic process equivalent to a fair coin toss, and losses are weighted more heavily than wins, but she knows nothing at all about the structure of the process. She can ask either Bob or Charlie what they think the next outcome will be. Bob is a \emph{per-realisation faithful} model of the process: he provides the correct information each time. As such, he always answers `I don't know', and provides even odds for each outcome. This uncertainty informs Alice's risk calculus, encouraging her to minimise her wager. Charlie is a \emph{faithful} model: his outputs average to the correct probability distribution of the future. However, he is not per-realisation faithful. In particular, each time he is queried, he confidently replies `definitely heads' or `definitely tails' at random. In spite of the fact that Charlie's output is correct on average, his behaviour resembles that of a charlatan, encouraging Alice to place higher wagers than she actually should. This information gap is the advantage of per-realisation faithfulness.

We begin with our loosest notion: a classical model is \emph{faithful} ($\texttt{F}$) if 
\begin{equation} \label{eq:classicalfaithfulness}
    \Phi_{\text{cl}} \circ e_{\text{cl}} (\overleftarrow{x}_{t-n':t-1}) = p(\overrightarrow{X}_{t:t+n-1}| \overleftarrow{x}_{t-n':t-1}),
\end{equation}
i.e. the model outputs the true probability distribution for the future given the past. Intuitively, this seems like all one needs from a model.

Our work is based on the assumption that classical computers intrinsically use definite memory states. This is not the only position one can reasonably take, but it does have utility in teasing out subtle distinctions in model operation. One motivation behind this assumption is: if one prints out the memory state of a model running on their classical computer, it will always be $\tilde{e}_{\text{cl}}(\overleftarrow{x})$, not $e_{\text{cl}}(\overleftarrow{x})$. Of course, one could store a probability distribution \emph{as} a definite state, but that is merely choosing a different model with a larger memory space such that $e_{\text{cl}}(\overleftarrow{x})$ becomes a delta function distribution (which may require unbounded memory in general). Given our assumption of definite states, there is no free lunch. Then, if not all $\mu^m_i$ output by $\tilde{e}_{\text{cl}}(\overleftarrow{x})$ yield the same future probability distribution as $e_{\text{cl}}(\overleftarrow{x})$, predictive errors can occur on a per-instance basis. As such, faithfulness only guarantees that the output of $\Phi_{\text{cl}}$ \emph{averages to} $p(\overrightarrow{X}_{t:t+n-1}| \overleftarrow{x}_{t-n':t-1})$ as in Eq.~\eqref{eq:classicalfaithfulness}. 

\emph{Per-realisation faithfulness}\footnote{In this section we use an exact, finite history and future version of $\texttt{pRF}$ for the sake of gently introducing the concept. We present the approximate, asymptotic, and probabilistic full definition in the quantum case.} ($\texttt{pRF}$) is the corresponding strengthening of faithfulness whereby the correct conditional future probability distribution---as in Eq.~\eqref{eq:classicalpredictionfunction}---is produced \emph{in every realisation} of a past $ \overleftarrow{x}_{t-n':t-1}$,
\begin{equation} \label{eq:classicalprfaithfulness}
    \Phi_{\text{cl}} \circ  \tilde{e}_{\text{cl}} (\overleftarrow{x}_{t-n':t-1}) = p(\overrightarrow{X}_{t:t+n-1}| \overleftarrow{x}_{t-n':t-1}).
\end{equation}

The cleanest case where Eq.~\eqref{eq:classicalprfaithfulness} holds is where $e_{\text{cl}}$ collapses to a degenerate probability distribution because the historical data is sufficient to fix the memory random variable to a single state. This behaviour is known as synchronisation~\cite{synchronizationandcontrolinintrinsic,asymptoticsynchronization}, and \emph{synchronising faithful} ($\texttt{SF}$) models alleviate the stochasticity of prediction because the input to the prediction function $\Phi_{\text{cl}} $ becomes a deterministic function of the past. By contrast, per-realisation faithfulness only provides that while $e_{\text{cl}}(\overleftarrow{x}_{t-n':t-1})$ need not be degenerate, if it is not, each memory state supported by $e_{\text{cl}}$ yields the same future.

The seminal result of classical computational mechanics is the $\epsilon$-machine~\cite{inferringstatisticalcomplexity,thecalculiofemergence,computationalmechanics}. This is an HMM with a special encoding function $\epsilon$ defined using the principle of causal equivalence:
\begin{equation} \label{eq:cefm}
\begin{aligned}
    \epsilon(\overleftarrow{x}) =&  \epsilon(\overleftarrow{x}') \\ 
    \text{iff}&  \\ 
    p(\overrightarrow{X}\big\vert \overleftarrow{x} ) 
   =& p(\overrightarrow{X}\big\vert \overleftarrow{x}' ).
    \end{aligned}
\end{equation}
$\epsilon$-machine memory states are equivalence classes of all histories with the same conditional future statistics, and hence in one-to-one correspondence with those futures. Note that any $\texttt{F}$ model satisfies that $I(\overleftarrow{X}:\overrightarrow{X}|M)=0$, because the past, memory probability distribution, and future form a Markov chain. Applying the chain rule to this yields $I(M:\overrightarrow{X}) - I(\overleftarrow{X}:\overrightarrow{X})=I(M:\overrightarrow{X}|\overleftarrow{X})$, where $I(M:\overrightarrow{X}|\overleftarrow{X})$ is known as \emph{oracular information} in the sense that it is future-relevant information contained in the memory but not in the past. The $\epsilon$-machine is minimal within the class of non-oracular faithful ($\texttt{nOF}$) models\footnote{Note that the formal notions of $\epsilon$-machines and $\texttt{nOF}$ are only obtained in the limit of infinite length pasts and futures; everything short of that is typically an approximation of the true behaviour.} (also known as `prescient' models~\cite{PhysRevE.59.275,computationalmechanics}) satisfying faithfulness and
\begin{equation} \label{eq:classicalprescience}
    \zeta_{\text{cl}} := I(M:\overrightarrow{X}|\overleftarrow{X})=0.
\end{equation}
$\zeta_{\text{cl}}$ is referred to as oracular information. $\texttt{nOF}$ models are defined by entropic quantifiers of their memory, while $\texttt{pRF}$ models are defined by their observable behaviour. We produce quantum generalisations of these model classes, and study what changes.

At face value, quantum processes are a closer analogue of input-output processes than standard stochastic processes, because quantum processes admit both inputs and outputs~\cite{quantumstochasticprocesses,anintroductiontooperationalquantumdynamics}. Indeed, both must be accounted for to reach a proper characterisation of the process. However, quantum mechanically we can obtain a more compact representation: the Choi isomorphism~\cite{anintroductiontooperationalquantumdynamics,quantumstochasticprocesses} between processes and states allows one to faithfully encode all possible input-output relations into a single quantum state, via a fixed maximally entangled input. This allows us to simplify our description of quantum processes in such a way that it is more directly comparable to the output-only stochastic processes that were presented in this section.

\section{Quantum Policies} \label{app:quantumpolicies}
As stated in the main text, a quantum policy is a collection of quantum combs $\mathbf{A}= \{ \mathbf{A}_{\overleftrightarrow{z}} \}_{\overleftrightarrow{z} \in \text{W}(\mathsf{Z}) }$, indexed by trajectories of measurement outcomes $\overleftrightarrow{z}$. Each comb $\mathbf{A}_{\overleftrightarrow{z}}$ represents a sequence of actions taken, and a corresponding sequence of measurement outcomes. The contraction between a particular $\mathbf{A}_{\overleftrightarrow{z}}$ and the process $\mathbf{T}$ yields the probability of the outcome sequence $\overleftrightarrow{z}$, given that policy $\mathbf{A}$ is being performed. As such, policies can be best understood as testers~\cite{theoreticalframeworkforquantumnetworks}. Testers---which, like POVMs, do not provide post-measurement states---are apt for our work because the `past' we care about is which sequence $\overleftarrow{z}$ occurred, given that we are applying policy $\mathbf{A}$. This is information that can be classically sent to a quantum computer simulating a model of the process. The main constraint on which testers are physically valid is that they must satisfy a hierarchy of trace conditions that generalises the completion relation of POVMs to the multitime case~\cite{theoreticalframeworkforquantumnetworks},
\begin{equation} \label{eq:conservationofprobability}
    \begin{aligned}
        1)& \quad \sum_{\overleftrightarrow{z}_{t-n':t+n-1}} \mathbf{A}_{\overleftrightarrow{z}_{t-n':t+n-1}} = \Xi_{t-n':t+n-1} \otimes \mathbbm{1}^{\text{out}_{t+n-1}}/d_s, \\
        2)& \quad \text{tr}_{\text{in}_{t+n-1}} \{ \Xi_{t-n':t+n-1} \} = \Xi_{t-n':t+n-2} \otimes \mathbbm{1}^{\text{out}_{t+n-2}}/d_s, \\
        \vdots &  \\
         n+n'+1)& \quad \text{tr}_{\text{in}_{t-n'}} \{ \Xi_{t-n':t-n'}  \} = 1,
    \end{aligned}
\end{equation}
where $\mathbf{A}_{\overleftrightarrow{z}_{t-n':t+n-1}}\geq 0$, for some set of positive semidefinite operators $\{ \Xi_{t-n':k}\}_{k=t-n'}^{t+n-1}$.

So long as this constraint is met, a tester can be as general as one desires. However, as mentioned in the main text, we also require that policies $\mathbf{A} \in \mathsf{A}^{*}$ only use a classical record for memory between episodes of evolution.

An important class of policies we utilise in this work is that of informationally complete policies.
\begin{definition}[Informationally complete policy] \label{def:informationallycomplete}
    Given a quantum process $\mathbf{T}$ with accessible subsystem $s$ of dimension $d_s$, accessed at times $t-n',\dots,t+n-1,$ a policy $\mathbf{A}=\{ \mathbf{A}_{\overleftrightarrow{z}} \}_{\overleftrightarrow{z} \in \text{W}(\mathsf{Z}) }$ satisfying Eq.~\eqref{eq:conservationofprobability} is informationally complete if any density matrix $\mathbf{S} \in \mathcal{D}\big( \mathcal{H}^{\text{\emph{in}}_{t-n'}} \otimes \mathcal{H}^{\text{\emph{out}}_{t-n'}} \otimes \dots \otimes \mathcal{H}^{\text{\emph{in}}_{t+n-1}} \otimes \mathcal{H}^{\text{\emph{out}}_{t+n-1}} \big)$ can be expressed as,
    \begin{equation}
        \mathbf{S} = \sum_{\overleftrightarrow{z}} c_{\overleftrightarrow{z}} \mathbf{A}_{\overleftrightarrow{z}},
    \end{equation}
    where $c_{\overleftrightarrow{z}}$ are real numbers.
\end{definition}
Intuitively, these are policies that probe every possible input-output relation that the process may exhibit. Practically they are policies that can be used as a basis for the vector space of Hermitian matrices that processes live on, hence enabling a tomographic representation of quantum processes.

While informationally complete policies do not require adaptivity, it is still worth detailing the full possibilities of $\mathsf{A}^{*}$. We generalise the state preparations and measurements to $\mathcal{P}^{sa}_{x_{\tau}|\overleftarrow{z}_{t-n':\tau-1}}$ and $\mathcal{M}^{sa}_{y_{\tau}|\overleftarrow{z}_{t-n':\tau-1},x_{\tau}}$, which have quantum memory within an episode facilitated by the experimenter's memory subsystem $a$, and can have classical memory between episodes, denoted by conditioning on the classical record. For $\mathbf{A} \in \mathsf{A}^{*}$ we can decompose $z_\tau=(x_\tau,y_\tau)$ where $\mathcal{P}^{sa}_{x_{\tau}|\overleftarrow{z}_{t-n':\tau-1}}$ prepares a unit-normalised state conditioned on the value of $x_\tau$ and everything prior, and an $sa$ measurement is performed conditioned on all of that information, with outcome $\mathcal{M}^{sa}_{y_{\tau}|\overleftarrow{z}_{t-n':\tau-1},x_{\tau}}$. Note that $x_\tau$ is trivial when the preparation is deterministic. The weights at a given episode can be decomposed as $w_{y_\tau|\overleftarrow{z}_{t-n':\tau-1},x_\tau} w_{x_\tau|\overleftarrow{z}_{t-n':\tau-1}}$, where $w_{x_\tau|\overleftarrow{z}_{t-n':\tau-1}}$ and $w_{y_\tau|\overleftarrow{z}_{t-n':\tau-1},x_\tau}$ are defined such that $\sum_{x_\tau} w_{x_\tau|\overleftarrow{z}_{t-n':\tau-1}} \mathcal{P}^{sa}_{x_{\tau}|\overleftarrow{z}_{t-n':\tau-1}}$ and $\sum_{y_\tau} w_{y_\tau|\overleftarrow{z}_{t-n':\tau-1},x_\tau} \mathcal{M}^{sa}_{y_{\tau}|\overleftarrow{z}_{t-n':\tau-1},x_{\tau}}$ are both trace preserving quantum maps. The weights of a full trajectory are given by the product of the weights of all individual times. Additionally, we enforce that quantum memory is confined within each episode by requiring that $w_{y_\tau|\overleftarrow{z}_{t-n':\tau-1},x_\tau} \mathcal{M}^{sa}_{y_{\tau}|\overleftarrow{z}_{t-n':\tau-1},x_{\tau}}$ is an element of a rank-1 measurement, with $\mathcal{M}^{sa}_{y_{\tau}|\overleftarrow{z}_{t-n':\tau-1},x_{\tau}}$ normalised such that $\text{tr}\{\mathcal{M}^{sa}_{y_{\tau}|\overleftarrow{z}_{t-n':\tau-1},x_{\tau}}(\mathbbm{1}^{sa})\}=1$.

Each element of a policy in $\mathsf{A}^{*}$ can be mathematically characterised via the Choi isomorphism as
\begin{equation} \label{eq:policychoi}
\begin{aligned}
   \mathbf{A}_{\overleftrightarrow{z}} = & w_{\overleftrightarrow{z}}{\text{tr}}_{a,\text{in}_{t+n}} \Bigg\{ \mathcal{M}^{\text{in}_{t+n},a}_{y_{t+n-1}|\overleftarrow{z}_{t-n':t+n-2},x_{t+n-1}}   \bigcirc_{\tau=t-n'}^{t+n-2}   \mathcal{P}^{{\text{in}}_{\tau+1},a}_{x_{\tau+1}|\overleftarrow{z}_{t-n':\tau}}  \circ \mathcal{M}^{{\text{in}}_{\tau+1},a}_{y_{\tau}|\overleftarrow{z}_{t-n':\tau-1},x_{\tau}} \\
    & \qquad \Bigg[ \bigotimes_{\tau=t-n'}^{t+n-1}  \Psi^{{\text{out}}_{\tau},{\text{in}}_{\tau+1}}   \otimes  P_{x_{t-n'}}^{\text{in}_{t-n'},a}  \Bigg] \Bigg\},
\end{aligned}
\end{equation}
and the policy $\mathbf{A}= \{ \mathbf{A}_{\overleftrightarrow{z}} \}_{\overleftrightarrow{z} \in \text{W}(\mathsf{Z}) }$ is the full collection of these individual action-outcome sequences $\mathbf{A}_{\overleftrightarrow{z}}$. $P_{x_{t-n'}}^{\text{in}_{t-n'},a} $ is the initial state on $s$ and $a$. $\Psi^{{\text{out}}_{\tau},{\text{in}}_{\tau+1}}$ is a unit-normalised maximally entangled state with subsystems corresponding to the output to the open evolution from $\mathbf{T}$ at step $\tau$ and the input at time $\tau+1$. We stress that the inputs (outputs) are those of the open evolution (which we formally define in Sec.~\ref{sec:physicalscenario} and Eq.~\eqref{eq:quantumprocessdef}), which are the outputs (inputs) of the policy $\mathbf{A}$. This makes it easier to identify common indices when the two objects are interlocked. The final trace on $a$ corresponds to discarding the memory of the experimenter at the conclusion of the experiment, and the trace of $\text{in}_{t+n}$ is performed because we do not care about the state of $s$ once the experiment has concluded.

\section{Choi Trace Distance and Operational Distinguishability} \label{app:operational_distinguishability}
In this section, we show that under the hypotheses relevant to our framework, a vanishing error defined in terms of trace norm difference between unit-normalised Choi states is equivalent to a vanishing error for observational statistics. This confirms that the trace norm is an operationally meaningful choice to define distance for Def.~\ref{def:per-realisationfaithfulness}, affirming that Thm.~\ref{thm:predictioncorrespondence} actually constrains observable statistics.

Firstly, we restate our error quantifier from Def.~\ref{def:per-realisationfaithfulness}, reminding the reader that $\ket{\sigma^m_{k,\mathbf{A}_{\overleftarrow{z}}}}$ is an element of some pure-state ensemble assigned to the memory state $\rho^m$ encoded by a history $\mathbf{A}_{\overleftarrow{z}}$, according to some policy $\mathbf{A}$, and that $\mathbf{T}^{\mathbf{A}_{\overleftarrow{z}}}_{t:t+n-1}$ is the true future of the process $\mathbf{T}$ having observed $\mathbf{A}_{\overleftarrow{z}}$:
\begin{equation} \label{eq:PAC_restated}
       E_{n,n'}(k,\overleftarrow{z},\mathbf{A}):= \left\Vert \Phi_n\big(\ket{\sigma^m_{k,\mathbf{A}_{\overleftarrow{z}}}} \bra{\sigma^m_{k,\mathbf{A}_{\overleftarrow{z}}}}  \big) - \mathbf{T}^{\mathbf{A}_{\overleftarrow{z}}}_{t:t+n-1} \right\Vert,
    \end{equation}
where $\Vert \cdot \Vert$ is the trace norm.

We now define the operational counterpart. Let $\mathbf{B}$ be a policy to be performed in the future, and let the number of future episodes $n \in \mathbbm{N}$ be arbitrary. Given a history $\mathbf{A}_{\overleftarrow{z}}$, the probability of a future trajectory $\mathbf{B}_{\overrightarrow{z}}$ is given by the Born rule as in Eq.~\eqref{eq:probabilityoftrajectory},
\begin{equation} \label{eq:future_only_born_rule}
p(\overrightarrow{z}|\mathbf{B},\mathbf{T}^{\mathbf{A}_{\overleftarrow{z}}}_{t:t+n-1})= d_s^{2n} \text{tr}\{\mathbf{B}_{\overrightarrow{z}}^{\text{T}} \mathbf{T}^{\mathbf{A}_{\overleftarrow{z}}}_{t:t+n-1} \}.
\end{equation}
Then, we can define the operational error as the supremum of summed differences in trajectory probabilities between the true future $\mathbf{T}^{\mathbf{A}_{\overleftarrow{z}}}_{t:t+n-1}$ and that produced by a given memory ensemble element $\ket{\sigma^m_{k,\mathbf{A}_{\overleftarrow{z}}}}$,
\begin{equation} \label{eq:PAC_operational}
       E^{\text{op}}_{n,n'}(k,\overleftarrow{z},\mathbf{A}):= \sup_{\mathbf{B}} \sum_{\overrightarrow{z}}\Bigg\vert p\Big(\overrightarrow{z}|\mathbf{B},\Phi_n\big(\ket{\sigma^m_{k,\mathbf{A}_{\overleftarrow{z}}}} \bra{\sigma^m_{k,\mathbf{A}_{\overleftarrow{z}}}}  \big) \Big) - p\Big(\overrightarrow{z}|\mathbf{B},\mathbf{T}^{\mathbf{A}_{\overleftarrow{z}}}_{t:t+n-1}\Big) \Bigg\vert.
    \end{equation}
The supremum is taken over all policies (testers), allowing arbitrary memory and adaptivity. A model is measurement $(\mathbf{A},\mathbf{T})$-per-realisation faithful ($\texttt{MpRF}$) if it satisfies Def.~\ref{def:per-realisationfaithfulness}, swapping out $ E_{n,n'}(k,\overleftarrow{z},\mathbf{A})$ for $E^{\text{op}}_{n,n'}(k,\overleftarrow{z},\mathbf{A})$, and correspondingly $q_{n,n'}(\delta)$ for $q^{\text{op}}_{n,n'}(\delta)$.

We show that $\texttt{pRF}$ is equivalent to $\texttt{MpRF}$. The key ingredient in establishing this is a lemma showing that there exists a finite future window length $n_0$ that allows trace distance error to define a uniform upper bound on operational error for all future window lengths $n \geq n_0$.
\begin{lemma}[Bound on Measurement Error] \label{lem:operationalPRF}
    Let $\mathscr{M}(\mathbf{T})$ be a faithful model of a stationary process $\mathbf{T}$ with $d_m<\infty$. Then, there exists an $n_0 \in \mathbbm{N}$ and a constant $C \in [1,\infty)$ independent of $n$ such that for all history lengths $n'$, for all histories $\mathbf{A}_{\overleftarrow{z}}$ of all policies $\mathbf{A}$ satisfying $p(\overleftarrow{z}|\mathbf{A})>0$, and every ensemble element $\ket{\sigma^m_{k,\mathbf{A}_{\overleftarrow{z}}}}$, 
    \begin{equation}
       E^{\text{\emph{op}}}_{n,n'}(k,\overleftarrow{z},\mathbf{A}) \leq C E_{n,n'}(k,\overleftarrow{z},\mathbf{A}),
    \end{equation}
    for all $n \geq n_0$.
\end{lemma}
\begin{proof}
Using Eq.~\eqref{eq:future_only_born_rule}, we define a quantum map $\Gamma^{\mathbf{B}}_n$ acting on $\mathcal{L}\big( \mathcal{H}^{m }\big)$ such that
    \begin{equation}
        \Gamma^{\mathbf{B}}_n(\cdot) := d_s^{2n}  \sum_{\overrightarrow{z}}\text{tr} \big\{ \mathbf{B}_{\overrightarrow{z}}^{\text{T}} \Phi_n(\cdot) \big\} \ket{\overrightarrow{z}_{t:t+n-1}} \bra{\overrightarrow{z}_{t:t+n-1}},
    \end{equation}
where $\ket{\overrightarrow{z}_{t:t+n-1}}$ are orthonormal vectors corresponding to trajectories of future classical outcomes. Since all elements of $\mathbf{B}$ are positive, and $\Phi_n$ is CP, $\Gamma^{\mathbf{B}}_n$ is CP. Since $\Phi_n$ is trace preserving, mapping input states to valid process tensors, and testers on process tensors yield valid probability distributions, $\Gamma^{\mathbf{B}}_n$ is trace preserving. Thus, $\Gamma^{\mathbf{B}}_n$ is a quantum channel.

Consider some arbitrary Hermitian linear operator $Y \in \mathcal{L}\big( \mathcal{H}^{m }\big)$. We decompose it into its positive and negative orthogonal contributions via its eigendecomposition: 
\begin{equation}
    Y=\sum_{\lambda_i>0} \lambda_i \ket{\psi^m_i}  \bra{\psi^m_i}- \sum_{\lambda_j<0} (-\lambda_j) \ket{\psi^m_j}\bra{\psi^m_j} ,
\end{equation}
where the first sum is denoted by $Y_{+}$ and the second by $Y_{-}$, both satisfying $Y_{\pm}\geq 0$. Then, we can write $Y=Y_{+}-Y_{-}$. We use this decomposition to show contractivity of $\Phi_n$ and $\Gamma^{\mathbf{B}}_n$ under the trace norm:
\begin{equation} \label{eq:trace_norm_bound_1}
    \big\Vert \Phi_n(Y) \big\Vert \leq \big\Vert \Phi_n(Y_{+}) \big\Vert + \big\Vert \Phi_n(Y_{-}) \big\Vert = \text{tr}\{ Y_{+} \} + \text{tr}\{ Y_{-} \} = \Vert Y \Vert,
\end{equation}
and by the same argument,
\begin{equation} \label{eq:trace_norm_bound_2}
    \big\Vert \Gamma^{\mathbf{B}}_n(Y) \big\Vert \leq  \Vert Y \Vert.
\end{equation}

Next, we study how this contractivity depends on $n$. It holds directly from the definition of $\Phi_n$ in Eq.~\eqref{eq:quantumprocessdef} that 
\begin{equation}
    \Phi_n= \text{tr}_{\text{in}_{t+n},\text{out}_{t+n}} \circ \Phi_{n+1},
\end{equation}
i.e. future Choi states for shorter futures are partial traces of those for longer futures. Combining this with the fact that the partial trace is itself a quantum channel, we have the contractive hierarchy,
\begin{equation} \label{eq:output_trace_norm_hierarchy}
    \big\Vert \Phi_n(Y) \big\Vert \leq \big\Vert \Phi_{n+1}(Y) \big\Vert.
\end{equation}

Let $\mathsf{G}$ be the $(d_m^2-1)$-dimensional real vector space of traceless Hermitian operators acting on the model's memory Hilbert space $\mathcal{H}^{m}$. Denote the kernel of $\Phi_n$ within $\mathsf{G}$ by $\mathsf{K}_n$. Due to Eq.~\eqref{eq:output_trace_norm_hierarchy}, $\mathsf{K}_{n+1} \subseteq \mathsf{K}_n$. Since $d_m<\infty$, $\mathsf{G}$ is finite dimensional, and because a non-increasing sequence of subspaces is eventually constant, there exists a finite $n_0$ such that for all $n\geq n_0$, $\mathsf{K}_n=\mathsf{K}_{n_0}:=\mathsf{K}$. Thus, for all $n\geq n_0$ and all $\kappa \in \mathsf{K}$, $\Phi_n(\kappa)=0$. 

For any $X \in \mathsf{G}$, consider adding any $\kappa \in \mathsf{K}$ to the above equations: $\Phi_n(X+\kappa)=\Phi_n(X)$ and $\Gamma^{\mathbf{B}}_n(X+\kappa)=\Gamma^{\mathbf{B}}_n(X)$. Contributions to the input that are within the kernel do not modify the resultant future, and hence either notion of error. What these contributions can influence, however, is the trace norm of the input, i.e. $\Vert X+\kappa \Vert \neq \Vert X \Vert$ in general, which impacts our bound in Eq.~\eqref{eq:trace_norm_bound_1}. 

Consider, for a fixed $X \in \mathsf{G}$, the smallest value one can obtain for the trace norm of the input to $\Phi_{n_0}$ over all added kernel elements, $W(X):=\min_{\kappa \in \mathsf{K}}\Vert X+\kappa \Vert$. We will address the existence of the minimum shortly. $W(X)$ can be thought of as the weight (norm) of $X$ as measured by trace norm, \emph{modulo everything in the kernel} $\mathsf{K}$, and is a seminorm. This enables us to `squash' our bound in Eq.~\eqref{eq:trace_norm_bound_1}, given that $X \in \mathsf{G} \subset \mathcal{L}(\mathcal{H}^{m})$,
\begin{equation} \label{eq:squashed_bound}
    \Vert \Phi_{n_0}(X) \Vert = \Vert \Phi_{n_0}(X+\kappa^*) \Vert \leq \Vert X+\kappa^* \Vert = W(X),
\end{equation}
where $\kappa^* \in \mathsf{K}$ is a kernel element that attains the minimum in $W(X)$.

Now consider the following set on $\mathsf{G}$ defined by $W$,
\begin{equation}
    \mathsf{S}_{W}:=\{ X \in \mathsf{G} : W(X)=1  \}.
\end{equation} 
For all $X \in \mathsf{S}_{W}$, due to Eq.~\eqref{eq:squashed_bound}, it holds that $\Vert \Phi_{n_0}(X) \Vert \leq 1$. We seek to find the minimum value of $\Vert \Phi_{n_0}(X) \Vert$ over this set:
\begin{equation} \label{eq:minimise_Q}
    Q:= \min \big\{ \Vert \Phi_{n_0}(X) \Vert : X \in \mathsf{S}_{W} \big\}.
\end{equation}
In the trivial case where $\mathsf{K}=\mathsf{G}$, the memory is invisible to the future, and we fix $Q=1$.

We seek to show that this minimum is attained. Firstly, we show that for all $X \in \mathsf{S}_{W}$ there exists a $\kappa^*$ that attains $W(X)$. If $\Vert \kappa \Vert > 2 \Vert X \Vert$, then  $\Vert X + \kappa  \Vert \geq \Vert \kappa \Vert - \Vert X \Vert > \Vert X \Vert$. Thus, for $\Vert \kappa \Vert > 2 \Vert X \Vert$, its addition increases the trace norm, rather than decreasing it. With this information, we can confine our minimisation to the ball of $\kappa \in \mathsf{K}$ such that $\Vert \kappa \Vert \leq 2 \Vert X \Vert$. This ball is compact because $\mathsf{K}$ is finite dimensional, and thus the minimum is attained for all $X \in \mathsf{G}$, which, in turn implies it is attained for $X \in \mathsf{S}_{W}$. 

Now, we turn our attention to the minimisation over $X \in \mathsf{S}_{W}$. We define $\tilde{\mathsf{S}}_{W} \subseteq \mathsf{S}_{W}$,
\begin{equation}
    \tilde{\mathsf{S}}_{W}:=\big\{ X \in \mathsf{G} : W(X)=1, \quad \Vert X \Vert =1  \big\},
\end{equation} 
which differs only in that all elements also satisfy $\Vert X \Vert =1$. The key point here is that every $X \in \mathsf{S}_{W}$ has a counterpart in $\tilde{\mathsf{S}}_{W}$ that obtains the same value in Eq.~\eqref{eq:minimise_Q}. To see this, observe that for all $X \in \mathsf{S}_W$, there exists an optimal $\kappa^*$ such that $1=W(X)=\Vert X + \kappa^* \Vert$, noting that $X+\kappa^* \in \mathsf{G}$ because $\mathsf{K} \subseteq \mathsf{G}$. Due to the fact that $\mathsf{K}$ is a vector space, adding $\kappa^*$ to every element of $\mathsf{K}$ does not change the value obtained by optimising over it, meaning that $W(X + \kappa^*)=W(X)=1$, which implies that $X + \kappa^* \in \tilde{\mathsf{S}}_{W}$.

This counterpart yields the same value in the optimisation: $\Phi_{n_0}(X)=\Phi_{n_0}(X+\kappa^*)$. Thus, the same value is obtained whether the optimisation is taken over $\mathsf{S}_{W}$ or $\tilde{\mathsf{S}}_{W}$.

We show that $W$ is continuous on $\mathsf{G}$ in order to establish the closedness of $\tilde{\mathsf{S}}_{W}$. For all $X,X' \in \mathsf{G}$, let $\kappa'$ attain $W(X')$. Then, it holds that $W(X) \leq \Vert X + \kappa' \Vert \leq \Vert X - X' \Vert+W(X')$, and the same holds for $X \leftrightarrow X'$. This implies that $\vert W(X)-W(X') \vert \leq \Vert X-X' \Vert $ for all $X,X' \in \mathsf{G}$, which in turn, means that $W$ is continuous on $\mathsf{G}$. Given that $\tilde{\mathsf{S}}_{W}$ is the intersection on $\mathsf{G}$ of two equality-defined sets, via the functions $W$ and $\Vert \cdot \Vert$ respectively, continuity of both $W$ and $\Vert \cdot \Vert$ provides that $\tilde{\mathsf{S}}_{W}$ is closed. Thus, $\tilde{\mathsf{S}}_{W}$ is a closed, bounded subset of a finite-dimensional space, and hence is compact. 

$\Vert \Phi_{n_0}(\cdot) \Vert$ is also a continuous function, meaning that its minimum is attained on $\tilde{\mathsf{S}}_{W}$. Finally, since $X \in \mathsf{S}_{W}$, it must be that $W(X)=1$, implying that $X$ cannot also be in $\mathsf{K}$, and hence that $Q>0$ strictly.

As shorthand, denote $\ket{\sigma^m_{k,\mathbf{A}_{\overleftarrow{z}}}} \bra{\sigma^m_{k,\mathbf{A}_{\overleftarrow{z}}}}$ by $\sigma^m_k$ for a given $\mathbf{A}$, $\mathbf{A}_{\overleftarrow{z}}$, and ensemble assignment.
Next we define the following difference: $\tilde{X}:=\sigma^m_k-e(\mathbf{A}_{\overleftarrow{z}}) \in \mathsf{G}$. Since $\Phi_n$ and $\Gamma^{\mathbf{B}}_n$ are both linear, and $\mathscr{M}(\mathbf{T})$ is faithful, it holds that 
    \begin{equation} \label{eq:trace_dist_error_phi}
        E_{n,n'}(k,\overleftarrow{z},\mathbf{A})=\Vert \Phi_n(\tilde{X}) \Vert,
    \end{equation}
    and
\begin{equation} \label{eq:operational_error_gamma}
         E^{\text{op}}_{n,n'}(k,\overleftarrow{z},\mathbf{A}) = \sup_{\mathbf{B}} \big\Vert \Gamma^{\mathbf{B}}_n(\tilde{X}) \big\Vert,
    \end{equation}
respectively. If $\tilde{X} \in \mathsf{K}$, then for all $n$ and $\mathbf{B}$, $\Vert \Phi_n(\tilde{X}) \Vert = \Vert \Gamma^{\mathbf{B}}_n(\tilde{X}) \big\Vert =0$. Otherwise, the operator $\tilde{X}':=\tilde{X}/W(\tilde{X})$ is a member of $\mathsf{S}_{W}$, yielding that
\begin{equation}
    Q \leq \Vert \Phi_{n_0}(\tilde{X}') \Vert \leq 1,
\end{equation}
where the second inequality is from Eq.~\eqref{eq:squashed_bound}. Multiplying by $W(\tilde{X})$, we obtain 
\begin{equation} \label{eq:weight_sandwich}
    Q W(\tilde{X}) \leq \Vert \Phi_{n_0}(\tilde{X}) \Vert \leq W(\tilde{X}),
\end{equation}
for all $\tilde{X} \in \mathsf{G}$, noting that all three quantities vanish for $\tilde{X} \in \mathsf{K}$.

The crucial point to observe here is that the value of $Q$ does not depend on the policy $\mathbf{A}$, the history $\mathbf{A}_{\overleftarrow{z}}$, the input memory state $\sigma^m_k$, or the future window size $n$. It only depends on the evolution channel $\mathcal{T}^{sm}$ of the model itself.

Finally, we combine the ingredients we have established in this proof. Invoking Eq.~\eqref{eq:trace_norm_bound_2} and `squashing' the bound as in Eq.~\eqref{eq:squashed_bound}, we obtain, for all $\mathbf{A}$, $\mathbf{A}_{\overleftarrow{z}}$ satisfying $p(\overleftarrow{z}|\mathbf{A})>0$, and $\sigma_k^m$,
\begin{equation} \label{eq:upper_bound_on_operational_error_1}
    \big\Vert \Gamma^{\mathbf{B}}_n(\tilde{X}) \big\Vert \leq W(\tilde{X}).
\end{equation}
This holds for all future policies $\mathbf{B}$, so it also holds for the supremum over those policies. We invoke Eq.~\eqref{eq:trace_dist_error_phi} and Eq.~\eqref{eq:operational_error_gamma} to convert $\Phi_n$ and $\Gamma^{\mathbf{B}}_n$ to errors, and combine Eq.~\eqref{eq:upper_bound_on_operational_error_1} with Eq.~\eqref{eq:weight_sandwich}. The resulting expression is
\begin{equation}
    E^{\text{op}}_{n,n'}(k,\overleftarrow{z},\mathbf{A}) \leq \frac{1}{Q}  E_{n_0,n'}(k,\overleftarrow{z},\mathbf{A}) \leq \frac{1}{Q}  E_{n,n'}(k,\overleftarrow{z},\mathbf{A}).
\end{equation}
For the second inequality, we invoked Eq.~\eqref{eq:output_trace_norm_hierarchy}. Letting $C=1/Q$, and noting that $0<Q \leq 1$, we arrive at
\begin{equation}
    E^{\text{op}}_{n,n'}(k,\overleftarrow{z},\mathbf{A}) \leq C  E_{n,n'}(k,\overleftarrow{z},\mathbf{A}),
\end{equation}
with $1 \leq C < \infty$.
\end{proof}

The core result follows directly.
\begin{proposition}[Equivalence of $\texttt{pRF}$ and $\texttt{MpRF}$]
    Let $\mathscr{M}(\mathbf{T})$ be a faithful model of a stationary process $\mathbf{T}$ with $d_m<\infty$. Then, $\mathscr{M}(\mathbf{T})$ is $(\mathbf{A},\mathbf{T})$-$\texttt{pRF}$ if and only if it is $(\mathbf{A},\mathbf{T})$-$\texttt{MpRF}$.
\end{proposition}
\begin{proof}
   For the `only if' direction, invoking Lem.~\ref{lem:operationalPRF}, for all $n \geq n_0$, $E^{\text{op}}_{n,n'}(k,\overleftarrow{z},\mathbf{A}) > C \delta$ implies $E_{n,n'}(k,\overleftarrow{z},\mathbf{A}) > \delta$. Then, $q_{n,n'}(\delta) \geq q^{\text{op}}_{n,n'}(C \delta)$. $\mathscr{M}(\mathbf{T})$ being $(\mathbf{A},\mathbf{T})$-$\texttt{pRF}$ implies that $q_{n,n'}(\delta) \rightarrow 0$ as $n=n' \rightarrow \infty$ for all values of $\delta>0$. Noting that $1 \leq C < \infty$, it then follows that $q^{\text{op}}_{n,n'}(C \delta) \rightarrow 0$ for all values of $C \delta >0$.

   The `if' direction follows directly: within the supremum in Eq.~\eqref{eq:PAC_operational}, we allow policies that construct the $n$-step Choi state, and perform a Helstrom measurement on that Choi state. This policy $\tilde{\mathbf{B}}$ feeds halves of maximally entangled pairs $\Psi^{{\text{in}}_{\tau},{\text{out}}_\tau}$ into each episode of the future, $t\leq \tau \leq t+n-1$, and all subsystems are retained (as in Eq.~\eqref{eq:quantumprocessdef})---we imposed no restrictions on the memory of the policies. We note that $\tilde{\mathbf{B}}$ is defined for all finite $n$, which we will later use to take the limit of $n=n'\rightarrow \infty$. At the conclusion of episode $t+n-1$, this yields one of a pair of states, $\Phi_n(\sigma^m_k)$ or $\mathbf{T}^{\mathbf{A}_{\overleftarrow{z}}}_{t:t+n-1}$, for the model and true process respectively. At this time, $\tilde{\mathbf{B}}$ enacts a Helstrom measurement on these quantum states, where trace distance is precisely the operational distinguishability under this optimal measurement, so,
   \begin{equation}
       \frac{1}{2}\sum_{\overrightarrow{z}}\Bigg\vert p\Big(\overrightarrow{z}|\tilde{\mathbf{B}},\Phi_n\big(\ket{\sigma^m_{k,\mathbf{A}_{\overleftarrow{z}}}} \bra{\sigma^m_{k,\mathbf{A}_{\overleftarrow{z}}}}  \big) \Big) - p\Big(\overrightarrow{z}|\tilde{\mathbf{B}},\mathbf{T}^{\mathbf{A}_{\overleftarrow{z}}}_{t:t+n-1}\Big) \Bigg\vert = \frac{1}{2}E_{n,n'}(k,\overleftarrow{z},\mathbf{A}).
   \end{equation} 
   Here $\overrightarrow{z}$ are the measurement outcomes, and the factors of $\frac{1}{2}$ are placed on both sides so that the RHS is precisely a trace distance. Observing that $\tilde{\mathbf{B}}$ is only one potential choice of $\mathbf{B}$, not necessarily the optimal one, we have that $E^{\text{op}}_{n,n'}(k,\overleftarrow{z},\mathbf{A}) \geq E_{n,n'}(k,\overleftarrow{z},\mathbf{A})$. 
   
   Then, $E_{n,n'}(k,\overleftarrow{z},\mathbf{A}) > \delta$ implies $E^{\text{op}}_{n,n'}(k,\overleftarrow{z},\mathbf{A}) >  \delta$, and $q_{n,n'}(\delta) \leq q^{\text{op}}_{n,n'}( \delta)$. The premise of $(\mathbf{A},\mathbf{T})$-$\texttt{MpRF}$ means that $q^{\text{op}}_{n,n'}(\delta) \rightarrow 0$ as $n=n' \rightarrow \infty$ for all values of $\delta>0$. As such, $q_{n,n'}(\delta) \rightarrow 0$ as $n=n' \rightarrow \infty$ too.
\end{proof}

\section{Proof of Thm.~\ref{thm:predictioncorrespondence}: Exact Case} \label{app:finite_case}
In this section we establish the conceptual core of Thm.~\ref{thm:predictioncorrespondence}, making our subsequent, more mathematically complex, asymptotic proof more transparent. Here we establish that per-realisation faithfulness and non-oracularity are equivalent when they are required to hold exactly at finite history length $n'$. Note that we still assume infinite future length to ensure that oracular information does not go undetected due to short-sightedness. The finite past versions of Def.~\ref{def:per-realisationfaithfulness} and non-oracularity are as follows.
\begin{definition}[Finite Past Per-Realisation Faithfulness] \label{def:finite-past-per-realisationfaithfulness}
    A faithful model $\mathscr{M}(\mathbf{T})$ is finite past $(\mathbf{A},\mathbf{T})$-per-realisation faithful if there exists a history length $n'_{0}<\infty$ such that, for all histories $\mathbf{A}_{\overleftarrow{z}}$ with lengths at least $n'_{0}$ with $p(\overleftarrow{z}|\mathbf{A})>0$, and for all pure quantum states $\ket{\psi^m}$ in the support of $\rho^m:=e(\mathbf{A}_{\overleftarrow{z}})$,
    \begin{equation}
        \Phi \big( \ket{\psi^m} \bra{\psi^m} \big) = \mathbf{T}^{\mathbf{A}_{\overleftarrow{z}}}_{t:t+n-1} .
    \end{equation}
\end{definition}
Recall that $\Phi$ refers to the family of $\Phi_n$ as in Eq.~\eqref{eq:quantumprocessdef} for all values of $n\geq 1$.

For simplification of the coming arguments, we introduce the following lemma.
\begin{lemma}[Ensembles and Support] \label{lem:ensemblesandsupport}
    Given a quantum state $\rho$ acting on a finite-dimensional Hilbert space, the unit state vector $\ket{\psi}$ appears as a non-zero weight member of some finite pure-state ensemble of $\rho$ if and only if it lies in the support of $\rho$.
\end{lemma}
\begin{proof}
    If $\rho$ is pure, the statement is immediate. For the case where that does not hold, we begin with the `if' direction. For all $\ket{\psi}$ in the support of $\rho$, it holds that $\rho-\lambda_{\min} \ket{\psi} \bra{\psi} \geq 0$ where $\lambda_{\min}$ is the smallest non-zero eigenvalue of $\rho$. This is true because $\rho \geq \lambda_{\min} \Pi_{\rho} \geq \lambda_{\min} \ket{\psi}\bra{\psi}$, where $\Pi_{\rho}$ is the projector onto the support of $\rho$. Then $\ket{\psi}$ is a member of the ensemble consisting of $\big\{ \lambda_{\min}, \ket{\psi} \bra{\psi} \big\}$ and $\Big\{ 1-\lambda_{\min}, \frac{\rho-\lambda_{\min} \ket{\psi} \bra{\psi}}{1-\lambda_{\min}} \Big\}$, implying that all rays in the support are also elements of some ensemble. If $\frac{\rho-\lambda_{\min} \ket{\psi} \bra{\psi}}{1-\lambda_{\min}}$ is not a pure state, replace it in the ensemble with its eigenvectors, weighted by $(1-\lambda_{\min})$ times their eigenvalues. This yields a pure-state ensemble containing $\ket{\psi}$ with a positive weight $\lambda_{\min}$. For the `only if' direction, consider any state $\ket{\eta}$ in the kernel of $\rho$. This implies that $\braket{\eta | \rho | \eta}=0$. Writing out $\rho$ as an explicit ensemble containing $\ket{\psi}$ with weight $p_{\psi}$, this can be re-expressed as
    \begin{equation}
      0 = p_{\psi} \big\vert \braket{\eta|\psi} \big\vert^2 + \sum_{\phi \neq \psi} p_{\phi}\big\vert \braket{\eta | \phi}\big\vert^2 \geq p_{\psi} \big\vert \braket{\eta|\psi} \big\vert^2.
    \end{equation}
    Hence $p_{\psi} \big\vert \braket{\eta|\psi} \big\vert^2=0$, and since $p_{\psi}>0$, $\braket{\eta|\psi}=0$. Thus, $\ket{\psi}$ is orthogonal to the kernel of $\rho$. Since $\rho$ is Hermitian, its support is the orthogonal complement of its kernel, implying that $\ket{\psi}$ is in the support of $\rho$.
\end{proof}

\begin{definition}[Finite Past Non-Oracularity] \label{def:finite-past-non-oracularity}
$\mathscr{M}(\mathbf{T})$ is finite past $(\mathbf{A},\mathbf{T})$-non-oracular if there exists a history length $n'_{0}<\infty$ such that for all history lengths $n'$ at least $n'_{0}$,
\begin{equation} \label{eq:oracular_information_finite}
    \zeta_{n'}\big(\mathbf{A},\mathscr{M}(\mathbf{T})\big):= \lim_{n \rightarrow \infty}\sum_{\overleftarrow{z}} p(\overleftarrow{z}|\mathbf{A})I_{\mathbf{A}_{\overleftarrow{z}}}(f:m')=0.
\end{equation}
\end{definition}

Now, we state and prove the finite past version of Thm.~\ref{thm:predictioncorrespondence}.
\begin{proposition}[Memory Correspondence (Exact Case)] \label{prop:finite-past-predictioncorrespondence}
    Let $\mathscr{M}(\mathbf{T})$ be a faithful model of a stationary process $\mathbf{T}$ with $d_m < \infty$, and let $\mathbf{A} \in \mathsf{A}^{*}$. Then, $\mathscr{M}(\mathbf{T})$ is finite past $(\mathbf{A},\mathbf{T})$-per-realisation faithful if and only if it is finite past $(\mathbf{A},\mathbf{T})$-non-oracular.
\end{proposition}
\begin{proof}
We begin by showing that Def.~\ref{def:finite-past-per-realisationfaithfulness} implies Def.~\ref{def:finite-past-non-oracularity}. For a past $\mathbf{A}_{\overleftarrow{z}}$ with $p(\overleftarrow{z}|\mathbf{A})>0$, let $\rho^{m}:=e(\mathbf{A}_{\overleftarrow{z}})$. Consider the ensemble $\{p_k,\ket{\psi^m_k}\}$ with $p_k>0$, defined from the eigendecomposition $\sum_k p_k \ket{\psi^m_k}\bra{\psi^m_k}=\rho^m$ such that $\ket{\psi^m_k}$ are mutually orthogonal. This defines a purification $\ket{\psi^{mm'}_{\rho}}=\sum_k \sqrt{p_k} \ket{\psi^m_k} \otimes \ket{\phi^{m'}_k}$ for arbitrary orthonormal $\ket{\phi^{m'}_k}$. We begin by writing the full expression
\begin{equation}
    \Upsilon_{\mathbf{A}_{\overleftarrow{z}}} =  \sum_{k,k'} \sqrt{p_k p_{k'}} \Phi^{m \rightarrow{f}} \big( \ket{\psi^{m}_k} \bra{\psi^{m}_{k'}} \big) \otimes  \ket{\phi^{m'}_k} \bra{\phi^{m'}_{k'}}.
\end{equation}
We remind the reader that by suppressing $n'$ on $e$ and $n$ on $\Phi$, we are referring to the families $e_{n'}$ $\Phi_n$ for all finite $n,n'$ values, where $n'\geq n'_0$.

The next step is to prove that per-realisation faithfulness causes $\Phi$ to annihilate the off-diagonal terms. Consider the vectors $\ket{u^m}=\frac{1}{\sqrt{2}}\big( \ket{\psi^{m}_k} + \ket{\psi^{m}_{k'}} \big)$ and $\ket{v^m}=\frac{1}{\sqrt{2}} \big( \ket{\psi^{m}_k} + i\ket{\psi^{m}_{k'}} \big)$. Then,
\begin{equation} \label{eq:uexpansion}
\begin{aligned}
    & \Phi \big(\ket{u^m} \bra{u^m} \big) \\
    =& \mathbf{T}^{\mathbf{A}_{\overleftarrow{z}}} 
     + \frac{1}{{2}} \Big[ \Phi\big( \ket{\psi^{m}_k} \bra{\psi^{m}_{k'}} \big) + \Phi\big( \ket{\psi^{m}_{k'}} \bra{\psi^{m}_{k}} \big) \Big] ,
\end{aligned}
\end{equation}
and
\begin{equation}\label{eq:vexpansion}
\begin{aligned}
   & \Phi \big(\ket{v^m} \bra{v^m} \big) \\
    =& \mathbf{T}^{\mathbf{A}_{\overleftarrow{z}}} 
     + \frac{1}{{2}} \Big[ -i\Phi\big( \ket{\psi^{m}_k} \bra{\psi^{m}_{k'}} \big) + i\Phi\big( \ket{\psi^{m}_{k'}} \bra{\psi^{m}_{k}} \big) \Big].
\end{aligned}
\end{equation}
Since $\ket{\psi^{m}_k}$ and $\ket{\psi^{m}_{k'}}$ are in the support of $\rho^m$, so are $\ket{u^m}$ and $\ket{v^m}$. Thus, per-realisation faithfulness implies that $\Phi \big(\ket{u^m} \bra{u^m} \big)=\Phi \big(\ket{v^m} \bra{v^m} \big)= \mathbf{T}^{\mathbf{A}_{\overleftarrow{z}}}$. Combining this fact with Eq.~\eqref{eq:uexpansion} and Eq.~\eqref{eq:vexpansion} yields
\begin{equation} \label{eq:uexpansion2}
     \Phi\big( \ket{\psi^{m}_k} \bra{\psi^{m}_{k'}} \big) =- \Phi\big( \ket{\psi^{m}_{k'}} \bra{\psi^{m}_{k}} \big)  ,
\end{equation}
and
\begin{equation}\label{eq:vexpansion2}
  \Phi\big( \ket{\psi^{m}_k} \bra{\psi^{m}_{k'}} \big) =\Phi\big( \ket{\psi^{m}_{k'}} \bra{\psi^{m}_{k}} \big) ,
\end{equation}
respectively, which can only be simultaneously satisfied if $\Phi\big( \ket{\psi^{m}_k} \bra{\psi^{m}_{k'}} \big) =\Phi\big( \ket{\psi^{m}_{k'}} \bra{\psi^{m}_{k}} \big)=0 $. Observe that the exact cancellation of off-diagonals used in the argument we employ relies on the fact that $\braket{\psi^{m}_{k'} | \psi^{m}_{k}}=0$, but the conclusion yielded by our argument does not, as we shall see shortly. 

Now we can re-express $\Upsilon_{\mathbf{A}_{\overleftarrow{z}}}$ as an explicit function of the future length,
\begin{equation}
    \Upsilon_{\mathbf{A}_{\overleftarrow{z}}}(n) =   \mathbf{T}_{t:t+n-1}^{\mathbf{A}_{\overleftarrow{z}}} \otimes \sum_{k} p_k  \ket{\phi^{m'}_k} \bra{\phi^{m'}_{k}},
\end{equation}
which is a product state for all $n$, implying that $I_{\mathbf{A}_{\overleftarrow{z}}}(f:m')$ is identically zero for all $n$, and hence in the limit $n \rightarrow \infty$.

Now, repeating this argument for every $\mathbf{A}_{\overleftarrow{z}} \in \mathbf{A}$ satisfying $p(\overleftarrow{z}|\mathbf{A})>0$, we get that $\zeta_{n'}\big(\mathbf{A},\mathscr{M}(\mathbf{T})\big)=0$ for all $n'\geq n'_{0}$, since all terms in the average over histories with non-zero probability have zero $f$-$m'$ mutual information. 

We have established that plugging the \emph{eigenensemble-derived} purification of the encoded memory state into the prediction function yields a product state. The mutual information between the output of the prediction function and a purification of its input is purification-independent because mutual information is invariant under isometries on one half. Thus, while our argument explicitly leveraged the orthogonality property of eigenensembles, the result it establishes is purification-independent.

We move on to the converse direction: Def.~\ref{def:finite-past-non-oracularity} implies Def.~\ref{def:finite-past-per-realisationfaithfulness}. We begin this direction with the assumption of faithfulness as in Eq.~\eqref{eq:faithfulness1}.

We note that finite past oracular information defined in Def.~\ref{def:finite-past-non-oracularity} still uses the limit of $n \rightarrow \infty$. However, since mutual information is non-increasing under tracing out subsystems, any finite $n$ approximation of finite past oracular information will be no greater than the true value. With this in mind, adding in non-oracularity as a premise for each $n'\geq n'_{0}$, average mutual information being zero over histories $\mathbf{A}_{\overleftarrow{z}}$ that have a non-zero probability of occurrence $p(\overleftarrow{z}|\mathbf{A})>0$ implies that $\Upsilon_{\mathbf{A}_{\overleftarrow{z}}}$ is a product state about the $f$-$m'$ partition for all such histories for all future lengths $n$. Thus, where $\Upsilon_{\mathbf{A}_{\overleftarrow{z}}}$ does have a product form, non-oracularity prevents the measurement on $m'$ from changing the future. Combining this with faithfulness yields that, for every measurement outcome on $m'$, the result on $f$ is the state $\mathbf{T}^{\mathbf{A}_{\overleftarrow{z}}}$ in particular, i.e. a faithful forecast.

By the HJW Theorem, we can transform between any pair of purifications of $\rho^m$ via a unitary (or isometric if we assume dimensions can change) transformation $U$ that acts on $m'$ alone. Then, by embedding $U$ into a choice of rank-1 measurement on $m'$, it is always possible to steer the $m$ input for $\Phi$ into any other ensemble.

Then what we have established so far is: for all finite $n'\geq n'_{0}$, for all $\mathbf{A}_{\overleftarrow{z}}$ with $p(\overleftarrow{z}|\mathbf{A})>0$, all ensemble elements in all ensembles of $\rho^m$ yield the same future. By Lem.~\ref{lem:ensemblesandsupport}, this is equivalent to the statement that for $\mathbf{A}_{\overleftarrow{z}}$ with $p(\overleftarrow{z}|\mathbf{A})>0$, all rays in the support of $\rho^m$ produce the same future, which yields Def.~\ref{def:finite-past-per-realisationfaithfulness}.
\end{proof}
Observe that Def.~\ref{def:finite-past-per-realisationfaithfulness} is the exact, deterministic, finite history analogue of Def.~\ref{def:per-realisationfaithfulness} (expressed as a support condition due to Lem.~\ref{lem:ensemblesandsupport}), implying that Prop.~\ref{prop:finite-past-predictioncorrespondence} is the exact, deterministic, finite history analogue of Thm.~\ref{thm:predictioncorrespondence}.

\section{Proof of Thm.~\ref{thm:predictioncorrespondence}: Asymptotic Case} \label{app:asymptotic_case}
Moving on to the asymptotic case, we begin with the heart of the argument: a continuity lemma that bounds prediction error in terms of oracular information for finite ensembles.
\begin{lemma}[Steering Continuity] \label{lem:steeringcontinuity}
Given a faithful model $\mathscr{M}(\mathbf{T})$ of a stationary process $\mathbf{T}$, a policy $\mathbf{A}$, and a history $\mathbf{A}_{\overleftarrow{z}}$, any finite pure-state ensemble decomposition $\{ p(k|\overleftarrow{z}), \ket{\sigma^m_{k,\mathbf{A}_{\overleftarrow{z}}}} \}$ of the encoded state $\rho^m$ satisfies
\begin{equation}
    \sum_{k} p(k|\overleftarrow{z}) E_{n,n'}(k,\overleftarrow{z},\mathbf{A}) \leq  \sqrt{ 2 \ln(2) I_{\mathbf{A}_{\overleftarrow{z}}}(f:m')}.
\end{equation}
\end{lemma}
\begin{proof}
    Starting with the eigendecomposition-derived purification $\ket{\psi^{mm'}_{\rho}}=\sum_j \sqrt{p_j} \ket{\psi^m_j} \otimes \ket{\phi^{m'}_j}$ of $\rho^m$ for arbitrary mutually orthogonal $\ket{\phi^{m'}_k}$, the HJW Theorem implies the existence of a POVM $\{ E^{m'}_k \}$ on $m'$ such that $\text{tr}_{m'} \big\{  \mathbbm{1}^{m} \otimes E^{m'}_k \ket{\psi^{mm'}_{\rho}} \bra{\psi^{mm'}_{\rho}} \big\}= p(k|\overleftarrow{z})\ket{\sigma^m_{k,\mathbf{A}_{\overleftarrow{z}}}} \bra{\sigma^m_{k,\mathbf{A}_{\overleftarrow{z}}}}$, yielding the ensemble $\{ p(k|\overleftarrow{z}), \ket{\sigma^m_{k,\mathbf{A}_{\overleftarrow{z}}}} \}$. Note that we define $m'$ such that $\mathcal{H}^{m'}:=\mathrm{span}\big( \{ \ket{\phi^{m'}_j} \} \big)$. This loses no generality for our purposes, because oracular information is purification-independent.
    
    Since $\Phi_n$ acts on $m$ alone, the conditional state on $f$ given ensemble outcome $k$ is $\Phi_n\big( \ket{\sigma^m_{k,\mathbf{A}_{\overleftarrow{z}}}} \bra{\sigma^m_{k,\mathbf{A}_{\overleftarrow{z}}}} \big)$. Now, consider the measure-and-replace channel $\Lambda^{m' \rightarrow m''}(\cdot)=\sum_{k}  \text{tr} \{  E^{m'}_k \cdot \} \ket{k^{m''}}\bra{k^{m''}}$, using an orthonormal set $\big\{ \ket{k^{m''}} \big\}_{k=1}^{N}$, where $N$ is the number of elements in the ensemble. This is realised by implementing the POVM from our invocation of the HJW Theorem. Upon applying $\Lambda^{m'\rightarrow m''}$ to each term, we take the trace norm of the difference between $\Upsilon^{fm'}_{\mathbf{A}_{\overleftarrow{z}}}$ and the product of its marginals $\Upsilon^{f}_{\mathbf{A}_{\overleftarrow{z}}} \otimes \Upsilon^{m'}_{\mathbf{A}_{\overleftarrow{z}}}$,
    \begin{equation}
       \left\Vert \Lambda^{m' \rightarrow m''}( \Upsilon^{fm'}_{\mathbf{A}_{\overleftarrow{z}}} ) - \Lambda^{m' \rightarrow m'' }( \Upsilon^{f}_{\mathbf{A}_{\overleftarrow{z}}} \otimes \Upsilon^{m'}_{\mathbf{A}_{\overleftarrow{z}}} ) \right\Vert  = \sum_{k} p(k|\overleftarrow{z}) E_{n,n'}(k,\overleftarrow{z},\mathbf{A}) ,
    \end{equation}
    using that $\text{tr} \{ \Upsilon^{m'}_{\mathbf{A}_{\overleftarrow{z}}} E^{m'}_k  \}= p(k|\overleftarrow{z})$, the mutual orthogonality of $\ket{k^{m''}}$, and faithfulness. Next, we use contractivity of trace distance under quantum channels, and invoke Pinsker's inequality,
    \begin{equation}
             \sum_{k} p(k|\overleftarrow{z}) E_{n,n'}(k,\overleftarrow{z},\mathbf{A}) 
            \leq  \left\Vert  \Upsilon^{fm'}_{\mathbf{A}_{\overleftarrow{z}}} - \Upsilon^{f}_{\mathbf{A}_{\overleftarrow{z}}} \otimes \Upsilon^{m'}_{\mathbf{A}_{\overleftarrow{z}}}  \right\Vert 
            \leq  \sqrt{ 2 \ln(2) I_{\mathbf{A}_{\overleftarrow{z}}}(f:m')}.
    \end{equation}
    Note that we are measuring information in bits.
\end{proof}

Thus, we can use Lem.~\ref{lem:steeringcontinuity} to directly establish the first direction of the proof of Thm.~\ref{thm:predictioncorrespondence} for all ensemble assignments.
\begin{proposition}[Memory Correspondence `If'] \label{prop:predictioncorrespondence_direction_1}
    Let $\mathscr{M}(\mathbf{T})$ be a faithful model of a stationary process $\mathbf{T}$, and let $\mathbf{A} \in \mathsf{A}^{*}$. Then, $\mathscr{M}(\mathbf{T})$ is $(\mathbf{A},\mathbf{T})$-per-realisation faithful if it is $(\mathbf{A},\mathbf{T})$-non-oracular.
\end{proposition}
\begin{proof}
    Fix an assignment and a value $\delta>0$. Denote the finite past and future oracular information by $\zeta_{n,n'}\big(\mathbf{A},\mathscr{M}(\mathbf{T})\big).$ We average the result of Lem.~\ref{lem:steeringcontinuity} over all histories,
    \begin{equation} \label{eq:expandsquareroot}
        \sum_{\overleftarrow{z}}\sum_{k} p(k|\overleftarrow{z})p(\overleftarrow{z}|\mathbf{A}) E_{n,n'}(k,\overleftarrow{z},\mathbf{A}) \leq \sum_{\overleftarrow{z}} p(\overleftarrow{z}|\mathbf{A}) \sqrt{ 2 \ln(2) I_{\mathbf{A}_{\overleftarrow{z}}}(f:m')} \leq  \sqrt{ 2 \ln(2)  \zeta_{n,n'}\big(\mathbf{A},\mathscr{M}(\mathbf{T})\big) },
    \end{equation}
observing that the average of square roots is a lower bound on the square root of the average. Finally, we use the fact that the LHS of Eq.~\eqref{eq:expandsquareroot} is an expectation value of the error $E_{n,n'}(k,\overleftarrow{z},\mathbf{A})$, allowing us to invoke Markov's inequality and bound the probability that the error exceeds our threshold $\delta$,
\begin{equation} \label{eq:markovsinequality}
        q_{n,n'}(\delta) \leq  \frac{1}{\delta}\sum_{\overleftarrow{z}}\sum_{k} p(k|\overleftarrow{z})p(\overleftarrow{z}|\mathbf{A}) E_{n,n'}(k,\overleftarrow{z},\mathbf{A}) \leq \frac{1}{\delta} \sqrt{2 \ln(2)  \zeta_{n,n'}\big(\mathbf{A},\mathscr{M}(\mathbf{T})\big) }.
    \end{equation}
Given that Eq.~\eqref{eq:markovsinequality} holds for all $n,n'$, and that  $\zeta_{n,n'}\big(\mathbf{A},\mathscr{M}(\mathbf{T})\big) \rightarrow 0$ is a premise, we conclude that $q_{n,n'}(\delta) \rightarrow 0$ for all $\delta>0$, as $n=n'\rightarrow \infty$.
\end{proof}

We move on to the setup for the converse direction. This requires finding an approximate form for our off-diagonal annihilation argument from the exact proof.
\begin{lemma}[Approximate Off-Diagonal Annihilation] \label{lem:approximateoffdiagonalannihilation}
    Given a quantum state $\rho$ acting on a finite-dimensional Hilbert space, with an orthonormal basis decomposition $\rho=\sum_k p_k \ket{\psi_k}\bra{\psi_k}$, consider the following superpositions: $\ket{u_{\pm}}:=\frac{1}{\sqrt{2}}(\ket{\psi_k} \pm \ket{\psi_{k'}})$, and $\ket{v_{\pm}}:=\frac{1}{\sqrt{2}}(\ket{\psi_k} \pm i\ket{\psi_{k'}})$ for any $k\neq k'$. Then, for any quantum channel $\mathcal{E}$ and any state $\sigma$, if
    \begin{equation} \label{eq:error_off_diagonal}
        \Big\Vert \mathcal{E} \big( \ket{w} \bra{w} \big) - \sigma \Big\Vert  \leq \delta
    \end{equation}
    for all $\ket{w} \in \big\{ \ket{u_{+}},\ket{u_{-}},\ket{v_{+}},\ket{v_{-}} \big\}$, then
    \begin{equation} \label{eq:lemma3consequence}
     \Big\Vert   \mathcal{E}\big( \ket{\psi_k} \bra{\psi_{k'}} \big) \Big\Vert \leq 2 \delta
    \end{equation}
\end{lemma}
\begin{proof}
    Using the orthogonality of our decomposition, we can write the identities,
    \begin{equation}
    \begin{aligned}
        \ket{u_{+}} \bra{u_{+}} - \ket{u_{-}} \bra{u_{-}} =&  \ket{\psi_k} \bra{\psi_{k'}} +  \ket{\psi_{k'}} \bra{\psi_k},\\
        \ket{v_{+}} \bra{v_{+}} - \ket{v_{-}} \bra{v_{-}} =& -i\big( \ket{\psi_k} \bra{\psi_{k'}} - \ket{\psi_{k'}} \bra{\psi_k} \big).
    \end{aligned}
    \end{equation}
    Then, by linearity
    \begin{equation}
        \begin{aligned}
            \mathcal{E} \big( \ket{\psi_k} \bra{\psi_{k'}} +  \ket{\psi_{k'}} \bra{\psi_k} \big) =& \mathcal{E}\big(   \ket{u_{+}} \bra{u_{+}}\big) - \mathcal{E}\big(   \ket{u_{-}} \bra{u_{-}}\big) ,\\
             -i\mathcal{E} \big( \ket{\psi_k} \bra{\psi_{k'}} -  \ket{\psi_{k'}} \bra{\psi_k} \big) =& \mathcal{E}\big(   \ket{v_{+}} \bra{v_{+}}\big) - \mathcal{E}\big(   \ket{v_{-}} \bra{v_{-}}\big).
        \end{aligned}
    \end{equation}
This is equivalent to
    \begin{equation} \label{eq:four_cases_off_diagonal}
        \begin{aligned}
           X:= \mathcal{E} \big( \ket{\psi_k} \bra{\psi_{k'}} +  \ket{\psi_{k'}} \bra{\psi_k} \big) =& \Big[ \mathcal{E}\big(   \ket{u_{+}} \bra{u_{+}}\big) - \sigma \Big]- \Big[ \mathcal{E}\big(   \ket{u_{-}} \bra{u_{-}}\big)  - \sigma \Big],\\
             -iY:=-i\mathcal{E} \big( \ket{\psi_k} \bra{\psi_{k'}} -  \ket{\psi_{k'}} \bra{\psi_k} \big) =& \Big[ \mathcal{E}\big(   \ket{v_{+}} \bra{v_{+}}\big)  - \sigma \Big] -  \Big[ \mathcal{E}\big(   \ket{v_{-}} \bra{v_{-}}\big)  - \sigma \Big].
        \end{aligned}
    \end{equation}

Observing that the four terms of Eq.~\eqref{eq:error_off_diagonal} are enumerated in the RHS of Eq.~\eqref{eq:four_cases_off_diagonal}, we invoke the premise that the trace norm of each item is bounded by $\delta$. We use this to apply the triangle inequality to the trace norm $\frac{1}{2}\Vert X+Y \Vert= \Big\Vert \mathcal{E} \big( \ket{\psi_k} \bra{\psi_{k'}} \big) \Big\Vert$, 
\begin{equation} 
    \Big\Vert \mathcal{E} \big( \ket{\psi_k} \bra{\psi_{k'}} \big) \Big\Vert \leq 2 \delta.
\end{equation}
\end{proof}

This enables us to show the other direction of the memory correspondence.
\begin{proposition}[Memory Correspondence `Only If'] \label{prop:predictioncorrespondence_direction_2}
    Let $\mathscr{M}(\mathbf{T})$ be a faithful model of a stationary process $\mathbf{T}$ with $d_m < \infty$, and let $\mathbf{A} \in \mathsf{A}^{*}$. Then, $\mathscr{M}(\mathbf{T})$ is $(\mathbf{A},\mathbf{T})$-per-realisation faithful only if it is $(\mathbf{A},\mathbf{T})$-non-oracular.
\end{proposition}
\begin{proof}
   Given a past $\mathbf{A}_{\overleftarrow{z}}$ with $p(\overleftarrow{z}|\mathbf{A})>0$, let $\rho^{m}:=e(\mathbf{A}_{\overleftarrow{z}})$. Consider the ensemble $\{p(k|\overleftarrow{z}),\ket{\psi^m_k}\}$ defined from the eigendecomposition $\sum_k p(k|\overleftarrow{z}) \ket{\psi^m_k}\bra{\psi^m_k}=\rho^m$ such that $\ket{\psi^m_k}$ are mutually orthogonal. Since $d_m < \infty$, there is a finite number of non-zero eigenvalues. Using the purification $\ket{\psi^{mm'}_{\rho}}=\sum_k \sqrt{p(k|\overleftarrow{z})} \ket{\psi^m_k} \otimes \ket{\phi^{m'}_k}$, with mutually orthogonal $\ket{\phi^{m'}_k}$, we can write
   \begin{equation} \label{eq:cross_and_not}
   \begin{aligned}
    \Upsilon_{\mathbf{A}_{\overleftarrow{z}}} = & \sum_{k} p(k|\overleftarrow{z})  \Phi_n^{m \rightarrow{f}} \big( \ket{\psi^{m}_k} \bra{\psi^{m}_{k}} \big) \otimes  \ket{\phi^{m'}_k} \bra{\phi^{m'}_{k}} \\ 
    + & \sum_{k \neq k'} \sqrt{p(k|\overleftarrow{z}) p(k'|\overleftarrow{z})} \Phi_n^{m \rightarrow{f}} \big( \ket{\psi^{m}_k} \bra{\psi^{m}_{k'}} \big) \otimes  \ket{\phi^{m'}_k} \bra{\phi^{m'}_{k'}}.
    \end{aligned}
\end{equation}
We also observe that the marginals of $\Upsilon^{fm'}_{\mathbf{A}_{\overleftarrow{z}}}$ are $\Upsilon^{f}_{\mathbf{A}_{\overleftarrow{z}}}=\Phi_n^{m \rightarrow{f}} \big( \rho^m \big)$ and $\Upsilon^{m'}_{\mathbf{A}_{\overleftarrow{z}}}= \sum_{k} p(k|\overleftarrow{z})   \ket{\phi^{m'}_k} \bra{\phi^{m'}_{k}}$. We consider (twice) the trace distance between the $\Upsilon^{fm'}_{\mathbf{A}_{\overleftarrow{z}}}$ and the product of its own marginals:
\begin{equation}
   \zeta_{\text{tr}}(n,n') := \sum_{\overleftarrow{z}} p(\overleftarrow{z}|\mathbf{A}) \left\Vert \Upsilon^{fm'}_{\mathbf{A}_{\overleftarrow{z}}} - \Upsilon^{f}_{\mathbf{A}_{\overleftarrow{z}}} \otimes \Upsilon^{m'}_{\mathbf{A}_{\overleftarrow{z}}} \right\Vert.
\end{equation}
Using Eq.~\eqref{eq:cross_and_not} and invoking faithfulness, we can re-express the argument of the trace norm as
\begin{equation}
\begin{aligned}
  \Delta_{n,n'}(\overleftarrow{z}) :=&  \sum_{k} p(k|\overleftarrow{z}) \left[ \Phi_{n}^{m \rightarrow{f}} \big( \ket{\psi^{m}_k} \bra{\psi^{m}_{k}} \big)  - \mathbf{T}^{\mathbf{A}_{\overleftarrow{z}}} \right] \otimes  \ket{\phi^{m'}_k} \bra{\phi^{m'}_{k}} \\
  +& \sum_{(k,k') | k \neq k'} \sqrt{p(k|\overleftarrow{z}) p(k'|\overleftarrow{z})} \Phi_{n}^{m \rightarrow{f}} \big( \ket{\psi^{m}_k} \bra{\psi^{m}_{k'}} \big) \otimes  \ket{\phi^{m'}_k} \bra{\phi^{m'}_{k'}}.
  \end{aligned}
\end{equation}
Invoking the triangle inequality allows us to decompose the norm into distinct contributions
\begin{equation}
\begin{aligned}
    \big\Vert\Delta_{n,n'}(\overleftarrow{z}) \big\Vert \leq & \sum_{k} p(k|\overleftarrow{z}) \left\Vert \left[ \Phi_n^{m \rightarrow{f}} \big( \ket{\psi^{m}_k} \bra{\psi^{m}_{k}} \big)  - \mathbf{T}^{\mathbf{A}_{\overleftarrow{z}}} \right] \otimes  \ket{\phi^{m'}_k} \bra{\phi^{m'}_{k}} \right\Vert \\
    +& \sum_{(k,k') |k \neq k'}  \sqrt{p(k|\overleftarrow{z}) p(k'|\overleftarrow{z})}  \left\Vert\Phi_n^{m \rightarrow{f}} \big( \ket{\psi^{m}_k} \bra{\psi^{m}_{k'}} \big) \otimes  \ket{\phi^{m'}_k} \bra{\phi^{m'}_{k'}} \right\Vert.
    \end{aligned}
\end{equation}
We take the average over histories, and invoke multiplicativity of the trace norm for tensor products, observing that $\big\Vert \ket{\phi^{m'}_k} \bra{\phi^{m'}_{k'}}  \big\Vert \leq 1$, simplifying the expression to 
\begin{equation} \label{eq:prop_only_if_main_goal}
\begin{aligned}
   \sum_{\overleftarrow{z}}p(\overleftarrow{z}|\mathbf{A}) \big\Vert \Delta_{n,n'}(\overleftarrow{z}) \big\Vert \leq &\sum_{k,\overleftarrow{z}}p(\overleftarrow{z}|\mathbf{A})p(k|\overleftarrow{z}) E_{n,n'}(k,\overleftarrow{z},\mathbf{A}) \\
   + & \sum_{(k,k')|k\neq k',\overleftarrow{z}}p(\overleftarrow{z}|\mathbf{A})  \sqrt{p(k|\overleftarrow{z}) p(k'|\overleftarrow{z})}  \left\Vert\Phi_n^{m \rightarrow{f}} \big( \ket{\psi^{m}_k} \bra{\psi^{m}_{k'}} \big) \right\Vert.
\end{aligned}
\end{equation}
We will proceed to use the eigenensemble assignment to bound the diagonal terms of Eq.~\eqref{eq:prop_only_if_main_goal}. 

There are two cases for what these errors can be: either they are upper bounded by some arbitrary $\delta>0$, or they are bounded by 2 due to their trace norm. Recalling that $q_{n,n'}(\delta)$ is precisely the probability that the error exceeds $\delta$, this implies that
\begin{equation} \label{eq:first_term_two_cases}
   \sum_{k,\overleftarrow{z}}p(\overleftarrow{z}|\mathbf{A}) p(k|\overleftarrow{z}) E_{n,n'}(k,\overleftarrow{z},\mathbf{A}) \leq 2  q_{n,n'}(\delta) + \delta \big[ 1 -  q_{n,n'}(\delta) \big].
\end{equation}

Now we consider a separate assignment to show that the off-diagonal terms can be bounded. This second assignment is defined as in Lem.~\ref{lem:approximateoffdiagonalannihilation}: $\ket{u^m_{k,k',\pm}}:=\frac{1}{\sqrt{2}}(\ket{\psi^m_k} \pm \ket{\psi^m_{k'}})$, and $\ket{v^m_{k,k',\pm}}:=\frac{1}{\sqrt{2}}(\ket{\psi^m_k} \pm i\ket{\psi^m_{k'}})$ for all $k\neq k'$. When $p(k|\overleftarrow{z})>0$ and $p(k'|\overleftarrow{z})>0$, these superpositions are members of the support of $\rho^m$. However, the contribution of a particular ensemble element to $q_{n,n'}(\delta)$ is constrained by its own weight with respect to $\rho^m$. The implication is that under per-realisation faithfulness, ensemble elements are permitted to behave poorly so long as their contribution to $\rho^m$ is sufficiently small. The sum of such contributions is controlled by the smallness of the coefficients $\sqrt{p(k|\overleftarrow{z}) p(k'|\overleftarrow{z})}$ directly, rather than by per-realisation faithfulness. Framing this complication quantitatively for a fixed history $\overleftarrow{z}$, define a set of ordered pairs of $(k,k')$ values which satisfy that $p(k|\overleftarrow{z}) p(k'|\overleftarrow{z})$ is large relative to some cutoff $\eta$ that we can vary,
\begin{equation}
    \mathfrak{S}_{\overleftarrow{z}}(\eta):= \big\{(k,k') \big\vert k \neq k', \quad p(k|\overleftarrow{z})>\eta \quad \text{and} \quad p(k'|\overleftarrow{z})>\eta \big\}.
\end{equation}
Even if one of $p(k|\overleftarrow{z})$ or $p(k'|\overleftarrow{z})$ is larger than $\eta$, it is still at most 1 due to being a probability. This implies that if $(k,k') \notin \mathfrak{S}_{\overleftarrow{z}}(\eta)$, then $p(k|\overleftarrow{z})p(k'|\overleftarrow{z})<\eta$. 

Now we seek to upper bound the contributions to the second term of Eq.~\eqref{eq:prop_only_if_main_goal} due to these small-weight eigenvalues $(k,k') \notin \mathfrak{S}_{\overleftarrow{z}}(\eta)$.
\begin{equation} \label{eq:prop_only_if_main_goal_small_eigenvalues}
\begin{aligned}
  & \sum_{(k,k') \notin \mathfrak{S}_{\overleftarrow{z}}(\eta) } \sqrt{p(k|\overleftarrow{z}) p(k'|\overleftarrow{z})}  \left\Vert\Phi_n^{m \rightarrow{f}} \big( \ket{\psi^{m}_k} \bra{\psi^{m}_{k'}} \big) \right\Vert \\
  \leq & \sum_{(k,k') \notin \mathfrak{S}_{\overleftarrow{z}}(\eta) } \sqrt{\eta} \\
  \leq & d_m(d_m-1) \sqrt{\eta}.
   \end{aligned}
\end{equation}
In the first inequality we used $p(k|\overleftarrow{z})p(k'|\overleftarrow{z})<\eta$ and $\left\Vert\Phi_n \big( \ket{\psi^{m}_k} \bra{\psi^{m}_{k'}} \big) \right\Vert \leq 1$. The latter inequality holds because $\Phi$ is a quantum channel, implying that it is contractive with respect to the trace norm (and the trace norm of $\ket{\psi^{m}_k} \bra{\psi^{m}_{k'}}$ is upper bounded by 1). In the second, we simply upper bound each probability by 1 and count the terms, noting that $\mathfrak{S}_{\overleftarrow{z}}(\eta)$ can at worst be empty. We now have means to drive this pathological contribution towards zero by making our arbitrary cutoff $\eta$ sufficiently small. 

Now we can treat the elements of our assignment that \emph{do} meaningfully contribute to $\rho^m$: $(k,k') \in \mathfrak{S}_{\overleftarrow{z}}(\eta)$, showing that per-realisation faithfulness does control the size of their off-diagonal contribution. We invoke the following argument only when $\mathfrak{S}_{\overleftarrow{z}}(\eta)$ is non-empty; when $\mathfrak{S}_{\overleftarrow{z}}(\eta)$ is empty, Eq.~\eqref{eq:prop_only_if_main_goal_small_eigenvalues} guarantees that they are controlled without the premise of per-realisation faithfulness, and the eigenensemble assignment suffices to control the diagonals. 

For any state in the assignment $\ket{w^m} \in \big\{ \ket{u^m_{k,k',+}},\ket{u^m_{k,k',-}},\ket{v^m_{k,k',+}},\ket{v^m_{k,k',-}} \big\}$ it holds that
\begin{equation}
    \rho^m-r \ket{w^m} \bra{w^m} \geq 0
\end{equation}
for some $r>0$. The largest such value is
\begin{equation}
    r_{\max} = \big(\braket{w|\rho^{-1}|w}\big)^{-1},
\end{equation}
where $\rho^{-1}$ refers to the inverse of $\rho^m$ restricted to its own support. Recalling that $\ket{w^m}$ are equal-weight sums of two eigenvectors, 
\begin{equation}
    \frac{1}{r_{\max}}=\frac{1}{2}\left( \frac{1}{p(k|\overleftarrow{z})} + \frac{1}{p(k'|\overleftarrow{z})} \right),
\end{equation}
and then 
\begin{equation}
    r_{\max}=\frac{2 p(k|\overleftarrow{z}) p(k'|\overleftarrow{z})}{ p(k|\overleftarrow{z}) + p(k'|\overleftarrow{z})} .
\end{equation}
Given this, $r_{\max} \geq \min\big\{ p(k|\overleftarrow{z}) ,p(k'|\overleftarrow{z})  \big\}$, and since $(k,k') \in \mathfrak{S}_{\overleftarrow{z}}(\eta)$ we have that $r_{\max} \geq \eta$. Thus,
\begin{equation} \label{eq:highweightnonneg}
    \rho^m-\eta \ket{w^m} \bra{w^m} \geq 0.
\end{equation}
For illustrative purposes, define $\sigma^m:= \frac{1}{1-\eta}\big( \rho^m-\eta \ket{w^m} \bra{w^m} \big)$. With this, we can write the ensemble
\begin{equation}
  \Big\{ \big( (1-\eta), \sigma^m \big), \big(\eta , \ket{w^m}\bra{w^m} \big) \Big\}.
\end{equation}
This allows us to conclude that for all $(k,k') \in \mathfrak{S}_{\overleftarrow{z}}(\eta)$, there exists an ensemble in which $\ket{w^m}$ contributes $\eta$.

Consider the further decomposition $\sigma^m=\sum_{j =1}^{\mathrm{rank}(\sigma)} s_j \ket{\nu^m_j}\bra{\nu^m_j}$, allowing us to write the \emph{pure-state} ensemble,
\begin{equation} \label{eq:assignmentforonew}
  \Big\{ \big( (1-\eta)s_j, \ket{\nu^m_j}\bra{\nu^m_j} \big)_{j =1}^{\mathrm{rank}(\sigma)}, \big(\eta , \ket{w^m}\bra{w^m} \big) \Big\}.
\end{equation} 
What we use is a further generalisation of this assignment: we turn it into an even mixture of all possible assignments of the form of Eq.~\eqref{eq:assignmentforonew}, and make explicit that each eigendecomposition gets its own version of $\ket{w^m} \in \big\{ \ket{u^m_{k,k',+}},\ket{u^m_{k,k',-}},\ket{v^m_{k,k',+}},\ket{v^m_{k,k',-}} \big\}$:
\begin{equation} \label{eq:assignmentformanyw}
  \left\{ \big( (1-\eta)s_j, \ket{\nu^m_j}\bra{\nu^m_j} \big)_{j =1}^{\mathrm{rank}(\bar{\sigma})}, \left(\frac{\eta}{L_{\overleftarrow{z}}} , \ket{w^m_{l|\overleftarrow{z}}}\bra{w^m_{l|\overleftarrow{z}}} \right)_{l=1}^{L_{\overleftarrow{z}}} \right\},
\end{equation} 
where $l$ is a flattening of the enumeration of all $\ket{w^m} \in \big\{ \ket{u^m_{k,k',+}},\ket{u^m_{k,k',-}},\ket{v^m_{k,k',+}},\ket{v^m_{k,k',-}} \big\}$ for all high eigenvalue weight pairs $(k,k') \in \mathfrak{S}_{\overleftarrow{z}}(\eta)$. The number of such $l$ values is $L_{\overleftarrow{z}}\leq 4d_m(d_m-1)$. Restricting $l$ to only high eigenvalue weight pairs allows Eq.~\eqref{eq:highweightnonneg} to hold in each case, implying that 
\begin{equation}
    \rho^m - \frac{\eta}{L_{\overleftarrow{z}}} \sum_{l=1}^{L_{\overleftarrow{z}}} \ket{w^m_{l|\overleftarrow{z}}}\bra{w^m_{l|\overleftarrow{z}}} \geq 0,
\end{equation}
which in turn means that Eq.~\eqref{eq:assignmentformanyw} is itself a proper ensemble of $\rho^m$. Note that this argument is only invoked if $L_{\overleftarrow{z}}>0$, i.e. $\mathfrak{S}_{\overleftarrow{z}}(\eta)$ is non-empty, so that diverging normalisation in Eq.~\eqref{eq:assignmentformanyw} never occurs. For histories where $L_{\overleftarrow{z}}=0$, we use the eigenensemble assignment instead. The corresponding alternative state in this ensemble is
\begin{equation}
    \bar{\sigma}_{\overleftarrow{z}}^{m}:=\frac{1}{1-\eta}\left( \rho^m - \frac{\eta}{L_{\overleftarrow{z}}} \sum_{l=1}^{L_{\overleftarrow{z}}} \ket{w^m_{l|\overleftarrow{z}}}\bra{w^m_{l|\overleftarrow{z}}} \right).
\end{equation}

We define a `bad set' of histories where the error threshold is exceeded for at least one $l$ value,
\begin{equation}
    B_{n,n'}(\delta,\eta) := \Big\{ \overleftarrow{z} : \exists l\leq L_{\overleftarrow{z}} \ \text{s.t.} \left\Vert \Phi_n^{m \rightarrow{f}} \big(\ket{w^m_{l|\overleftarrow{z}}}\bra{w^m_{l|\overleftarrow{z}}} \big)  - \mathbf{T}^{\mathbf{A}_{\overleftarrow{z}}} \right\Vert > \delta  \Big\}.
\end{equation}
Then we define the probability of being in this bad set as 
\begin{equation}
    \epsilon_{n,n'}(\delta,\eta) := \sum_{\overleftarrow{z} \in B_{n,n'}(\delta,\eta)} p(\overleftarrow{z}|\mathbf{A}).
\end{equation}
Now consider a sum over all violations for all values of $l$,
\begin{equation} \label{eq:all_l_values_that_are_bad}
       V_{n,n'}(\delta,\eta) := \sum_{\overleftarrow{z}} p(\overleftarrow{z}|\mathbf{A})   \sum_{l=1}^{L_{\overleftarrow{z}}} \texttt{I} \Big[ \left\Vert \Phi_n^{m \rightarrow{f}} \big(\ket{w^m_{l|\overleftarrow{z}}}\bra{w^m_{l|\overleftarrow{z}}} \big)  - \mathbf{T}^{\mathbf{A}_{\overleftarrow{z}}} \right\Vert > \delta \Big]  .
\end{equation}
Clearly, there can be more than one contributing term per history, so
\begin{equation} \label{eq:epsilon_V_bound}
    \epsilon_{n,n'}(\delta,\eta) \leq V_{n,n'}(\delta,\eta) .
\end{equation}
Furthermore, the probability of a violation is given by 
\begin{equation} \label{eq:probability_terms_L_sigma}
\begin{aligned}
        \tilde{q}_{n,n'}(\delta)=& \sum_{\overleftarrow{z}} p(\overleftarrow{z}|\mathbf{A}) \Bigg( \frac{\eta}{L_{\overleftarrow{z}}} \sum_{l=1}^{L_{\overleftarrow{z}}} \texttt{I} \Big[ \left\Vert \Phi_n^{m \rightarrow{f}} \big(\ket{w^m_{l|\overleftarrow{z}}}\bra{w^m_{l|\overleftarrow{z}}} \big)  - \mathbf{T}^{\mathbf{A}_{\overleftarrow{z}}} \right\Vert > \delta \Big] \\
        +& (1-\eta) \sum_{j =1}^{\mathrm{rank}(\bar{\sigma})} s_j  \texttt{I} \Big[ \left\Vert \Phi_n^{m \rightarrow{f}} \big( \ket{\nu^m_j} \bra{\nu^m_j} \big)  - \mathbf{T}^{\mathbf{A}_{\overleftarrow{z}}} \right\Vert > \delta \Big] \Bigg) ,
\end{aligned}
\end{equation}
where the tilde on $\tilde{q}_{n,n'}(\delta)$ is used to distinguish the probability of a violation under this assignment from that under the eigenensemble assignment, as was used in Eq.~\eqref{eq:first_term_two_cases}. Now, considering the fact that for all pasts $\overleftarrow{z}$, it holds that $\frac{L_{\overleftarrow{z}}}{\eta}
    \leq  \frac{4d_m(d_m-1)}{\eta}$, and ignoring the non-negative contribution of the $\overline{\sigma}$ terms in Eq.~\eqref{eq:probability_terms_L_sigma}, we can write,
\begin{equation} \label{eq:inequalityhierarchyconvergence}
    \epsilon_{n,n'}(\delta,\eta) \leq  V_{n,n'}(\delta,\eta)  
    \leq  \frac{4d_m(d_m-1)}{\eta} \tilde{q}_{n,n'}(\delta).
\end{equation}
The first  inequality is a re-statement of Eq.~\eqref{eq:epsilon_V_bound}. We note that the existence of a purely dimensional bound for the second inequality that works for all histories uses that $d_m<\infty$.

We have now established that there are three cases for the behaviour of the off-diagonal terms in Eq.~\eqref{eq:prop_only_if_main_goal}. Firstly, for a given $\overleftarrow{z}$, a value of $(k, k')$ is either in $\mathfrak{S}_{\overleftarrow{z}}(\eta)$ or not. Then, if $(k, k') \in \mathfrak{S}_{\overleftarrow{z}}(\eta)$, we need to further consider whether or not the history is in the bad set $B_{n,n'}(\delta,\eta)$.
\begin{enumerate}
    \item $(k, k') \in \mathfrak{S}_{\overleftarrow{z}}(\eta)$ and $\overleftarrow{z} \in B_{n,n'}(\delta,\eta)$: Magnitude $\left\Vert\Phi_n^{m \rightarrow{f}} \big( \ket{\psi^{m}_k} \bra{\psi^{m}_{k'}} \big) \right\Vert$ is bounded only by 1, the number of terms is at most $d_m(d_m-1)$, and probability of histories that fall under this category is $\epsilon_{n,n'}(\delta,\eta)$.
    \item $(k, k') \in \mathfrak{S}_{\overleftarrow{z}}(\eta)$ and $\overleftarrow{z} \notin B_{n,n'}(\delta,\eta)$: Lem.~\ref{lem:approximateoffdiagonalannihilation} applies, so magnitude $\left\Vert\Phi_n^{m \rightarrow{f}} \big( \ket{\psi^{m}_k} \bra{\psi^{m}_{k'}} \big) \right\Vert$ is bounded by $2\delta$, while probability of histories that fall under this category is $1-\epsilon_{n,n'}(\delta,\eta)$.
    \item $(k, k') \notin \mathfrak{S}_{\overleftarrow{z}}(\eta)$: Contribution is bounded by $d_m(d_m-1)\sqrt{\eta}$ as in Eq.~\eqref{eq:prop_only_if_main_goal_small_eigenvalues}.
\end{enumerate}

Thus, the bound of Eq.~\eqref{eq:prop_only_if_main_goal} resolves to
\begin{equation} \label{eq:putting_everything_together}
    \sum_{\overleftarrow{z}}p(\overleftarrow{z}|\mathbf{A}) \big\Vert \Delta_{n,n'}(\overleftarrow{z}) \big\Vert \leq  2  q_{n,n'}(\delta) + \delta \big[ 1 -  q_{n,n'}(\delta) \big] + d_m(d_m-1)\Big[ \epsilon_{n,n'}(\delta,\eta) + 2\delta\big(1-\epsilon_{n,n'}(\delta,\eta)\big) + \sqrt{\eta} \Big],
\end{equation}
where the first two terms arise from the diagonal case of Eq.~\eqref{eq:first_term_two_cases}.
Now we apply the limit $n=n'\rightarrow \infty$ and invoke per-realisation faithfulness,
\begin{equation} \label{eq:putting_everything_together_time_limit}
   \limsup_{n=n'\rightarrow \infty} \sum_{\overleftarrow{z}}p(\overleftarrow{z}|\mathbf{A}) \big\Vert \Delta_{n,n'}(\overleftarrow{z}) \big\Vert \leq   \delta + d_m(d_m-1)\Big[  2\delta + \sqrt{\eta} \Big].
\end{equation}
The LHS side of Eq.~\eqref{eq:putting_everything_together_time_limit} does not depend on $\delta$ or $\eta$, meaning that taking the infimum (in either order) leads to
\begin{equation} \label{eq:putting_everything_together_tolerance_limit}
  \limsup_{n=n'\rightarrow \infty} \sum_{\overleftarrow{z}}p(\overleftarrow{z}|\mathbf{A}) \big\Vert \Delta_{n,n'}(\overleftarrow{z}) \big\Vert \leq   0.
\end{equation}
Since the sum in Eq.~\eqref{eq:putting_everything_together_tolerance_limit} is non-negative for all $n,n'$, the limit superior being non-positive forces the corresponding limit to exist and converge to zero:
\begin{equation} \label{eq:putting_everything_together_weight_limit}
  \lim_{n=n'\rightarrow \infty} \sum_{\overleftarrow{z}}p(\overleftarrow{z}|\mathbf{A}) \big\Vert \Delta_{n,n'}(\overleftarrow{z}) \big\Vert =   0.
\end{equation}
We stress that $\eta$ and $\delta$ must be held fixed while taking the limit of $n=n'$. Firstly, taking a limit on $\eta$ before $n,n'$ would render the bound of Eq.~\eqref{eq:inequalityhierarchyconvergence} meaningless due to the factor of $\frac{1}{\eta}$ in the term enforcing the bound. Additionally, per-realisation faithfulness is that the probability of error exceeding $\delta$ goes to zero for a \emph{fixed} $\delta$ as $n=n' \rightarrow \infty$.

All that remains is to show that vanishing trace distance to a state's own marginals implies vanishing mutual information. Let us begin this process by finding an alternate expression for the relative entropy contained in oracular information. Since $I_{\mathbf{A}_{\overleftarrow{z}}}(f:m')$ is the relative entropy between $ \Upsilon^{fm'}_{\mathbf{A}_{\overleftarrow{z}}}$ and the product of its own marginals, we can write,
\begin{equation} 
I_{\Upsilon^{fm'}_{\mathbf{A}_{\overleftarrow{z}}}}(f:m')=I_{\Upsilon^{fm'}_{\mathbf{A}_{\overleftarrow{z}}}}(f:m')- I_{\Upsilon^{f}_{\mathbf{A}_{\overleftarrow{z}}} \otimes \Upsilon^{m'}_{\mathbf{A}_{\overleftarrow{z}}}}(f:m')=S_{\Upsilon^{f}_{\mathbf{A}_{\overleftarrow{z}}} \otimes \Upsilon^{m'}_{\mathbf{A}_{\overleftarrow{z}}}}(m'|f) -  S_{\Upsilon^{fm'}_{\mathbf{A}_{\overleftarrow{z}}}}(m'|f).
\end{equation}
Given that this is a non-negative quantity, we take the modulus without loss of generality, and invoke Proposition 4 of Ref.~\cite{Shirokov_2023},
\begin{equation} \label{eq:alicki-fannes}
    I_{\Upsilon^{fm'}_{\mathbf{A}_{\overleftarrow{z}}}}(f:m') \leq  \Vert \Delta_{n,n'}(\overleftarrow{z}) \Vert \ln(d_m) + g\left(\frac{1}{2} \Vert \Delta_{n,n'}(\overleftarrow{z}) \Vert \right),
\end{equation}
where $g(x)=(x+1) \ln(x+1) - x \ln(x)$. This is a bound of the Alicki-Fannes type~\cite{Winter_2016}, but the specific case where the state has no classicality constraints, and with one subsystem having finite dimension, while the other is unrestricted---which is what our finite $d_m$ assumption yields. Note that the bound of Ref.~\cite{Shirokov_2023} applies to the dimension of $m'$, not to $d_m$ directly. However, we use $m'$ that are defined by a minimal purification of the encoded memory state $\rho^m$, implying that $d_{m'} \leq \mathrm{rank}(\rho^m) \leq d_m$, so Eq.~\eqref{eq:alicki-fannes} still holds. Additionally, the bound of Ref.~\cite{Shirokov_2023} is stated in nats rather than bits, so we maintain that convention here for the sake of comparability to the original source. This has no impact on the proof itself, since we care only about convergence, not numerical value.

The first term of Eq.~\eqref{eq:alicki-fannes} only differs from $\Vert \Delta_{n,n'}(\overleftarrow{z}) \Vert$ by a scaling factor, so Eq.~\eqref{eq:putting_everything_together_weight_limit} implies that its contribution vanishes. For the second term, we observe that $g(x)$ is concave for non-negative values, and satisfies $\lim_{x \rightarrow 0^+}g(x)=0$. Then,
\begin{equation}
    \sum_{\overleftarrow{z}}p(\overleftarrow{z}|\mathbf{A}) g\left(\frac{1}{2} \Vert \Delta_{n,n'}(\overleftarrow{z}) \Vert \right) \leq  g\left(  \sum_{\overleftarrow{z}}p(\overleftarrow{z}|\mathbf{A}) \frac{1}{2} \Vert \Delta_{n,n'}(\overleftarrow{z}) \Vert \right)  \rightarrow 0.
\end{equation}
Since both terms of Eq.~\eqref{eq:alicki-fannes} vanish as $n=n' \rightarrow \infty$, $\lim_{n=n' \rightarrow \infty }\zeta_{n,n'}\big(\mathbf{A},\mathscr{M}(\mathbf{T})\big)=\zeta\big(\mathbf{A},\mathscr{M}(\mathbf{T})\big)=0$.
\end{proof}

The proof of Thm.~\ref{thm:predictioncorrespondence} is now merely a restatement of results.
\begin{proof}
    Combine the `if' direction of Prop.~\ref{prop:predictioncorrespondence_direction_1} with the `only if' direction of Prop.~\ref{prop:predictioncorrespondence_direction_2}.
\end{proof}

\section{Number of Linearly Independent Futures} \label{app:algorithm}
Given some set of futures $\mathsf{F}$, Alg.~\ref{alg:GS} approximately finds how many of the futures are actually linearly independent as Hermitian matrices. It does so by orthonormalising $\mathsf{F}$ into a set $\mathsf{B}$ (under specified tolerance), which defines a vector space $\mathrm{span}(\mathsf{B})=\mathsf{V}$. Specifically, $\mathsf{V}$ is a subspace of the space of Hermitian linear operators on the field of real numbers, with the same shape as elements of $\mathsf{F}$. We stress that $\mathsf{B}$ and $\mathsf{V}$ are only mathematical tools to characterise the structure of $\mathsf{F}$, and their elements need not all be physical.

\SetKwInput{KwInput}{Input}                
\SetKwInput{KwOutput}{Output}              

\begin{algorithm}[ht] 
\DontPrintSemicolon
  
  \KwInput{A set $\mathsf{F}=\{ \mathbf{F}_k \}_{k=1}^{K}$ of futures, which all have some arbitrary length $n$, and a tolerance $\xi$ for when to reject near linear dependences.} 
  \KwOutput{A basis $\mathsf{B}$ for the vector space $\mathsf{V}$ of Hermitian matrices approximating the span of $\mathsf{F}$ at tolerance $\xi$, and the size of that basis $D(\mathsf{F},\xi)=|\mathsf{B}|$.}

Set $\mathsf{B}=\emptyset$ and $i=1$.

\For{$k \in \{ 1, \dots , K \}$}
{
Compute
\begin{equation}
   \mathbf{R}_k=\mathbf{F}_k - \sum_{\mathbf{B}_l \in \mathsf{B}} \frac{\text{tr}\{ \mathbf{F}_k \mathbf{B}_l  \}}{\text{tr}\{ \mathbf{B}_l \mathbf{B}_l  \}}  \mathbf{B}_l.
\end{equation}

\If{$\text{\emph{tr}}\{ \mathbf{R}_k \mathbf{R}_k  \} > \xi$}
{

$\mathbf{B}_i \leftarrow \frac{\mathbf{R}_k}{\sqrt{\text{tr}\{ \mathbf{R}_k \mathbf{R}_k  \}}}$

Append $\mathbf{B}_i$ to $\mathsf{B}$.

$i \leftarrow i+1$.

}
}

Set $D(\mathsf{F},\xi)=|\mathsf{B}|=i-1$.

Return $\mathsf{B},D(\mathsf{F},\xi)$.

\caption{Gram-Schmidt on Futures}
\label{alg:GS}
\end{algorithm} 
Intuitively, Alg.~\ref{alg:GS} is iteratively taking candidate futures for given pasts, and for each one, asking: \emph{is this future causally redundant with respect to the existing set of futures?} This is the global quantum stand-in for the pairwise causal equivalence test that allows a classical $\epsilon$-machine to trim redundant memory states. 

Applying Alg.~\ref{alg:GS} to some set of futures $\mathsf{F}$ yields a finite-tolerance approximation $D(n,n',\xi)$. Note that $\mathsf{B}$ is a basis for $\mathsf{V}$, which need not consist of valid density matrices. We employ the Hilbert-Schmidt inner product in Alg.~\ref{alg:GS}. While Alg.~\ref{alg:GS} pertains to an arbitrary set of futures $\mathsf{F}$, it can be used to characterise a process. Additionally, any future $\mathbf{R} \in \mathsf{F}$ can be written (for $\xi=0$) as a real linear combination of the $D$ elements of $\mathsf{B}$
\begin{equation} \label{eq:sumoverbasisforfuturemorphs}
    \mathbf{R}= \sum_{l=1}^{|\mathsf{B}|} c_l \mathbf{B}_l.
\end{equation}

We move on from studying some arbitrary set of futures to studying a set of futures \emph{exhibited by a process}. Let $\mathbf{T}$ be a stationary quantum process. Consider the set of all possible histories $\mathbf{A}_{\overleftarrow{z}}$ given an informationally complete policy $\mathbf{A}$. Alg.~\ref{alg:NLDFM} takes each $\mathbf{A}_{\overleftarrow{z}}$ and computes its corresponding future, producing the set $\mathsf{F}=\{\mathbf{T}^{\mathbf{A}_{\overleftarrow{z}}}_{t:t+n-1} : \mathbf{A}_{\overleftarrow{z}} \in \mathbf{A},\ p(\overleftarrow{z}|\mathbf{A})>\chi\}$ whose span at $\chi=0$ contains the length $n$ futures of every length $n'$ history of every policy. We apply Alg.~\ref{alg:GS} as a subroutine, and the resultant $D$ is the number of linearly independent futures that process $\mathbf{T}$ can exhibit. 

\SetKwInput{KwInput}{Input}                
\SetKwInput{KwOutput}{Output}              

\begin{algorithm}[ht] 
\DontPrintSemicolon
  
  \KwInput{A quantum process $\mathbf{T}$, lengths $n,n'$ of futures and histories to consider, a tolerance $\xi$ for when to reject near linear dependences, and a tolerance $\chi$ for rejecting unlikely histories.} 
  \KwOutput{An approximation $D(n,n',\xi,\chi)$ of number of linearly independent futures $D$ of $\mathbf{T}$.}

Find an informationally complete policy $\mathbf{A}$ of length $n'$.

Compute the set $\mathsf{F}$ of length $n$ futures for all $\  \mathbf{A}_{\overleftarrow{z}} \in \mathbf{A}$ such that $p(\overleftarrow{z}|\mathbf{A})>\chi$.

Apply Alg.~\ref{alg:GS} to $\mathsf{F}$.

Return $D(n,n',\xi,\chi)$.

\caption{Number of Linearly Independent Futures}
\label{alg:NLDFM}
\end{algorithm} 

The raw algorithm outputs an approximation $D(n,n',\xi,\chi)$, and the true number of linearly independent futures $D$ that a process $\mathbf{T}$ exhibits is formally defined as the idealisation of $\xi=\chi=0$ and arbitrarily long pasts/futures, 
\begin{equation} \label{eq:definitionofD}
    D:= \lim_{n=n' \rightarrow \infty } D(n,n',0,0) = \lim_{n=n' \rightarrow \infty } |\mathsf{B}(n,n',0,0)|.
\end{equation}
Here we have explicitly written the dependence of $\mathsf{B}$ on $(n,n',\xi,\chi)$ that results from Alg.~\ref{alg:GS} and Alg.~\ref{alg:NLDFM}, and taken the limit. One should not expect that this limit always exists, because some processes have infinite memory. However, non-convergence is still constrained by the fact that shorter futures are partial traces of longer ones, implying that $D(n,n',0,0)$ is non-decreasing with $n$. Similarly, shorter histories are averages over longer histories, implying that $D(n,n',0,0)$ is also non-decreasing with $n'$. Then, convergence is guaranteed when including the possibility of infinity $D \in \mathbbm{N}\cup\{\infty\}$.

Of course, in practice, one cannot compute infinite pasts and futures, which will lead to errors when the process has large or infinite Markov order. Another practical consideration is that it is useful to accept a finite tolerance $\xi$ for when futures are considered to be linearly dependent. Similarly, rejecting unlikely histories by truncating at finite $\chi$ may be a valuable simplification as well. We also point out that longer histories and futures will inherently put more strain on $\chi$ and $\xi$ respectively. These approximations will generically result in $D(n,n',\xi,\chi) \leq D$.

\section{Proof of Prop.~\ref{prop:schmidt}} \label{app:choi_schmidt}
Here, we prove that the number of linearly independent futures $D$ of a stationary process is equal to the operator Schmidt rank of its Choi state about any past-future bipartition.
\begin{proof}
    We start by writing a tomographic representation (Ref.~\cite{anintroductiontooperationalquantumdynamics} Eq.~16) of the process, thinking of $\mathbf{T}_{t-n':t+n-1}$ as a map from history operators $\mathbf{A}_{\overleftarrow{z}}$ to the Choi state of the corresponding future. Without loss of generality, we pick a well-behaved policy to express this tomographic representation. This is because any history of any policy can be expressed as a linear combination of elements from any informationally complete policy, allowing a single such policy to be used to test the span of the futures of all possible histories. We require that the policy $\mathbf{A}$ is informationally complete, has a product form as in Eq.~\eqref{eq:prepareandmeasurepolicy}, and all its elements are linearly independent. This provides that $\mathbf{A}$ specifies a tomographic representation, that the past-future split is well-defined, and that its dual set is linearly independent, respectively. The corresponding expression is
    \begin{equation} \label{eq:newtomographic}
        \mathbf{T}_{t-n':t+n-1} =\frac{1}{d_s^{2n'}} \sum_{\overleftarrow{z}} p(\overleftarrow{z}|\mathbf{A}) \mathbf{D}^{*}_{\overleftarrow{z}_{t-n':t-1}} \otimes \mathbf{T}^{\mathbf{A}_{\overleftarrow{z}}}_{t:t+n-1},
    \end{equation}
    where $\mathbf{D}_{\overleftarrow{z}_{t-n':t-1}}$ is the dual of the history $\mathbf{A}_{\overleftarrow{z}_{t-n':t-1}}$. Now we fix a Hermitian matrix (not necessarily density matrix) basis for futures $\{ \mathbf{B}^{(i)}_{t:t+n-1} \}_{i=1}^{d_s^{4n}}$. Then, each future can be written as
    \begin{equation}
        \mathbf{T}^{\mathbf{A}_{\overleftarrow{z}}}_{t:t+n-1} = \sum_{i=1}^{d_s^{4n}} q_i(\overleftarrow{z}) \mathbf{B}^{(i)}_{t:t+n-1},
    \end{equation}
    where the coefficients $q_i(\overleftarrow{z})$ are real numbers such that $q_i(\overleftarrow{z})=0$ when $p(\overleftarrow{z}|\mathbf{A})=0$. Then, we can re-write Eq.~\eqref{eq:newtomographic} as
    \begin{equation} \label{eq:newtomographic2}
        \mathbf{T}_{t-n':t+n-1} =\frac{1}{d_s^{2n'}} \sum_{\overleftarrow{z}} p(\overleftarrow{z}|\mathbf{A}) \sum_{i=1}^{d_s^{4n}} q_i(\overleftarrow{z}) \mathbf{D}^{*}_{\overleftarrow{z}_{t-n':t-1}} \otimes \mathbf{B}^{(i)}_{t:t+n-1}.
    \end{equation}
    Let $D(n,n'):=D(n,n',0,0)$ and $R(n,n')$ be the finite-step approximations of $D$ and $R$ respectively at future and past lengths $n,n'$ respectively, with $\xi=\chi=0$. Let $Q$ be the matrix whose entries are $Q(\overleftarrow{z},i)=q_i(\overleftarrow{z})$. Then, the rank of $Q$---the dimension of the vector space spanned by its rows---is precisely the number of linearly independent futures $D(n,n')$. 

   Turning our attention to $R(n,n')$, we can write 
    \begin{equation} \label{eq:newtomographic3}
        \mathbf{T}_{t-n':t+n-1} = \sum_{i,\overleftarrow{z}} C(\overleftarrow{z},i) \mathbf{D}^{*}_{\overleftarrow{z}_{t-n':t-1}} \otimes \mathbf{B}^{(i)}_{t:t+n-1},
    \end{equation}
    where $C(\overleftarrow{z},i)=\frac{ p(\overleftarrow{z}|\mathbf{A}) q_i(\overleftarrow{z}) }{d_s^{2n'}}$. Since $q_i(\overleftarrow{z})=0$ when $p(\overleftarrow{z}|\mathbf{A})=0$, defining $C$ as the matrix whose entries are $C(\overleftarrow{z},i)$, we have that $\mathrm{rank}(C)=\mathrm{rank}(Q)=D(n,n')$. While Eq.~\eqref{eq:newtomographic3} is not in general an operator Schmidt decomposition, the past and future basis sets $\{ \mathbf{D}^{*}_{\overleftarrow{z}_{t-n':t-1}} \}$ and $\{ \mathbf{B}^{(i)}_{t:t+n-1} \}$ are both linearly independent, so the rank of the coefficient matrix $C$ is basis invariant. This implies that $\mathrm{rank}(C)=R(n,n')$, which in turn establishes that $R(n,n')=D(n,n')$.

    Finally, we take the limit of $n=n'\rightarrow \infty$, and $R(n,n')=D(n,n')$ holds for all values of $n=n'$. As mentioned in App.~\ref{app:algorithm}, $D(n,n')$ is monotonically non-decreasing in both $n$ and $n'$, so $D(n,n')$ converges when including infinity: $R=D \in \mathbbm{N} \cup \{\infty\}$.
\end{proof}

\section{Proof of Thm.~\ref{thm:probdistinct}} \label{supp:SPDproof}
In this section we provide the proof of Thm.~\ref{thm:probdistinct}, that strong probabilistic distinctness and reachability imply that a model is dimensionally minimal.
\begin{proof}
Thm.~\ref{thm:lbfm} provides a lower bound $\sqrt{D} \leq d$. We will show that strong probabilistic distinctness implies saturation of this lower bound $d_m=\sqrt{D}$, which in turn implies that the model must be minimal. Consider the vector space of Hermitian linear operators $\mathcal{L}(\mathcal{H}^{m})$ that act on the memory Hilbert space $\mathcal{H}^{m}$, which contains the density operators $\mathcal{D}(\mathcal{H}^{m})$. 

Consider some Hermitian matrix $X \in \mathcal{L}(\mathcal{H}^{m})$ (not necessarily a valid memory density matrix) in the kernel of the prediction function $\Phi$. Recall that our definition of $\Phi$ with no subscript refers to a family, not a specific function. As such, $X$ is in the intersection of the kernels of $\Phi_n$ for all future lengths $n$, i.e. $X$ is a vector such that $\Phi_n(X)=0$ for all $n$. We assume $X \neq 0$ as a contradiction premise, and apply the Jordan-Hahn decomposition: $X=X_{+}-X_{-}$, where $X_{\pm}:=\frac{\sqrt{X^{\dag}X}\pm X}{2}$. Then, $X_{+}\geq 0$ and $X_{-}\geq 0$. Furthermore, $X_{+} \neq X_{-}$ so long as $X \neq 0$, which is our premise. Since $\Phi$ is trace preserving, and $X$ is in its kernel, $X$ must be traceless, implying that both of $X_{+}$ and $X_{-}$ must have equal trace $\text{tr} \{X_{\pm} \}=\alpha$. This trace must be non-zero, since a positive semidefinite operator with zero trace is the zero operator, which would imply that $X_{+}=X_{-}=X=0$. Thus, $X_{\pm}/\alpha$ are distinct positive semidefinite unit trace Hermitian matrices, i.e. valid quantum states satisfying
\begin{equation}
    \Phi\left( \frac{X_{+}}{\alpha} \right)= \Phi\left( \frac{X_{-}}{\alpha} \right),
\end{equation}
violating strong probabilistic distinctness. Thus, we reject the contradiction premise, implying that the kernel of $\Phi$ is only the zero vector, meaning that $\Phi$ is injective on $\mathcal{L}(\mathcal{H}^{m})$.

$D= \lim_{n=n' \rightarrow \infty}D(n,n')$ is the dimension of the span of the futures yielded by all pasts in the limit of infinite pasts and futures. We invoke the theorem premise that there exists a policy $\mathbf{A}$ where the encoded memory states $\{ e(\mathbf{A}_{\overleftarrow{z}}) \}_{\overleftarrow{z} \in \text{W}(\overleftarrow{\mathsf{Z}})}$ of all pasts span $\mathcal{L}(\mathcal{H}^{m})$. Due to linearity of the un-normalised encoding function $\hat{e}$ (Eq.~\eqref{eq:dilationencodingfunction} without dividing by the probability), the encoded states of any informationally complete policy span those of all other policies for any given value of $n'$. As such, reachability under the policy provided by the theorem premise implies reachability under any informationally complete policy. 

We now establish that $D(n,n')$ is a monotonically non-decreasing function of $n$ and $n'$. Recall that $\omega^m$ is the state that the model is initialised in at the beginning of every experiment. Stationarity and faithfulness jointly imply that evolution under the background $\mathbf{K}$ for one step must leave the predictions of the model unchanged. By injectivity of $\Phi$---as established above---$\omega^m$ itself must be invariant under this evolution under $\mathbf{K}$. This then implies that appending a step of evolution under the background $\mathbf{K}_{t-n'-1}$ prior to some finite history $\mathbf{A}_{\overleftarrow{z}_{t-n':t-1}}$ will not alter which memory state is reached, i.e. $e(\mathbf{K}_{t-n'-1},\mathbf{A}_{\overleftarrow{z}_{t-n':t-1}})=e(\mathbf{A}_{\overleftarrow{z}_{t-n':t-1}})$. Since the background's single-step element lies in the span of the elements of a single-step informationally complete policy, $\mathbf{K}_{t-n'-1},\mathbf{A}_{\overleftarrow{z}_{t-n':t-1}}$ is contained in the span of length $n'+1$ histories, and we can conclude that the span of the memory states reachable under it---which is the span of the memory states reachable under length $n'$ histories---is contained in those reachable under length $n'+1$ histories. Thus, $D(n,n')$ is a monotonically non-decreasing function of $n'$. We can see that it is a monotonically non-decreasing function of $n$ simply by the fact that $\Phi_{n}(\cdot)$ is a marginal of $\Phi_{n+1}(\cdot)$ for any input. Given monotonicity and boundedness above by $d_m^2$, as well as faithfulness, the limit 
\begin{equation}
    D= \lim_{n=n' \rightarrow \infty} \mathrm{dim} \Bigg[ \mathrm{span} \Big[ \Phi_n \circ e_{n'}\big( \{ \mathbf{A}_{\overleftarrow{z}} \}_{\overleftarrow{z}} \big) \Big] \Bigg]
\end{equation}
exists.

Now we establish what the limit is equal to. The span of all encoded states is a non-decreasing sequence of subspaces of the $d_m^2$-dimensional space $\mathcal{L}(\mathcal{H}^{m})$, so the sequence is eventually constant. Reachability then implies that there is some value $n^{*}$ of $n'$ such that for all $n'\geq n^{*}$, these subspaces have dimension $d_m^2$. Similarly, the kernels of each $\Phi_n$ form a non-increasing sequence of subspaces due to each length $n$ future being the marginal of the corresponding length $n+1$ future. Thus, due to injectivity and finite dimension, the sequence is eventually constant and equal to the zero matrix (as a subspace) itself. Hence, there is some value $n^{\star}$ of $n$ such that for all $n \geq n^{\star}$, $\Phi_n$ is injective. Then, by faithfulness, for all $n=n' \geq \max \{ n^{*},n^{\star}\}$, it holds that $D=D(n,n')=d_m^2$. Since Thm.~\ref{thm:lbfm} shows that $ \sqrt{D} \leq d$ for all models, the given model must be minimal.

\end{proof}

\twocolumngrid
\bibliography{refs}

\end{document}